\documentclass[12pt]{article}
\usepackage{float}
\usepackage{makeidx}
\usepackage{amssymb}
\usepackage{amsmath}
\usepackage{amsfonts}
\usepackage{inputenc}
\usepackage{version}
\usepackage{rotating}
\usepackage{lscape}
\usepackage[nodisplayskipstretch]{setspace}
\usepackage{geometry}
\usepackage[authoryear]{natbib}
\usepackage[toc,page,title,titletoc,header]{appendix}
\usepackage{color}
\usepackage{epsfig}
\usepackage{hyperref}

\graphicspath{{C:/HetroARfig/}}
\newtheorem{theorem}{Theorem}
\numberwithin{equation}{section}

\newtheorem{lemma}{Lemma}

\newtheorem{assumption}{Assumption}

\newtheorem{proposition}{Proposition}
\newtheorem{remark}{Remark}

\input{tcilatex}
\begin{document}

\title{Heterogeneous Autoregressions in Short $T$ Panel Data Models\thanks{%
We are grateful to two anonymous reviewers for most helpful and constructive
comments. We have also benefited greatly from helpful comments and
suggestions by Alexander Chudik, Ron Smith and Hayun Song.}}
\author{M. Hashem Pesaran\thanks{%
Department of Economics, University of Southern California, and Trinity
College, University of Cambridge. Email: \texttt{pesaran@usc.edu}.} \and %
Liying Yang\thanks{%
The working paper version of this paper was completed when Liying Yang was a
Ph.D. student at Department of Economics, University of Southern California.
She is now a postdoctoral research fellow at the Sauder School of Business
at the University of British Columbia. Email: \texttt{%
liying.yang@sauder.ubc.ca}.} }
\maketitle

\begin{abstract}
{\setstretch{1} This paper considers a first-order autoregressive panel data
model with individual-specific effects and heterogeneous autoregressive
coefficients defined on the interval $(-1,1]$, thus allowing for some of the
individual processes to have unit roots. It proposes estimators for the
moments of the cross-sectional distribution of the autoregressive (AR)
coefficients, assuming a random coefficient model for the autoregressive
coefficients without imposing any restrictions on the fixed effects. It is
shown the standard generalized method of moments estimators obtained under
homogeneous slopes are biased. Small sample properties of the proposed
estimators are investigated by Monte Carlo experiments and compared with a
number of alternatives, both under homogeneous and heterogeneous slopes. It
is found that a simple moment estimator of the mean of heterogeneous AR
coefficients performs very well even for moderate sample sizes, but to
reliably estimate the variance of AR coefficients much larger samples are
required. It is also required that the true value of this variance is not
too close to zero. The utility of the heterogeneous approach is illustrated
in the case of earnings dynamics. }


\end{abstract}

\thispagestyle{empty}

\newpage \setcounter{page}{1}

\section{Introduction}

\doublespacing%

The importance of cross-sectional heterogeneity in panel regressions is
becoming increasingly recognized in the literature. When the time dimension
of the panel, $T$, is short, significant advances have been made in the case
of random coefficient models with strictly exogenous regressors, for
example, \cite{Chamberlain1992}, \cite{Wooldridge2005}, and \cite%
{GrahamPowell2012}. A trimmed version of the mean group estimator proposed
by \cite{PesaranSmith1995} can also be applied to ultra short $T$ panels
when the regressors are strictly exogenous. See \cite{PesaranYang2024}. In
contrast, there are only a few papers that consider the estimation of
heterogeneous dynamic panels when the time dimension is short.

There are some limitations to applying existing estimation methods to such
heterogeneous short $T$ dynamic panels. The generalized method of moments
(GMM) estimators applied after first differencing by \cite%
{AndersonHsiao1981,AndersonHsiao1982}, \cite{ArellanoBond1991}, \cite%
{BlundellBond1998}, and \cite{ChudikPesaran2021}, allow for intercept
heterogeneity but not for possible heterogeneity in the autoregressive (AR)
coefficients, and as shown in this paper, can lead to biased estimates and
distorted inference. \cite{GuKoenker2017} and \cite{Liu2023} consider the
estimation of panel AR(1) models with exogenous regressors using Bayesian
techniques. While they assume random coefficients on strictly exogenous
regressors, they still impose homogeneity on the AR coefficients. The mean
group estimator and the hierarchical Bayesian estimator proposed by \cite%
{HsiaoEtal1999} allow for heterogeneity but require that $T$ is reasonably
large relative to the cross section dimension, $n$.

For moderate values of $T$, analytical, Bootstrap and Jackknife bias
correction approaches have also been proposed to deal with the small sample
bias of the mean group and other related estimators. See, for example, \cite%
{PesaranZhao1999}, \cite{OkuiYanagi2019} and \cite{OkuiYanagi2020}. Even
with bias corrections, $n$ cannot be too large compared with $T$, since a
valid inference based on the asymptotic distribution often requires $%
nT^{-c}\rightarrow 0,$ for some constant $c>2$. In short, none of the above
approaches are appropriate and can lead to seriously biased estimates and
distorted inference when $T$ is small and fixed as $n\rightarrow \infty $.
Nonetheless, heterogeneity in dynamics can play an important role in many
empirical studies using short $T$ panel data models. Examples include   
earnings dynamics studied by \cite{MeghirPistaferri2004}, unemployment
dynamics by \cite{BrowningCarro2014}, and firm's growth by \cite{Liu2023}

This paper considers a relatively simple panel AR(1) model, but allows for
both individual fixed effects and heterogeneous AR coefficients, $\phi _{i}$%
, where some of the individual processes, $\{y_{it}\},$ could have unit
roots, $\phi _{i}=1$. We eliminate the fixed effects by first differencing, $%
\Delta y_{it}=y_{it}-y_{i,t-1}$, and establish conditions under which the
mean and variance of $\phi _{i}$ can be identified from the autocovariances
of $\Delta y_{it}$, averaged over $i$. We show that existing GMM estimators
of $E(\phi _{i})=\mu _{\phi }$ are asymptotically biased, and derive
analytical expressions for their bias in simple cases. We then propose
estimators for the moments of $\phi _{i}$, in particular, $E(\phi _{i})$ and 
$E(\phi _{i}^{2})$, using cross-sectional averages of the autocorrelation
coefficients of the first differences. In terms of the estimation approach,
the most relevant paper to ours is by \cite{Robinson1978}, who considered a
random coefficient AR(1) model without fixed effects. Assuming the
\textquotedblleft usual" stationary conditions, he proposed identifying the
moments of $\phi _{i}$ as functions of autocovariances of $y_{it}$.

In particular, we propose two new estimators for the moments $\theta
_{s}=E(\phi _{i}^{s})$ for $s=1,2,...,T-3$. A relatively simple estimator
based on autocorrelations of first differences, denoted by FDAC, and a
generalized method of moments (GMM) estimator based on autocovariances of
first differences, which we denote by HetroGMM. We also consider estimation
of $Var(\phi _{i})=\sigma _{\phi }^{2}=\theta _{2}-\theta _{1}^{2}$, when
the true value of $\sigma _{\phi }^{2}$ is not too close to zero. 
We do not make any assumptions about the fixed effects and allow them to
have arbitrary correlations with $\phi _{i}$, but require the underlying
AR(1) processes to be stationary after first differencing and assume $\phi
_{i}$ and the error variances are independently distributed. It is possible
to extend our analysis to higher-order panel AR processes and dynamic panels
with exogenous regressors. However, these important extensions are outside
the scope of the present paper.

We compare FDAC and HetroGMM estimators to a kernel-weighted likelihood
estimator proposed by \cite{MavroeidisEtal2015}, MSW. Assuming independently
distributed Gaussian errors with cross-sectional heteroskedasticity, MSW
show that the unknown distribution of heterogeneous coefficients can be
identified, provided the linear operator that maps the unknown distribution
to the joint distribution of data is complete (or \textquotedblleft
invertible"). They provide an estimation algorithm for the parametric
version of their estimator assuming the heterogeneous coefficients,
including the intercepts and $\phi _{i}$, follow a multivariate normal
distribution. The estimation algorithm becomes computationally very
demanding if the parametric assumption about the distribution of $\phi _{i}$
is relaxed.

We investigate small sample properties of FDAC and HetroGMM estimators using
Monte Carlo (MC) experiments. The simulations show that the relatively
simple FDAC estimator performs better than the HetroGMM estimator uniformly
across different sample sizes, and is robust to non-Gaussian errors and
conditional error heteroskedasticity.

We also compare the small sample properties of the FDAC estimator of $\mu
_{\phi }$ with several GMM estimators proposed in the literature for
homogenous AR panels, including the popular \cite{ArellanoBond1991}, AB, and 
\cite{BlundellBond1998}, BB, estimators. We refer to these as HomoGMM
estimators, to be distinguished from the HetroGMM estimator proposed in this
paper. The simulation results confirm the neglected heterogeneity bias of
the HomoGMM estimators, and show that the FDAC estimator of $\mu _{\phi }$
performs well for all values of $T=4,6,10$ and $n=100,1,000$ and $5,000$, so
long as the underlying processes are stationary after first differencing.
This is true for bias, root mean square errors, and size. Both FDAC and
HetroGMM estimators are robust to the presence of unit roots and
non-Gaussian errors, but can be subject to bias and size distortions if the
distribution of the initial values, $y_{i0}$, significantly depart from
stationarity. Similar comparative outcomes are also obtained when estimating 
$\sigma _{\phi }^{2}$, except that much larger sample sizes ($n$ and/or $T$)
are required for reliable estimation and inference. In addition, it is
important that the true value of $\sigma _{\phi }^{2}$ is not too close to
the boundary value of $0$. When $n$ and $T$ are not sufficiently large,
estimates of $\sigma _{\phi }^{2}$ obtained using the plugging estimator, $%
\hat{\sigma}_{\phi }^{2}=\hat{\theta}_{2}-\hat{\theta}_{1}^{2}$, can be
negative. This occurs with a high frequency when $n=100$, and $T=5$. The
occurrence of negative estimates declines rapidly when $T=10$ and $n\geq
1,000$.

Using Monte Carlo experiments we also provide a limited comparison of the
MSW and FDAC estimators of $\mu_{\phi}$, and find that in general, the MSW
estimator does not have satisfactory small-sample performance under the data
generating process in the paper. As the MSW estimator depends on the assumed
Gaussian distribution of $\phi _{i}$, it can be severely biased with
uniformly and categorically distributed $\phi _{i}$ that we consider in our
MC experiments.

Finally, we provide an empirical application using five and ten yearly
samples from the Panel Study of Income Dynamics (PSID) dataset over the 1976--1995 period to estimate the
persistence of real earnings. To this end, we extend the basic panel AR(1)
model to allow for linear trends. Following the empirical literature we
report estimates for three educational categories (high school dropouts,
high school graduates and college graduates) and all three categories
combined. We find comparable estimates for the linear trend coefficients
across sub-periods and educational categories, around 2 per cent per annum.
The FDAC estimates of mean persistence $(\mu _{\phi })$ for the sub-periods
1991--1995 and 1986--1995 fall in the range of $0.570-0.734$, and tend to
rise with the level of educational attainment, with college graduates
showing the highest degree of persistence. No such patterns are observed for
other estimates, which are around $0.3,0.9$ and $0.41$ for the AB, BB and
MSW estimators, respectively. The FDAC estimates of $\sigma _{\phi }^{2}$
for all three categories combined are statistically significant and are
given by $0.100$ $(0.042)$ and $0.129$ $(0.023)$ for the sub-periods
1991--1995 ($n=1,366$) and 1986--1995 $(n=1,139)$, respectively, providing
further evidence of heterogeneity in real earnings persistence.

The rest of the paper is set out as follows. Section \ref{model} sets out
the model and assumptions. Section \ref{AutoDif} derives the autocovariances
of the first differences, $\Delta y_{it}=y_{it}-y_{i,t-1}$, and establishes
conditions under which they are stationary. Section \ref{bias} shows that
the HomoGMM estimators are biased in the heterogeneous panel AR(1) model.
Section \ref{moments} establishes conditions under which the moments of $%
\phi _{i}$ can be identified from the autocorrelation functions of first
differences. Section \ref{estimation} proposes FDAC and HetroGMM estimators
of the moments of $\phi _{i}$. The respective asymptotic distributions are
also derived. Section \ref{simulation} evaluates the performance of FDAC,
HetroGMM, HomoGMM, and MSW estimators by Monte Carlo simulations. Section %
\ref{application} presents the empirical application, and Section \ref%
{conclusion} concludes. Some of the mathematical derivations, Monte Carlo
evidence and additional empirical results are provided in an online
supplement.

\vspace{-3ex}

\section{Model and assumptions\label{model}}

\vspace{-1.5ex}

We consider the following first-order autoregressive panel data model 
\vspace{-1.5ex} 
\begin{equation}
y_{it}=\alpha _{i}+\phi _{i}y_{i,t-1}+u_{it},\text{ for }i=1,2,...,n,
\label{Par1}
\end{equation}%
where the fixed effects, $\alpha _{i}$, are restricted, $\alpha _{i}=\mu
_{i}\left( 1-\phi _{i}\right) $. This restriction is necessary for $y_{it}$
to have a fixed mean irrespective of whether $\phi _{i}=1$ or $\left\vert
\phi _{i}\right\vert <1$. If $\alpha _{i}$ is unrestricted, a linear trend
is introduced in $y_{it}$ when $\phi _{i}=1$. The restriction on $\alpha
_{i} $ is not binding when $\left\vert \phi _{i}\right\vert <1$. We impose
the restriction since we will be considering a mixture of processes with and
without unit roots. With $\alpha _{i}=\mu _{i}(1-\phi _{i})$, (\ref{Par1})
can be written equivalently as \vspace{-1.5ex} 
\begin{equation}
y_{it}-\mu _{i}=\phi _{i}\left( y_{i,t-1}-\mu _{i}\right) +u_{it},\text{ for 
}i=1,2,...,n.  \label{Par2}
\end{equation}

Suppose that $y_{it}$ is generated starting at time $t=-M_{i}\leq 0$ with
the initial value, $y_{i,-M_{i}}$. We assume observations on all the $n$
units are available over the periods $t=1,2,3,...,T$, yielding a total of $%
nT $ observations $\left\{ y_{i1},y_{i2},...,y_{iT}\text{, }%
i=1,2,...,n\right\} $. The parameters of interest are first and higher order
moments of $\phi _{i}$, which we denote by $\theta _{s}=E(\phi _{i}^{s})$, $%
s=1,2,...,T-2$. The key feature of our analysis is to allow for a high
degree of parameter heterogeneity when $T$ is short as $n\rightarrow \infty $%
. We allow $\phi _{i}$ to take any values in the non-explosive interval $%
[-1+\epsilon ,1]$ for some $\epsilon > 0$, which includes the unit root case, 
$\phi _{i}=1$ for some of the units, but rules out a negative unit root,
namely it is required that $\inf_{i}(1+\phi _{i})>0$. We are able to
accommodate distributions of $\phi _{i}$ with a non-zero mass on $\phi _{i}=1
$, by basing our estimation of $\theta _{s}$ on autocorrelations of first
differences, $\Delta y_{it}=y_{it}-y_{i,t-1}$, rather than the
autocovariances of $y_{it}$ considered by \cite{Robinson1978}. As examples,
we consider a uniform distribution of $\phi _{i}$ defined over the interval $%
(-1,1-\epsilon ]$ with $\epsilon \geq 0$, and a categorical distribution where $%
\phi _{i}$ takes two values, $\phi _{H}$ (high) and $\phi _{L}$ (low), with
probabilities $(1-\pi )$ and $\pi $, respectively. The unit root case arises
when $\epsilon =0$ (for the uniform distribution), and $\phi _{H}=1$ with $%
0<\phi _{L}<1$ (for the categorical distribution). Our analysis does not
allow for a negative unit root, namely when $\phi _{i}=-1$.

The key identification assumption is the stationarity of the first
differences. First differencing of (\ref{Par1}) eliminates the fixed
effects, $\alpha _{i}=\mu _{i}(1-\phi _{i}),$ but does not remove the
effects of initial values, $y_{i,-M_{i}}$, on the first differences when $T$
is small. Under slope heterogeneity, the effects of initial values on first
differences do not vanish for processes whose $\phi _{i}$ falls in the
stable region, $-1<\phi _{i}<1$, unless they are all initialized at a
distant past, namely only if $M_{i}\rightarrow \infty $, otherwise the
realized values $y_{it}$ and/or their first differences $\left\{ \Delta
y_{it}\text{, for }t=2,3,...,T\right\} $ will depend on $y_{i,-M_{i}}-\mu
_{i}$. Including the observations $y_{i0}$ amongst the realizations does not
resolve the problem, since we move one period backward and the distribution
of $y_{i,-1}$ must still be specified and so on. For processes with unit
roots, $\phi _{i}=1$, we have $\Delta y_{it}=u_{it}$, and initialization
will not be an issue, at least not for the unit-root AR(1) process.

To accommodate the possible mixture of stationary and unit-root processes
and achieve identification of the moments of $\phi _{i}$, we make the
following assumptions regarding unit-specific parameters, $\boldsymbol{\psi }%
_{i}=(\mu _{i},\phi _{i},\sigma _{i}^{2})^{\prime }$, the error terms, $%
u_{it}$, and the initial value deviations, $y_{i,-M_{i}}-\mu _{i}$ for $%
i=1,2,...,n$, where $\sigma _{i}^{2}=Var(u_{it})$.

\begin{assumption}
\label{fe} (individual effects) The individual specific means, $\mu _{i}$,
are bounded, $sup_{i}\left\vert \mu _{i}\right\vert <C$.
\end{assumption}

\begin{assumption}
\label{errors}(errors) Conditional on $\boldsymbol{\psi }_{i}=(\mu _{i},\phi
_{i},\sigma _{i}^{2})^{\prime }$, the errors, $u_{it}$, are
cross-sectionally and serially independent over $i$ and $t$, with zero
means, $E(u_{it}^{2})=\sigma _{i}^{2}$, and $\sup_{i,t}$E$\left\vert
u_{it}\right\vert ^{4}<C<\infty $.
\end{assumption}

\begin{assumption}
\label{error variances}(error variances) (a) The error variances, $\sigma
_{i}^{2}$, are independent draws from a common probability distribution such
that $E\left( \sigma _{i}^{2}\right) =\sigma ^{2}$, where $0<c<\sigma
_{i}^{2}$, and $\sigma ^{2}<C<\infty $. (b) $\sigma _{i}^{2}$ are
distributed independently of $\phi _{i}$.
\end{assumption}

\begin{assumption}
\label{hetro} (autoregressive coefficients) (a) The autoregressive
coefficients, $\phi _{i}$, for $i=1,2,...,n$, are independent draws from a
common probability distribution, defined on the closed interval $\phi
_{i}\in \lbrack -1+\epsilon ,1]$, for some $\epsilon >0$, with mean E$\left(
\phi _{i}\right) =\mu _{\phi }$ and variance Var$(\phi _{i})=\sigma _{\phi
}^{2}\geq 0$.
\end{assumption}

\begin{assumption}
\label{initial} (initialization) The process $\left\{ y_{it}\right\} $ is
initialized with $y_{i,-M_{i}}$, where $M_{i}\in 
\mathbb{N}
=\left\{ 0,1,2,...\right\} $, and $y_{i,-M_{i}}-\mu _{i}$ is given and
bounded, $sup_{i}\left\vert y_{i,-M_{i}}-\mu _{i}\right\vert <C$.
\end{assumption}

Assumption \ref{fe} imposes minimal restrictions on $\mu _{i}$ or on the
fixed effects $\alpha _{i}$ for $\left\vert \phi _{i}\right\vert <c<1$. But
as noted earlier, to ensure that $y_{it}$ is not subject to a drift, as it
is standard in the unit root literature, $\alpha _{i}$ is set to $0$ when $%
\phi _{i}=1$. Assumptions \ref{errors} and \ref{error variances} are
standard in the literature on short $T$ dynamic panels. They allow for
cross-sectional as well as conditional time series heteroskedasticity, such
as GARCH effects, but rule out unconditional time series heteroskedasticity.
Denoting the available information at time $t-1$ by $\mathcal{I}_{i,t-1}$, $%
E(u_{it}^{2}\left\vert \mathcal{I}_{i,t-1}\right.)$ could be time-varying,
so long as $E(u_{it}^{2})=\sigma _{i}^{2}$ as required by Assumption \ref%
{errors}.

Assumptions \ref{hetro} and \ref{initial} ensure that $\Delta y_{it}$ is
covariance stationary if $M_{i}\rightarrow \infty $, without requiring $%
y_{it}$ to be stationary for all $n$ units in the panel.

\vspace{-3ex}

\section{Autocovariances of first differences\label{AutoDif}}

\vspace{-1.5ex}

Before setting out our approach to the identification of $\theta _{s}=E(\phi
_{i}^{s}),$ we need to derive expressions for the autocovariances of $\Delta
y_{it}$. Given the available data and after first differencing (\ref{Par1}),
we have 
\begin{equation}
\Delta y_{it}=\phi _{i}\Delta y_{i,t-1}+\Delta u_{it},\text{ for }%
t=2,3,...,T.  \label{Fdif}
\end{equation}%
Also setting $t=1$ and using (\ref{Par2}) we obtain 
\begin{equation}
\Delta y_{i1}=-(1-\phi _{i})\left( y_{i0}-\mu _{i}\right) +u_{i1}.
\label{Dif1}
\end{equation}%
Iterating (\ref{Fdif}) forward from $t=2$ and using the above expression for 
$\Delta y_{i1}$, we obtain 
\begin{equation}
\Delta y_{it}=u_{it}-(1-\phi _{i})\sum_{\ell =1}^{t-1}\phi _{i}^{\ell
-1}u_{i,t-\ell }-\phi _{i}^{t-1}(1-\phi _{i})\left( y_{i0}-\mu _{i}\right) .
\label{SdyG}
\end{equation}%
It is clear that in general, $\Delta y_{it}$ depends on $y_{i0}-\mu _{i}$,
and Assumption \ref{initial} is required if we are to eliminate the impact
of initial values on the autocovariances of $\Delta y_{it}$. Iterating
equation (\ref{Par2}) forward from $y_{i,-M_{i}}$ to $t=0$ we have 
\begin{equation}
y_{i0}-\mu _{i}=\phi _{i}^{M_{i}}\left( y_{i,-M_{i}}-\mu _{i}\right)
+\sum_{\ell =0}^{M_{i}-1}\phi _{i}^{\ell }u_{i,-\ell }.  \label{IniVal}
\end{equation}%
Substituting $y_{i0}-\mu _{i}$ from (\ref{IniVal}) in (\ref{SdyG}) now yields%
\vspace{-1ex} 
\begin{equation}
\Delta y_{it}=u_{it}-(1-\phi _{i})\sum_{\ell =1}^{M_{i}+t-1}\phi _{i}^{\ell
-1}u_{i,t-\ell }+\phi _{i}^{t}R_{i}\left( y_{i,-M_{i}}\right) ,\text{ for }%
t=2,3,...,T,  \label{DyitG}
\end{equation}%
where $R_{i}\left( y_{i,-M_{i}}\right) =-\phi _{i}^{M_{i}-1}(1-\phi
_{i})\left( y_{i,-M_{i}}-\mu _{i}\right) $. For a fixed $T$, the remainder
term, $R_{i}$, does not vanish unless $M_{i}\rightarrow \infty $. Note that
under Assumption \ref{initial} $\sup_{i}\left\vert y_{i,-M_{i}}-\mu
_{i}\right\vert <C$, and $\left\vert R_{i}\left( y_{i,-M_{i}}\right)
\right\vert \leq \left\vert \phi _{i}\right\vert ^{M_{i}-1}\left\vert 1-\phi
_{i}\right\vert \left\vert y_{i,-M_{i}}-\mu _{i}\right\vert \leq C\left\vert
\phi _{i}\right\vert ^{M_{i}-1}\left\vert 1-\phi _{i}\right\vert ,$ and $%
\left\vert R_{i}\left( y_{i,-M_{i}}\right) \right\vert \rightarrow 0$, for 
\textit{all} $i$ (irrespective of whether $\phi _{i}=1$ or $\left\vert \phi
_{i}\right\vert <1$), \textit{if and only if} $M_{i}\rightarrow \infty $.
Under this condition \vspace{-1ex} 
\begin{equation}
\Delta y_{it}=u_{it}-(1-\phi _{i})\sum_{\ell =1}^{\infty }\phi _{i}^{\ell
-1}u_{i,t-\ell },  \label{Sdy1}
\end{equation}%
and the available first differences, $\Delta y_{it}$ for $t=2,3,...,T,$ do
not depend on $y_{i0}$, and can be used to derive expressions for $\gamma
_{\Delta }(h)=E\left( \Delta y_{it}\Delta y_{i,t-h}\right) $\ for $%
h=0,1,...,T-2$. But first, we need to establish that these autocovariances
do exist, particularly given that we allow for some $y_{it}$
processes to have unit roots. This requirement is easily established when
the distribution of $\phi _{i}$ is categorical. In this case we have \vspace{%
-1ex} 
\begin{equation*}
\gamma _{\Delta }(h)=\pi E\left( \Delta y_{it}\Delta y_{i,t-h}\left\vert
\left\vert \phi _{i}\right\vert <c<1\right. \right) +(1-\pi )E\left( \Delta
y_{it}\Delta y_{i,t-h}\left\vert \phi _{i}=1\right. \right) ,
\end{equation*}%
where $0<\pi \leq 1$. By application of Minkowski's inequality to (\ref{Sdy1}%
) we have (for $p\geq 1$) \vspace{-1.5ex} 
\begin{equation*}
\left\Vert \Delta y_{it}\right\Vert _{p}\leq \left\Vert u_{it}\right\Vert
_{p}+\sum_{\ell =1}^{\infty }\left[ \left\Vert \phi _{i}^{\ell
-1}\right\Vert _{p}+\left\Vert \phi _{i}^{\ell }\right\Vert _{p}\right]
\left\Vert u_{i,t-\ell }\right\Vert _{p},
\end{equation*}%
where $\left\Vert \Delta y_{it}\right\Vert _{p}=E\left( \left\vert \Delta
y_{it}\right\vert ^{p}\right) ^{1/p}$. By Assumption \ref{errors} $%
sup_{i,t}\left\Vert u_{it}\right\Vert _{4}<C$, and for units with $%
\left\vert \phi _{i}\right\vert <c<1$, we have $\left\Vert \phi _{i}^{\ell
}\right\Vert _{p}=\left\vert \phi _{i}\right\vert ^{\ell }<c^{\ell }$.
Hence, conditional on $\left\vert \phi _{i}\right\vert <c<1$, we have $%
\left\Vert \Delta y_{it}\right\Vert _{4}\leq \frac{2C}{1-c}<\infty $. Also
by Cauchy-Schwarz inequality $\left\vert E\left( \Delta y_{it}\Delta
y_{i,t-h}\right) \right\vert \leq \left[ E\left( \Delta y_{it}\right)
^{2}E\left( \Delta y_{i,t-h}\right) ^{2}\right] ^{1/2}$, and $%
\sup_{i}\left\vert E\left( \Delta y_{it}\Delta y_{i,t-h}\left\vert
\left\vert \phi _{i}\right\vert <c<1\right. \right) \right\vert <\infty $.
In the unit root case \vspace{-1.5ex} 
\begin{eqnarray*}
E\left( \Delta y_{it}\Delta y_{i,t-h}\left\vert \phi _{i}=1\right. \right)
&=&E(\sigma _{i}^{2})<C,\text{ for }h=0, \\
&=&0,\text{ for }h>0\text{,}
\end{eqnarray*}%
and overall $\left\vert \gamma _{\Delta }(h)\right\vert <\infty $, for $%
h\leq T-2$. Existence of $\gamma _{\Delta }(h)$ when $\phi _{i}$ is
distributed uniformly over the closed interval $[0,1]$ involves some algebra
and is established in Section \ref{ExU} of the online supplement.

General expressions for the mean, variance and autocovariances of the first
differences (covering unit root processes) are given in the following lemma
and will be used in our subsequent analysis.

\begin{lemma}
\label{VarAuto}Consider the panel AR(1) model given by (\ref{Par2}), and
suppose that Assumptions \ref{fe}-\ref{initial} hold, and $M_{i}\rightarrow
\infty $ for all units with $\left\vert \phi _{i}\right\vert <1$. Then for
all $i=1,2,...,n$\vspace{-1ex} 
\begin{align}
E\left( \Delta y_{it}\right) &=0,\quad E\left( \Delta y_{it}^{2}\right)
=\sigma ^{2}E\left( \frac{2}{1+\phi _{i}}\right) ,  \label{M&V} \\
E\left( \Delta y_{it}\Delta y_{i,t-h}\right) &=-\sigma ^{2}E\left[ \left( 
\frac{1-\phi _{i}}{1+\phi _{i}}\right) \phi _{i}^{h-1}\right] \text{, for }%
h=1,2,...,T-2,  \label{Cov_h}
\end{align}%
and\vspace{-1ex} 
\begin{equation}
E\left( \phi _{i}\Delta y_{it}\Delta y_{i,t-h}\right) =-\sigma ^{2}E\left[
\left( \frac{1-\phi _{i}}{1+\phi _{i}}\right) \phi _{i}^{h}\right] \text{,
for }h=1,2,...,T-2.  \label{phiCov_h}
\end{equation}
\end{lemma}

A proof is provided in Section \ref{ExiAutocov} of the online supplement.

\begin{remark}
Assumption \ref{initial} which in effect requires all processes $\left\{
y_{it},i=1,2,...,n\right\} $ with $\vert \phi_{i} \vert < 1$ are initialized
from a distant past, could be restrictive. Although this assumption is
required for our theoretical derivations, we do investigate the implications
of relaxing it using Monte Carlo experiments. See sub-sections \ref{Initial}
and \ref{InMC} in the online supplement.
\end{remark}

Our identification and estimation strategy is based on matching sample
estimates of autocorrelations of first differences (denoted as $\rho _{h}$)
with first and higher order moments of $\phi _{i}$. But before providing the
details of our proposed estimators, we first show that the HomoGMM
estimators of $E(\phi _{i})$ that neglect heterogeneity of $\phi _{i}$ over $%
i$ are biased even as $n\rightarrow \infty $, for any fixed $T$, and
inferences based on them could be misleading. It is recognized that
neglecting heterogeneity in dynamic panels can lead to biased estimates, but
to the best of our knowledge, there is no formal analysis of the extent of
the bias for short $T$ panels. In the case of heterogeneous dynamic panels
when both $n$ and $T$ are large, \cite{PesaranSmith1995} provide expressions
for asymptotic bias of fixed effects estimators.

\vspace{-3ex}

\section{Neglected heterogeneity bias \label{bias}}

\vspace{-1.5ex}

Under homogeneity where $\phi _{i}=\phi $ for all $i$, $\phi $ can be
consistently estimated by the method of moments after eliminating $\alpha
_{i}$, for example by first differencing. We begin our analysis by showing
the HomoGMM estimators are biased when $\phi _{i}$ are heterogeneous. The
extent of the bias depends on the degree of heterogeneity. To simplify the
exposition, without loss of generality, we consider the case where $T=4$,
the minimum value required for identification of $\mu _{\phi }=E(\phi _{i})$
under heterogeneity established in Section \ref{moments}\textbf{.} For the
Anderson-Hsiao (AH) estimator, $\hat{\phi}_{AH}=\left( \sum_{i=1}^{n}\Delta
y_{i4}\Delta y_{i2}\right) /\left( \sum_{i=1}^{n}\Delta y_{i3}\Delta
y_{i2}\right) $, and using (\ref{Fdif}) for $t=4$ we have \vspace{-1ex} 
\begin{equation}
\hat{\phi}_{AH}=\frac{n^{-1}\sum_{i=1}^{n}\phi _{i}\Delta y_{i3}\Delta y_{i2}%
}{n^{-1}\sum_{i=1}^{n}\Delta y_{i3}\Delta y_{i2}}+\frac{n^{-1}\sum_{i=1}^{n}%
\Delta u_{i4}\Delta y_{i2}}{n^{-1}\sum_{i=1}^{n}\Delta y_{i3}\Delta y_{i2}}.
\label{AH1}
\end{equation}%
Since $E\left( \Delta u_{i4}\Delta y_{i2}\right) =0,$ then under Assumptions %
\ref{fe} to \ref{initial} and assuming $M_{i}\rightarrow \infty $ for units
with $|\phi _{i}|<1$, we have (as $n\rightarrow \infty $)\vspace{-1.5ex} 
\begin{equation*}
\hat{\phi}_{AH}\rightarrow _{p}\frac{\lim_{n \rightarrow \infty } n^{-1}\sum_{i=1}^{n}E\left( \phi _{i}\Delta
y_{i3}\Delta y_{i2}\right) }{\lim_{n \rightarrow \infty } n^{-1} \sum_{i=1}^{n}E\left( \Delta y_{i3}\Delta
y_{i2}\right) },
\end{equation*}%
where $E\left( \Delta y_{i3}\Delta y_{i2}\right) $ and $E\left( \phi
_{i}\Delta y_{i3}\Delta y_{i2}\right) $ are given by (\ref{Cov_h}) and (\ref%
{phiCov_h}), respectively. Using these results \vspace{-1ex} 
\begin{equation}
\hat{\phi}_{AH}\rightarrow _{p}\frac{E\left[ \left( \frac{1-\phi _{i}}{%
1+\phi _{i}}\right) \phi _{i}\right] }{E\left[ \left( \frac{1-\phi _{i}}{%
1+\phi _{i}}\right) \right] }.  \label{AH2}
\end{equation}%
In the homogeneous case ($\phi _{i}=\phi $), we have $\hat{\phi}%
_{AH}\rightarrow _{p}\mu _{\phi }=\phi $, as expected. Under heterogeneity, $%
\hat{\phi}_{AH}$ is clearly not a consistent estimator of $E(\phi _{i})$.
The extent of the asymptotic bias of the AH estimator depends on the
distribution of $\phi _{i}$. The expression for the neglected
heterogeneity bias of the AH estimator is summarized in the following proposition.

\begin{proposition}
\label{NegBiasAH} Consider the Anderson-Hsiao estimator of $\mu _{\phi }$, $%
\hat{\phi}_{AH}$, given by (\ref{AH1}), and suppose $\mu _{\phi }=E(\phi
_{i})$ in the heterogeneous panel AR(1) model given by (\ref{Par2}). Suppose
Assumptions \ref{fe}--\ref{initial} hold, and $M_{i}\rightarrow \infty $ for
all $i$ with $\left\vert \phi _{i}\right\vert <1$. Then $\hat{\phi}_{AH}$ is
asymptotically biased as an estimator of $\mu _{\phi }$. For $T=4$, the
asymptotic bias of the AH estimator is given by \vspace{-1.5ex} 
\begin{equation}
plim_{n\rightarrow \infty }\left( \hat{\phi}_{AH}-\mu _{\phi }\right) =\frac{%
2\left( 1+\mu _{\phi }\right) \left[ \frac{1}{1+\mu _{\phi }}-E\left( \frac{1%
}{1+\phi _{i}}\right) \right] }{E\left( \frac{1-\phi _{i}}{1+\phi _{i}}%
\right) },  \label{AH3}
\end{equation}%
and $plim_{n\rightarrow \infty }\hat{\phi}_{AH}\leq \mu _{\phi }$. The
equality holds if and only if $\phi _{i}=\phi =\mu _{\phi }$, for all $i$.
\end{proposition}

A proof is provided in Section \ref{BiasAH} of the online supplement.

The asymptotic biases of the AB and BB estimators under heterogeneous slopes
are derived in Section \ref{NegBias} of the online supplement. The magnitude
of the asymptotic bias of AH, AB and BB estimators depends on the
distribution of $\phi _{i}$. For example, suppose that $\phi _{i}$ are
random draws from a uniform distribution centered at $E(\phi _{i})=\mu
_{\phi }>0$, with $\phi _{i}=\mu _{\phi }+v_{i}$, where $v_{i}$ $\sim
IIDU[-a,a],$ $a>0$.\footnote{%
To ensure that $\left\vert \phi _{i}\right\vert \leq 1$ we also require that 
$a\leq 1-\mu _{\phi }$.} Then \vspace{-1.5ex} 
\begin{equation}
plim_{n\rightarrow \infty }\left( \hat{\phi}_{AH}-\mu _{\phi }\right) =\frac{%
2(1+\mu _{\phi })\left[ \delta -\frac{1}{2}ln\left( \frac{1+\delta }{%
1-\delta }\right) \right] }{ln\left( \frac{1+\delta }{1-\delta }\right) -a},
\label{AsyBiasAH}
\end{equation}%
where $\delta =a/(1+\mu _{\phi })\leq \left( 1-\mu _{\phi }\right) /(1+\mu
_{\phi })<1$. It is easily seen that $\hat{\phi}_{AH}-\mu _{\phi
}\rightarrow 0$ with $a\rightarrow 0$. The magnitudes of the asymptotic bias
of the AH estimator for $\mu _{\phi }\in \{0.4,0.5\}$ and $a=0.5$ are around 
$-0.186$ and $-0.204$, respectively, which are very close to the
corresponding simulated bias in Tables \ref{tab:hetro_u1_a} and \ref%
{tab:hetro_u2_a} in the online supplement.

In the case where $\phi _{i}$ follows a categorical distribution, $\phi
_{i}=\phi _{L}$ $\left( 0<\phi _{L}<1\right) $ with probability $\pi $ and $%
\phi _{i}=\phi _{H}>\phi _{L}$ with probability $1-\pi $, we have \vspace{%
-1.5ex} 
\begin{equation*}
plim_{n\rightarrow \infty }\left( \hat{\phi}_{AH}-\mu _{\phi }\right) =\frac{%
-2\pi (1-\pi )(\phi _{H}-\phi _{L})^{2}}{\pi (1-\phi _{L})(1+\phi
_{H})+(1-\pi )(1+\phi _{L})(1-\phi _{H})}.
\end{equation*}%
As to be expected the asymptotic bias is negative, and its magnitude depends
on the degree of dispersion of $\phi _{i}$ which is given by $Var(\phi
_{i})=\sigma _{\phi }^{2}=\pi (1-\pi )(\phi _{H}-\phi _{L})^{2}$. The unit
root case arises for the units with $\phi _{H}=1$.

Asymptotic bias, even if small, can lead to substantial size distortions
when $n$ is sufficiently large. See sub-section \ref{MChomo} for Monte Carlo
evidence on the bias and size distortions of AH and other HomoGMM estimators.

\vspace{-3ex}

\section{Identification of moments of the AR coefficients \label{moments}}

\vspace{-1.5ex}

In this section, we formally establish conditions necessary for
identification of $E(\phi _{i}^{s})$ without making any specific
distributional assumptions on $\phi _{i}$. Suppose Assumptions \ref{fe} to %
\ref{initial} hold. We consider the minimum number of periods needed to
consistently estimate $E(\phi _{i}^{s}),$ for $s=1,2,...,S$. Denote the $%
h^{th}$-order autocorrelation coefficients of $\Delta y_{it}$ as $\rho _{h} $
given by \vspace{-1ex} 
\begin{equation}
\rho _{h}=\frac{E\left( \Delta y_{it}\Delta y_{i,t-h}\right) }{E\left[
\left( \Delta y_{it}\right) ^{2}\right] },  \label{rhoh0}
\end{equation}%
for $h=1,2,...$, with $\left\vert \rho _{h}\right\vert \leq 1$. Since by
assumption $\phi _{i}$ and $\sigma _{i}^{2}$ are independently distributed
(see part (b) of Assumption \ref{error variances}), then using the results
in Lemma \ref{VarAuto} we have\vspace{-1ex} 
\begin{equation}
\rho _{h}=\frac{E\left( \Delta y_{it}\Delta y_{i,t-h}\right) }{E\left[
\left( \Delta y_{it}\right) ^{2}\right] }=-\frac{E\left[ \left( \frac{1-\phi
_{i}}{1+\phi _{i}}\right) \phi _{i}^{h-1}\right] }{2E\left( \frac{1}{1+\phi
_{i}}\right) },  \label{rhoh}
\end{equation}%
for $h=1,2,...$, with $\left\vert \rho _{h}\right\vert \leq 1$.

Suppose that $\rho _{h}$ can be consistently estimated by the moment
estimators of $E\left( \Delta y_{it}\Delta y_{i,t-h}\right) $ and $E\left[
\left( \Delta y_{it}\right) ^{2}\right] $. Then the identification condition
of $E(\phi _{i}^{s})$ can be derived by the system of equations in (\ref%
{rhoh}). For $h=1$, $2E\left( \frac{1}{1+\phi _{i}}\right) \rho
_{1}=-E\left( \frac{1-\phi _{i}}{1+\phi _{i}}\right) =1-2E\left( \frac{1}{%
1+\phi _{i}}\right) $, which can be equivalently written as $2E\left( \frac{1%
}{1+\phi _{i}}\right) =\frac{1}{1+\rho _{1}}$. Using this result and noting
that for $h=2$, $2E\left( \frac{1}{1+\phi _{i}}\right) \rho _{2} =-2+E\left(
\phi _{i}\right) +2E\left( \frac{1}{1+\phi _{i}}\right) $, we have \vspace{%
-1ex} 
\begin{equation}
E\left( \phi _{i}\right) =\frac{1+2\rho _{1}+\rho _{2}}{1+\rho _{1}}.
\label{Ephi1}
\end{equation}%
Similarly, for $h=3$ we have $2E\left( \frac{1}{1+\phi _{i}}\right) \rho
_{3}=-E\left( 2\phi _{i}-2-\phi _{i}^{2}+\frac{2}{1+\phi _{i}}\right) $,
which yields\vspace{-1ex} 
\begin{equation}
E\left( \phi _{i}^{2}\right) =\frac{1+2\rho _{1}+2\rho _{2}+\rho _{3}}{%
1+\rho _{1}}.  \label{Ephi2}
\end{equation}%
For $h=4$, $2E\left( \frac{1}{1+\phi _{i}}\right) \rho _{4}=-E\left( 2\phi
_{i}^{2}-\phi _{i}^{3}-2\phi _{i}+2-\frac{2}{1+\phi _{i}}\right)$, and upon
using the results of the lower-order moments we obtain \vspace{-1.5ex} 
\begin{equation}
E\left( \phi _{i}^{3}\right) =\frac{1+2\rho _{1}+2\rho _{2}+2\rho _{3}+\rho
_{4}}{1+\rho _{1}}.  \label{Ephi3}
\end{equation}%
Higher-order moments of $\phi _{i}$ can be obtained similarly. To identify
the $s^{th}$ order moment of $\phi _{i}$ requires consistent estimation of $%
\rho _{h}$ for $h=1,2,...,s+1$. In general, we must have $T\geq s+3$, as $%
n\rightarrow \infty $ to identify $E\left( \phi _{i}^{s}\right) $.

\begin{remark}
Note that under homogeneity where $\phi _{i}=\phi $ for all $i$, using (\ref%
{rhoh}) we have \vspace{-1.5ex} 
\begin{equation}
\rho _{h}=\frac{E\left( \Delta y_{it}\Delta y_{i,t-h}\right) }{E\left[
\left( \Delta y_{it}\right) ^{2}\right] } =-\frac{1}{2}\phi ^{h-1}\left(
1-\phi \right) , \text{ for }h=1,2,...,T-2.  \label{homoRes}
\end{equation}%
For $h=1$ under homogeneity, $\rho _{1}=-(1-\phi )/2$ and $\phi $ can be
estimated by $\hat{\phi}_{Homo}=1+2\hat{\rho}_{1,nT}$. In this case for
identification of $\phi $, we need $T\geq 2$. This result also follows if we
let $\rho _{h}=\phi \rho _{h-1}$ in (\ref{Ephi1}) $E\left( \phi _{i}\right)
=\phi =1+\frac{\rho _{1}+\phi \rho _{1}}{1+\rho _{1}},$ which is satisfied
when $\rho _{1}=-(1-\phi )/2$.
\end{remark}

\vspace{-3.5ex}

\section{Estimation of the moments of the AR coefficients\label{estimation}}

\vspace{-1ex}

We now turn our attention to consistent estimation of the moments of $\phi
_{i}$, namely $\theta _{s}=E\left( \phi _{i}^{s}\right) $, for $s=1,2,$ and $%
3$. We consider a simple moment estimator which we refer to as the first
differenced autocorrelation (FDAC) estimator, and a GMM-type estimator that
we refer to as HetroGMM to be distinguished from the GMM estimators proposed
in the literature for estimation of the homogeneous AR coefficient assuming $%
\phi _{i}=\phi $ for all $i$.

\vspace{-2ex}

\subsection{First differenced autocorrelation (FDAC) estimator \label%
{fdacestm}}

The FDAC estimator uses the sample analogs of autocorrelations of the first
differences, $\rho _{h}$ given by (\ref{rhoh0}), in equations (\ref{Ephi1}),
(\ref{Ephi2}) and (\ref{Ephi3}) to obtain consistent estimators of $\theta
_{s}=E(\phi _{i}^{s})$ for $s=1,2$ and 3, respectively. Specifically, using $%
\left\{ \Delta y_{it}\text{, }t=2,3,...,T;i=1,2,...,n\right\} $, $\rho _{h}$
can be consistently estimated by\vspace{-1ex} 
\begin{equation}
\hat{\rho}_{h,nT}=\frac{(T-h-1)^{-1}\sum_{t=h+2}^{T}\left[
n^{-1}\sum_{i=1}^{n}\Delta y_{it}\Delta y_{i,t-h}\right] }{%
(T-1)^{-1}\sum_{t=2}^{T}\left[ n^{-1}\sum_{i=1}^{n}\left( \Delta
y_{it}\right) ^{2}\right] }\text{, for }h=1,2,...,T-2.  \label{EstMom}
\end{equation}%
Then plugging these estimators in (\ref{Ephi1})--(\ref{Ephi3}) we have the
following FDAC estimators \vspace{-1ex} 
\begin{equation}
\hat{\theta}_{1,FDAC}=\widehat{E\left( \phi _{i}\right) }=\frac{1+2\hat{\rho}%
_{1,nT}+\hat{\rho}_{2,nT}}{1+\hat{\rho}_{1,nT}},\text{ for }T\geq 4,
\label{Estm1}
\end{equation}%
\vspace{-1.5ex} 
\begin{equation}
\hat{\theta}_{2,FDAC}=\widehat{E\left( \phi _{i}^{2}\right) }=\frac{1+2\hat{%
\rho}_{1,nT}+2\hat{\rho}_{2,nT}+\hat{\rho}_{3,nT}}{1+\hat{\rho}_{1,nT}},%
\text{ for }T\geq 5,  \label{Estm2}
\end{equation}%
and\vspace{-1.5ex} 
\begin{equation}
\hat{\theta}_{3,FDAC}=\widehat{E\left( \phi _{i}^{3}\right) }=\frac{1+2\hat{%
\rho}_{1,nT}+2\hat{\rho}_{2,nT}+2\hat{\rho}_{3,nT}+\hat{\rho}_{4,nT}}{1+\hat{%
\rho}_{1,nT}},\text{ for }T\geq 6.  \label{Estm3}
\end{equation}%
These estimators can also be viewed as moment estimators that place equal
weights on the cross-section averages, $n^{-1}\sum_{i=1}^{n}\Delta
y_{it}\Delta y_{i,t-h}$, for different $t$. This makes sense since under our
assumptions for each $t$, $\Delta y_{it}\Delta y_{i,t-h}$ are
cross-sectionally independent with finite second-order moments, and by the
law of large numbers $n^{-1}\sum_{i=1}^{n}\Delta y_{it}\Delta
y_{i,t-h}\rightarrow _{p}E\left( \Delta y_{it}\Delta y_{i,t-h}\right) $, and
hence $\hat{\rho}_{h,nT}\rightarrow _{p}E\left( \Delta y_{it}\Delta
y_{i,t-h}\right) /E\left( \Delta y_{it}\right) ^{2}=\rho _{h}$ as $n
\rightarrow \infty$. Using this result and noting that $1+\hat{\rho}%
_{1,nT}\rightarrow _{p}1+\rho _{1}>0$, it then readily follows that $\hat{%
\theta}_{1,FDAC}\rightarrow \frac{1+2\rho _{1}+\rho _{2}}{1+\rho _{1}}%
=\theta _{1}=E(\phi _{i})$. Similarly, $\hat{\theta}_{s,FDAC}\rightarrow
_{p}E(\phi _{i}^{s})$, for $s=2$ and $3$. \ Since $\Delta y_{it}\Delta
y_{i,t-h}$ for $h=1,2,...,T-2$ have second order moments, it also follows
that the convergence of $\hat{\theta}_{s,FDAC}$ to $\theta _{s}$ is in the
mean squared error sense which is stronger than convergence in probability.

\vspace{-2ex}

\subsection{Generalized method of moments estimator based on autocovariances 
\label{hetrogmmestm}}

The FDAC estimator is a plug-in type estimator and needs not be efficient.
An alternative and arguably more efficient approach would be to base the
estimation of $\theta _{s}$ directly on the sample moments of $E\left(
\Delta y_{it}\Delta y_{i,t-h}\right) $ and then use standard results from
the GMM literature to obtain asymptotically optimum weighted moment
conditions rather than equally weighted moments which might not be
efficient. In practice, the differences between the two approaches could
depend on the degree of heterogeneity and the sampling uncertainty
associated with the GMM weights. The relative performance of FDAC and
heterogeneous GMM estimators of $\theta _{s}$ will be investigated by Monte
Carlo simulations.

\vspace{-2ex}

\subsubsection{Heterogeneous generalized method of moments (HetroGMM)
estimator of $E(\protect\phi _{i})$}

Given (\ref{rhoh0}), the moment condition (\ref{Ephi1}) can be written
equivalently as\vspace{-1.5ex} 
\begin{equation}
\theta _{1}[E\left( \Delta y_{it}\right) ^{2}+E\left( \Delta y_{it}\Delta
y_{i,t-1}\right) ]=E\left( \Delta y_{it}\right) ^{2}+2E\left( \Delta
y_{it}\Delta y_{i,t-1}\right) +E\left( \Delta y_{it}\Delta y_{i,t-2}\right) ,
\label{mit}
\end{equation}%
which yields $T-3$ moment conditions for $t=4,5,...,T$, requiring that $%
T\geq 4$. These moment conditions can be written more compactly as \vspace{%
-1.5ex} 
\begin{equation}
E\left[ M_{nt}(\theta _{1,0})\right] =0\text{, for }t=4,5,...,T,  \label{Mnt}
\end{equation}%
where $M_{nt}(\theta _{1,0})=n^{-1}\sum_{i=1}^{n}m_{it}(\theta _{1,0})$, $%
m_{it}(\theta _{1})=\theta _{1}h_{it}-g_{it}$, \vspace{-1.5ex} 
\begin{equation}
h_{it}=\left( \Delta y_{it}\right) ^{2}+\Delta y_{it}\Delta y_{i,t-1}\text{,
and }g_{it}=\left( \Delta y_{it}\right) ^{2}+2\Delta y_{it}\Delta
y_{i,t-1}+\Delta y_{it}\Delta y_{i,t-2}.  \label{hgit}
\end{equation}%
To optimally combine the moment conditions in (\ref{Mnt}) set \vspace{-1.5ex}
\begin{align}
\mathbf{h}_{iT}& =\left( h_{i4},h_{i5},...,h_{iT}\right) ^{\prime },
\label{h_iT} \\
\text{and }\mathbf{g}_{iT}& =\left( g_{i4},g_{i5},...,g_{iT}\right) .
\label{g_iT}
\end{align}%
Then $\mathbf{M}_{nT}(\theta _{1})=\left( m_{n,4}(\theta
_{1}),m_{n,5}(\theta _{1}),...,m_{n,T}(\theta _{1})\right) ^{\prime }=%
\mathbf{G}_{nT}-\mathbf{H}_{nT}\theta _{1}$, where $\mathbf{G}_{nT}=\frac{1}{%
n}\sum_{i=1}^{n}\mathbf{g}_{iT}$ and $\mathbf{H}_{nT}=n^{-1}\sum_{i=1}^{n}%
\mathbf{h}_{iT}$. Using (\ref{mit}), it readily follows that $E\left[ 
\mathbf{M}_{nT}(\theta _{1,0})\right] =\mathbf{0}$. The HetroGMM estimator
of $\theta _{1}$ is given by 
\vspace{-1.5ex} 
\begin{equation*}
\hat{\theta}_{1,HetroGMM}=\func{argmin}_{\theta _{1}}\left( \mathbf{G}_{nT}-%
\mathbf{H}_{nT}\theta _{1}\right) ^{\prime }\mathbf{A}_{nT}\left( \mathbf{G}%
_{nT}-\theta _{1}\mathbf{H}_{nT}\right) ,
\end{equation*}%
where $\mathbf{A}_{nT}$ is a $\left( T-3\right) \times (T-3)$ positive
definite stochastic weight matrix, and for any $T\geq 4$, it tends to a
non-stochastic positive definite matrix \thinspace $\mathbf{A}_{T}$ as $%
n\rightarrow \infty $. The most efficient HetroGMM estimator is given by 
\vspace{-1.5ex} 
\begin{equation}
\hat{\theta}_{1,HetroGMM}\left( \mathbf{A}_{T}^{\ast }\right) =\left( 
\mathbf{H}_{nT}^{\prime }\mathbf{A}_{T}^{\ast }\mathbf{H}_{nT}\right) ^{-1}%
\mathbf{H}_{nT}^{\prime }\mathbf{A}_{T}^{\ast }\mathbf{G}_{nT},
\label{GMMm*}
\end{equation}%
where $\mathbf{A}_{T}^{\ast }=\mathbf{S}_{T}^{-1}(\theta _{1})$ is the
optimal weight matrix with \vspace{-1.5ex} 
\begin{equation*}
\mathbf{S}_{T}(\theta _{1})=Var\left( \sqrt{n}\mathbf{M}_{nT}(\theta
_{1})\right) =nVar\left( \mathbf{G}_{nT}-\theta _{1}\mathbf{H}_{nT}\right)
=nVar\left[ n^{-1}\sum_{i=1}^{n}\left( \mathbf{g}_{iT\ }-\theta _{1}\mathbf{h%
}_{iT}\right) \right] .
\end{equation*}%
Given (\ref{mit}), $E\left( \mathbf{g}_{iT}-\theta _{1,0}\mathbf{h}%
_{iT}\right) =\boldsymbol{0}$, and $\mathbf{g}_{iT}-\theta _{1,0}\mathbf{h}%
_{iT}$ are cross-sectionally independent, then \vspace{-1.5ex} 
\begin{equation}
\mathbf{S}_{T}(\theta _{1,0})=\frac{1}{n}\sum_{i=1}^{n}E\left[ \left( 
\mathbf{g}_{iT\ }-\theta _{1,0}\mathbf{h}_{iT}\right) \left( \mathbf{g}_{iT\
}-\theta _{1,0}\mathbf{h}_{iT}\right) ^{\prime }\right] .  \label{STtheta}
\end{equation}%
It is difficult to derive an analytical expression for $\mathbf{S}%
_{T}(\theta _{1,0})$, but for a given value of $\theta _{1}$, $\mathbf{S}%
_{T}(\theta _{1})$ can be consistently estimated by its sample mean given by 
\vspace{-1.5ex} 
\begin{equation}
\mathbf{\hat{S}}_{T}\left( \theta _{1}\right) =\frac{1}{n}%
\sum_{i=1}^{n}\left( \mathbf{g}_{iT\ }-\theta _{1}\mathbf{h}_{iT}\right)
\left( \mathbf{g}_{iT\ }-\theta _{1}\mathbf{h}_{iT}\right) ^{\prime }\text{,
for }n>T-3.  \label{Shat}
\end{equation}%
A standard two-step GMM estimator of $\theta _{1}$ can now be obtained using 
$\hat{\theta}_{1,FDAC}$ given by (\ref{Estm1}) as an initial estimate to
consistently estimate the optimal weight matrix, $\mathbf{S}_{T}^{-1}(\theta
_{1,0})$, in the first step. Substituting $\hat{\theta}_{1,FDAC}$ into (\ref%
{Shat}) yields the following two-step HetroGMM estimator \vspace{-1.5ex} 
\begin{equation}
\hat{\theta}_{1,HetroGMM}=\left[ \mathbf{H}_{nT}^{\prime }\mathbf{\hat{S}}%
_{T}^{-1}\left( \hat{\theta}_{1,FDAC}\right) \mathbf{H}_{nT}\right] ^{-1}%
\left[ \mathbf{H}_{nT}^{\prime }\mathbf{\hat{S}}_{T}^{-1}\left( \hat{\theta}%
_{1,FDAC}\right) \mathbf{G}_{nT}\right] ,  \label{theta1GMM}
\end{equation}%
where \vspace{-1.5ex} 
\begin{equation}
\mathbf{\hat{S}}_{T}\left( \hat{\theta}_{1,FDAC}\right) =\frac{1}{n}%
\sum_{i=1}^{n}\left( \mathbf{g}_{iT\ }-\hat{\theta}_{1,FDAC}\mathbf{h}%
_{iT}\right) \left( \mathbf{g}_{iT\ }-\hat{\theta}_{1,FDAC}\mathbf{h}%
_{iT}\right) ^{\prime }.  \label{Shathat}
\end{equation}%
It is also possible to obtain an iterated version of the above, where $\hat{%
\theta}_{1,HetroGMM}$ is used to obtain a new estimate of $\mathbf{\hat{S}}%
_{T}\left( \theta _{1}\right) $, namely $\mathbf{\hat{S}}_{T}(\hat{\theta}%
_{1,HetroGMM})$, and so on. But there seems little gain in doing so since $%
\hat{\theta}_{1,HetroGMM}$ is asymptotically efficient.

The above results are summarized in the following theorem.

\vspace{-1ex}

\begin{theorem}
Consider the panel AR(1) model given by (\ref{Par2}) and suppose that
Assumptions \ref{fe}--\ref{initial} hold, $T\geq 4$, and $M_{i}\rightarrow
\infty $ for all $i$ with $|\phi _{i}|<1$. Then the HetroGMM estimator of $%
\theta _{1}=E(\phi _{i})$ given by (\ref{theta1GMM}) is asymptotically
efficient. The asymptotic distribution of $\hat{\theta}_{1,HetroGMM}$ is
given by \vspace{-1.5ex} 
\begin{equation}
\sqrt{n}\left( \hat{\theta}_{1,HetroGMM}-\theta _{1,0}\right) \rightarrow
_{d}N\left( 0,V_{\theta _{1}}\right) ,  \label{AsyDis1}
\end{equation}%
where $\theta _{1,0}$ is the true value of $\theta _{1}$, $V_{\theta
_{1}}^{-1}=plim_{n\rightarrow \infty }\left( \mathbf{H}_{nT}^{\prime }%
\mathbf{S}_{T}^{-1}(\theta _{1,0})\mathbf{H}_{nT}\right) $, $\mathbf{H}_{nT}=%
\frac{1}{n}\sum_{i=1}^{n}\mathbf{h}_{iT}$, and $\mathbf{S}_{T}^{-1}(\theta
_{1,0})$ and $\mathbf{h}_{iT}$ are defined by (\ref{STtheta}) and (\ref{h_iT}%
), respectively. The asymptotic variance of $\hat{\theta}_{1,HetroGMM}$ can
be estimated consistently by $n^{-1}\left[ \mathbf{H}_{nT}^{\prime }\mathbf{%
\hat{S}}_{T}^{-1}\left( \hat{\theta}_{1,FDAC}\right) \mathbf{H}_{nT}\right]
^{-1}$, where $\mathbf{\hat{S}}_{T}\left( \hat{\theta}_{1,FDAC}\right) $ is
given by (\ref{Shathat}).
\end{theorem}

Our use of the FDAC estimator as an initial estimator for the two-step GMM
estimator is based on the observations that FDAC exploits the stationarity
properties of moments in the first differences and is based on more
information as compared to the first step GMM estimator. As an example,
consider the exact identified case when $T=4$. Then (see (\ref{hgit})) 
\begin{equation*}
\hat{\theta}_{1,HetroGMM}=\frac{n^{-1}\sum_{i=1}^{n}\left( \Delta
y_{i4}\right) ^{2}+2n^{-1}\sum_{i=1}^{n}\Delta y_{i4}\Delta
y_{i,3}+n^{-1}\sum_{i=1}^{n}\Delta y_{i4}\Delta y_{i,2}}{n^{-1}%
\sum_{i=1}^{n}\left( \Delta y_{i4}\right) ^{2}+n^{-1}\sum_{i=1}^{n}\Delta
y_{i4}\Delta y_{i,3}},
\end{equation*}%
as compared to $\hat{\theta}_{1,FDAC}$ given by (\ref{Estm1}) which can be
written equivalently 
\begin{equation*}
\hat{\theta}_{1,FDAC}=\frac{\left( \frac{1}{3}\right) \sum_{t=2}^{4}\left[ 
\frac{1}{n}\sum_{i=1}^{n}\left( \Delta y_{it}\right) ^{2}\right]
+\sum_{t=3}^{4}\left[ \frac{1}{n}\sum_{i=1}^{n}\Delta y_{it}\Delta y_{i,t-1}%
\right] +\left[ \frac{1}{n}\sum_{i=1}^{n}\Delta y_{i4}\Delta y_{i,4-2}\right]
}{\left( \frac{1}{3}\right) \sum_{t=2}^{4}\left[ \frac{1}{n}%
\sum_{i=1}^{n}\left( \Delta y_{it}\right) ^{2}\right] +\left( \frac{1}{2}%
\right) \sum_{t=3}^{4}\left[ \frac{1}{n}\sum_{i=1}^{n}\Delta y_{it}\Delta
y_{i,t-1}\right] }.
\end{equation*}%
Both estimators converge to $\theta _{1,0}$ at the rate of $\sqrt{n}$, but $%
\hat{\theta}_{1,FDAC}$ exploits the stationary properties of the $\left(
\Delta y_{it}\right) ^{2}$ and $\Delta y_{it}\Delta y_{i,t-1}$ more
effectively. Specifically, $n^{-1}\sum_{i=1}^{n}\left( \Delta y_{i4}\right)
^{2}$ and $(1/3)\sum_{t=2}^{4}\left[ n^{-1}\sum_{i=1}^{n}\left( \Delta
y_{it}\right) ^{2}\right] $ converge to the same limit, but the latter makes
use of $\left( \Delta y_{i2}\right) ^{2}$ and $\left( \Delta y_{i3}\right)
^{2}$ obervations as well as $\left( \Delta y_{i4}\right) ^{2}$. Similarly, $%
2n^{-1}\sum_{i=1}^{n}\Delta y_{i4}\Delta y_{i3}$ and $\sum_{t=3}^{4}\left[
n^{-1}\sum_{i=1}^{n}\Delta y_{it}\Delta y_{i,t-1}\right] $ converge to the
same limit, but the latter makes use of $\Delta y_{i3}\Delta y_{i2}$ in
addition to $\Delta y_{i4}\Delta y_{i3}$.

\vspace{-1ex}
\begin{remark}
Both FDAC and HetroGMM estimators should work fine asymptotically under $%
E(u_{it}^{2})=\sigma _{it}^{2}$, so long as the time variations of $\sigma
_{it}^{2}$ is stationary, in a sense that $E(\sigma _{it}^{2})=\sigma
_{i}^{2}$. One important example is when $u_{it}$ has a stationary GARCH
specification. This property is illustrated in the Monte Carlo simulations
where we consider the properties of the proposed estimators with and without
GARCH effects.
\end{remark}
\vspace{-2ex}

\subsubsection{Generalized method of moments estimator of $E(\protect\phi%
_{i}^{2})$}

Similarly, the HetroGMM estimator of $\theta _{2}=E\left( \phi
_{i}^{2}\right) $ can be obtained based on the equation below for $%
t=5,6,...,T$,
\vspace{-1.5ex} 
\begin{align}
& \theta _{2}\left[ E\left[ \left( \Delta y_{it}\right) ^{2}\right] +E\left(
\Delta y_{it}\Delta y_{i,t-1}\right) \right]  \label{theta2} \\
=& E\left[ \left( \Delta y_{it}\right) ^{2}\right] +2E\left( \Delta
y_{it}\Delta y_{i,t-1}\right) +2E\left( \Delta y_{it}\Delta y_{i,t-2}\right)
+E\left( \Delta y_{it}\Delta y_{i,t-3}\right) .  \notag
\end{align}%
Let \vspace{-1.5ex} 
\begin{align}
\mathbf{h}_{2,iT}& =\left( h_{2,i5},h_{2,i6},...,h_{2,iT}\right) ^{\prime }
\label{h_2iT} \\
\text{and }\mathbf{g}_{2,iT}& =\left( g_{2,i5},g_{2,i6},...,g_{2,iT}\right)
^{\prime }  \label{g_2iT}
\end{align}%
with $h_{2,it}=( \Delta y_{it}) ^{2}+\Delta
y_{it}\Delta y_{i,t-1}$ and $g_{2,it}=\left( \Delta y_{it}\right)
^{2}+2\Delta y_{it}\Delta y_{i,t-1}+2\Delta y_{it}\Delta y_{i,t-2}+\Delta
y_{it}\Delta y_{i,t-3}$. Denote $\mathbf{G}_{2,nT}=n^{-1}\sum_{i=1}^{n}%
\mathbf{g}_{2,iT}$, and $\mathbf{H}_{2,nT}=n^{-1}\sum_{i=1}^{n}\mathbf{h}%
_{2,iT}$, where $\mathbf{G}_{2,nT}$ and $\mathbf{H}_{2,nT}$ are $\left(
T-4\right) \times 1$ vectors (with $T>4$). Then, the two-step HetroGMM
estimator of the second moment can be derived as \vspace{-1.5ex} 
\begin{equation}
\hat{\theta}_{2,HetroGMM}=\left[ \mathbf{H}_{2,nT}^{\prime }\mathbf{\hat{S}}%
_{2,T}^{-1}\left( \hat{\theta}_{2,FDAC}\right) \mathbf{H}_{2,nT}\right] ^{-1}%
\left[ \mathbf{H}_{2,nT}^{\prime }\mathbf{\hat{S}}_{2T}^{-1}\left( \hat{%
\theta}_{2,FDAC}\right) \mathbf{G}_{2,nT}\right] ,  \label{theta2GMM}
\end{equation}%
where the initial estimator can be the FDAC estimator of $\theta _{2}$ given
by equation (\ref{Estm2}), and $\mathbf{\hat{S}}_{2,T}\left( \theta
_{2}\right) =\frac{1}{n}\sum_{i=1}^{n}\left( \mathbf{g}_{2,iT\ }-\theta _{2}%
\mathbf{h}_{2,iT}\right) \left( \mathbf{g}_{2,iT\ }-\theta _{2}\mathbf{h}%
_{2,iT}\right) ^{\prime }$. Finally, the asymptotic distribution of $\hat{%
\theta}_{2,HetroGMM}$ is given by 
\vspace{-1.5ex} 
\begin{equation}
\sqrt{n}\left( \hat{\theta}_{2,HetroGMM}-\theta _{2,0}\right) \rightarrow
_{d}N\left( 0,V_{\theta _{2}}\right) ,  \label{Asytheta2}
\end{equation}%
where $\theta _{2,0}$ is the true value of $\theta _{2}$, and $V_{\theta
_{2}}$ can be consistently estimated by\vspace{-1.5ex} 
\begin{equation}
\hat{V}_{\theta _{2}}=\left[ \mathbf{H}_{2,nT}^{\prime }\mathbf{\hat{S}}%
_{2,T}^{-1}\left( \hat{\theta}_{2,HetroGMM}\right) \mathbf{H}_{2,nT}\right]
^{-1}.  \label{Vartheta2}
\end{equation}

\vspace{-2ex}

\subsection{Plug-in estimator of $\protect\sigma_{\protect\phi}^{2}$ \label%
{estzig}}

Consider now the estimation of $\sigma _{\phi }^{2}=Var(\phi _{i})$, and
recall that in terms of $\boldsymbol{\theta }=(\theta _{1},\theta
_{2})^{\prime }$ we have $\sigma _{\phi }^{2}=\theta _{2}-\theta _{1}^{2}$.
Therefore, a plug-in estimator of $\sigma _{\phi }^{2}$ is given by \vspace{%
-1.5ex} 
\begin{equation}
\hat{\sigma}_{\phi }^{2}=\hat{\theta}_{2}-\left( \hat{\theta}_{1}\right)
^{2},  \label{Estvar}
\end{equation}%
which is an asymptotically valid estimator of $\sigma _{\phi }^{2}$ if $\hat{%
\theta}_{2}-\left( \hat{\theta}_{1}\right) ^{2}>0$. This condition will be
met for $n$ sufficiently large, noting that $\boldsymbol{\hat{\theta}}%
=\left( \hat{\theta}_{1},\hat{\theta}_{2}\right) ^{\prime }$ is a consistent
estimator of $\boldsymbol{\theta }_{0}=(\theta _{1,0},\theta _{2,0})^{\prime
}$. The asymptotic distribution of $\boldsymbol{\hat{\theta}}=(\hat{\theta}%
_{1},\hat{\theta}_{2})^{\prime }$, $\sqrt{n}(\boldsymbol{\hat{\theta}}-%
\boldsymbol{\theta }_{0})\rightarrow _{d}N\left( 0,\mathbf{V}_{\boldsymbol{%
\theta }}\right) $ is derived in Section \ref{Asyvar} of the online
supplement. Then using the Delta method it follows that $\sqrt{n}\left( \hat{%
\sigma}_{\phi }^{2}-\sigma _{\phi ,0}^{2}\right) \rightarrow _{d}N\left(
0,V_{\sigma ^{2}}\right) $, where $\sigma _{\phi ,0}^{2}=\theta
_{2,0}-\theta _{10}^{2}$ denotes the true value of $\sigma _{\phi }^{2}$,
and $V_{\sigma ^{2}}=\left( -2\theta _{1,0},1\right) \mathbf{V}_{\boldsymbol{%
\theta }}\left( -2\theta _{1,0},1\right) ^{\prime }$. $V_{\sigma ^{2}}$ can
be consistently estimated by $\hat{V}_{\sigma }=\left( -2\hat{\theta}%
_{1},1\right) \mathbf{\hat{V}}_{\boldsymbol{\theta }}\left( -2\hat{\theta}%
_{1},1\right) ^{\prime }$, where $\mathbf{\hat{V}}_{\boldsymbol{\theta }}$ $%
\ $is a consistent estimator of $\mathbf{V}_{\boldsymbol{\theta }}$ given by
(\ref{Vartheta}) in the online supplement. However, it is important to bear
in mind that the asymptotic distribution of $\hat{\sigma}_{\phi }^{2}$ is
valid only in the locality of the true value of $\sigma _{\phi }^{2}$, and
only if this true value is sufficiently away from the boundary value of $0$.
In practice, we recommend using the plug-in estimator of $\sigma _{\phi
}^{2} $ only when $n$ is large, in excess of $1,000,$ judging by the Monte
Carlo evidence to be discussed below.

\vspace{-3ex}

\section{Monte Carlo experiments \label{simulation}}

\vspace{-2ex}

\subsection{Data generating process of Monte Carlo experiments\label{DGP}}

For each $i=1,2,...,n$, the process $\{y_{it}\}$ is generated starting at
time $t=-M_{i}+1$, with the initial value $y_{i,-M_{i}}$ using \vspace{-1.5ex%
} 
\begin{equation}
y_{it}=\mu _{i}(1-\phi _{i})+\phi _{i}y_{i,t-1}+h_{it}\varepsilon _{it},%
\text{ for }t=-M_{i}+1,-M_{i}+2,...,0,1,2,...,T.  \label{DGPmc}
\end{equation}%
\vspace{-2ex} We experiment with two distributions to generate $\phi _{i}\in
(-1,1]$: (a) uniform and (b) categorial. Under the former we set $\phi
_{i}=\mu _{\phi }+v_{i}$, with $v_{i}\thicksim IIDU[-a,a]$. To distinguish
between cases when $\left\vert \phi _{i}\right\vert <1$ for all $i$ and when 
$\phi_{i} \in [-1+\epsilon , 1]$ for some $\epsilon>0$ with $\phi _{i}=1$ for some $i$, we fix $a=0.5$ and
consider the values of $\mu _{\phi }=0.4$ and $0.5$, with $E(\phi _{i})=\mu
_{\phi }$ and $\sigma _{\phi }^{2}=a^{2}/3=0.083$. Under case (b), we
generate $\phi _{i}=\phi _{H}$ (high) and $\phi _{i}=\phi _{L}$ (low) with
probabilities $1-\pi $ and $\pi $, respectively. Two sets of parameter
values for $(\phi _{H},\phi _{L},\pi )$ are considered: $(0.8,0.5,0.85)$
with $|\phi _{i}|<1$ for all $i$, and $(1,0.5,0.95)$ with $\phi_{i} \in (-1, 1]$ for all $i$. Then $\mu _{\phi }=E(\phi _{i})=\phi _{L}\pi +\phi _{H}(1-\pi
)=0.545$ and $0.525$, and $\sigma _{\phi }^{2}=\left[ \phi _{L}^{2}\pi +\phi
_{H}^{2}(1-\pi )\right] -\mu _{\phi }^{2}=0.011$ and $0.012$, respectively.
The individual-specific means of $\{y_{it}\}$ are generated as $\mu
_{i}=\phi _{i}+\eta _{i}$ with $\eta _{i}\thicksim IIDN(0,1)$, allowing for
a non-zero correlation between $\mu _{i}$ and $\phi _{i}$.

We consider two choices when generating $\varepsilon_{it}$: Gaussian $%
\varepsilon _{it}\sim IIDN(0,1)$, and non-Gaussian $\varepsilon _{it}=\left(
e_{it}-2\right) /2$, with $e_{it}\sim IID\chi _{2}^{2}$, where $\chi
_{2}^{2} $ is a chi-squared variate with two degrees of freedom, for all $i$
and $t$. $\left\{ h_{it}\right\} $ is generated as a $GARCH(1,1)$ process,
namely $h_{it}^{2}=\sigma _{i}^{2}(1-\psi _{0}-\psi _{1})+\psi
_{0}h_{i,t-1}^{2}+\psi _{1}(h_{i,t-1}\varepsilon _{i,t-1})^{2}$, with $%
\sigma _{i}^{2}\sim IID\left( 0.5+0.5z_{i}^{2}\right) $ and $z_{i}\sim
IIDN(0,1)$. We set $\psi _{0}=0.6$ and $\psi _{1}=0.2$, with the initial
values $h_{i,-M_{i}}=\sigma _{i}$.\footnote{%
Our approach also allows the coefficients of the $GARCH(1,1)$ model to be
heterogeneous across $i$, so long as they are drawn from the same common
distribution. But to keep the MC design simple, we are only reporting for
the case where $\psi _{0}$ and $\psi _{1}$ are homogeneous.} The case where
errors are conditionally homoskedastic over time is obtained as a special
case by setting $\psi _{0}=\psi _{1}=0$.

We generate the initial values of $\{y_{it}\}$ as $(y_{i,-M_{i}}-\mu
_{i})\sim IIDN(b,\kappa \sigma _{i}^{2})$ with $b=1$ and $\kappa =2$ for all 
$i$. Again the choice of $M_{i}$ is set depending on whether $\left\vert
\phi _{i}\right\vert <1$ or $\phi _{i}=1$. For the former case, we set $%
M_{i}=100$, which applies to all the units when $\phi _{i}$ is uniformly
distributed as it is not known which $\phi _{i}=1$, and units with $\phi
_{i}<1$ in the case of categorical-distributed $\phi _{i}$. For draws with $%
\phi _{i}=1$ in the categorical distribution, we set $M_{i}=1$ such that $%
y_{it}$ for $t=1,2,...,T$ has finite moments as $T$ is fixed in our design.

To check the robustness of the results to non-stationary initialization for $%
|\phi _{i}|<1$, when the processes start from a finite date in the past, we
conduct two sets of experiments, one set with $M_{i}=1$, and another set
with $M_{i}=3$ for all $i$.

The estimation of the moments of $\phi _{i}$, $\mu _{\phi } = E(\phi _{i})$
and $\sigma _{\phi }^{2} = Var(\phi_{i})$, are based on $\{y_{it}^{(r)},$
for $i=1,2,...,n;t=1,2,...,T\},$ where $r$ denotes the $r^{th}$ replication
of DGP in (\ref{DGPmc}). We carry out $2,000$ replications for the
experiments that compare the small sample performances of FDAC, HetroGMM,
and a number of estimators proposed in the literature for the homogeneous
slope case (denoted by HomoGMM), specifically, the estimators proposed by 
\cite{AndersonHsiao1981,AndersonHsiao1982} (AH), \cite{ArellanoBond1991}
(AB), \cite{BlundellBond1998} (BB), and the augmented Anderson-Hsiao (AAH)
estimator proposed by \cite{ChudikPesaran2021}, as well as the FDLS
estimator due to \cite{HanPhillips2010}.\footnote{%
We have downloaded the codes of the AH, AB, BB, and AAH estimators from the
supplementary materials of \cite{ChudikPesaran2021} using the link: %
\url{https://www.econ.cam.ac.uk/people-files/emeritus/mhp1/fp21/CP_AAH_paper_July_2021_codes_and_data.zip}%
. We are grateful to Alexander Chudik for making the codes publicly
available.} For experiments that compare our proposed estimator with the MSW
estimator in \cite{MavroeidisEtal2015}, we use $1,000$ replications as it
takes a substantial amount of time to compute the MSW estimator.\footnote{%
We have downloaded the codes of the MSW estimator used in empirical
applications from the supplementary materials of \cite{MavroeidisEtal2015}
using the link: %
\url{https://drive.google.com/file/d/1hdRFpcWo3r88YV_5Kc40ur-siCYGSBDN/view?usp=sharing}%
. We are grateful to Yuya Sasaki for also sharing the codes of the MSW
estimator used in their Monte Carlo experiments by private correspondence.}
To save space, the tables summarize the results of the MC experiments are
all included in the online supplement.

\vspace{-2ex}

\subsection{Comparison of FDAC and HetroGMM estimators\label{MChetro}}

\vspace{-1.5ex}

\subsubsection{MC results for estimation of $\protect\mu _{\protect\phi }$}

Bias, root mean square errors (RMSE), and size of tests of FDAC and HetroGMM
estimators of $\mu _{\phi }=E(\phi _{i})$ with uniformly distributed $\phi
_{i}$ are summarized in Table \ref{tab:MCAm1u} in the online supplement. The
results with categorically distributed $\phi _{i}$ are shown in Table \ref%
{tab:MCAm1c} in the online supplement. These tables provide results for the
sample size combinations $T=4,5,6,10$ and $n=100,1000,5000$, in the case of
Gaussian errors without GARCH effects. The parameters of distributions are
chosen to distinguish between cases where $|\phi _{i}|<1$ and $ \phi
_{i} \in (-1, 1]$, with the related results displayed in the left and right
panels of the tables, respectively.

In line with our theoretical results, both FDAC and HetroGMM estimators
offer reliable estimates for $\mu _{\phi }$ in the case of heterogeneous
short $T$ panels under both uniform and categorical distributions. The
categorical distribution yields marginally lower RMSEs, which is largely due
to the fact that $\sigma _{\phi }^{2}$ is much smaller under the categorical
distribution around $0.012$, as compared to $0.083$ under the uniform
distribution. More importantly, the magnitudes of bias, RMSE, and size are
very similar irrespective of whether $\left\vert \phi _{i}\right\vert <1$ or 
$\phi _{i} \in (-1,1]$. This result holds even if a fixed proportion of units
have unit roots, as is the case with the categorical distribution where $%
\phi _{i}=1$ in the case of 5 per cent of all units in the sample. The
empirical power functions for FDAC and HetroGMM estimators of $\mu _{\phi }$
are displayed in Figure \ref{fig:pw_fdac_hetrogmm_u2_a_m1} of the online
supplement for the uniformly distributed AR coefficients with $\phi _{i} \in (-1,1]$
in the baseline case (Gaussian errors and no GARCH effects).
The power functions for the other experiments are very similar and can be
obtained from the authors upon request.

Compared with the HetroGMM estimator, the FDAC estimator has uniformly
smaller biases across all sample size combinations, lower RMSE, and greater
power for $T=4,5,6$, and $n=100,1000,$ and $5000$. The differences between
the two estimators of $\mu _{\phi }$ become negligible only when $T=10$. In
the light of our discussion in sub-section \ref{hetrogmmestm}, this could be
because the FDAC estimator uses averages of the individual sample moments
both over time and across all units given the stationary properties of the
autocovariances of the first differences, and thus it is not subject to the
many moment problem that could adversely impact the HetroGMM estimator.
Consequently, tests based on the FDAC estimator are not adversely affected
as $T$ is increased with $n$ small, and its size is mostly around the
nominal size of five per cent. However, tests based on the HetroGMM
estimator tend to over-reject slightly as $T$ is increased when $n$ is
relatively small ($n=100$). For example, for the uniform distribution with $\phi _{i} \in (-1,1]$ 
and $n=100$, the size of the tests of $\mu _{\phi }=0.5$
based on the HetroGMM estimator rises from $5.7$ to $10.5$ per cent when $T$
is increased from $4$ to $10$. These findings are in line with the results
obtained in the literature when GMM is applied to homogeneous dynamic panels.%
\footnote{%
For GMM estimators with many moment conditions, some of the moment
conditions can be weak. The small-sample bias associated with the weak
moments will result in substantial size distortions, which become more
severe with greater weights on the weak moments. See also Section 6 of \cite%
{ChudikPesaran2021}.}

As can be seen from the empirical power functions in Figure \ref%
{fig:pw_fdac_hetrogmm_u2_a_m1}, the tests based on FDAC and HetroGMM
estimators can not reject $\mu _{\phi }=1$ with 100 per cent certainty due
to the small sample sizes with $n=100$. But as $n$ and $T$ increase, the
empirical power functions become steeper, illustrating an enhanced ability
to discern deviations from the null hypothesis.

\subsubsection{MC results for estimation of $\protect\sigma _{\protect\phi %
}^{2}$ \label{MCzig}}

As discussed in sub-section \ref{estzig}, the FDAC and HetroGMM estimators
of $\sigma _{\phi }^{2}$ are consistent so long as the true value of $\sigma
_{\phi }^{2}$, namely $\theta _{2,0}-\theta _{0,1}^{2}$ is not too close to
the boundary value of zero. Also to avoid negative estimates of the plug-in
estimator of $\sigma _{\phi }^{2}$ given by (\ref{Estvar}) we need $n$ to be
sufficiently large. Table \ref{tab:npzig} in the online supplement
summarizes the number of replications, out of $2,000$, with negative or
close to zero estimates (defined as estimates below $0.0001$) for the
baseline experiments and sample size combinations $n=100,1000,2500,5000$ and 
$T=5,6,10$. The frequencies of the HetroGMM estimator are noticeably higher
than those of the FDAC estimator for small $T$ and $n$. When $n=100$, a
sizeable proportion of the estimates of $\sigma _{\phi }^{2}$ are negative,
suggesting that $n=100$ is not sufficiently large for the asymptotic
properties to hold. However, as to be expected, the number of negative
estimates declines rapidly as $n$ and $T$ are increased. Accordingly, we
only focus on samples with $n\geq 1000$, and report the bias and RMSE of the
estimates of $\sigma _{\phi }^{2}$ for sample size combinations $%
n=1000,2500,5000$ and $T=5,6,10$. The results for the positive estimates are
summarized in Table \ref{tab:MCAvu} of the online supplement for uniformly
distributed $\phi _{i}$. For these sample size combinations, we only
encounter very few negative estimates and none when $n=5000$ and $T\geq 6$.%
\footnote{%
We did not consider estimating $\sigma _{\phi }^{2}$ under the categorical
distributions of $\phi _{i}$ since the associated true values of $\sigma
_{\phi }^{2}$ are too close to zero.}

Overall, both FDAC and HetroGMM estimators of $\sigma _{\phi }^{2}$ perform
well when $T=10$ or $n$ is large, with comparable performances whether $%
|\phi |<1$ or $\phi _{i} \in (-1,1]$. However, the FDAC estimator performs much
better for smaller values of $T$ and $n$, as can be seen from the larger
bias and RMSE of the HetroGMM estimator.

The empirical power functions for FDAC and HetroGMM estimators of $\sigma
_{\phi }^{2}$ are shown in Figure \ref{fig:pw_fdac_hetrogmm_u2_a_var} of the
online supplement for the uniformly distributed AR coefficients with $\phi _{i} \in (-1,1]$ 
in the baseline case (Gaussian errors and no GARCH effects).
The empirical power functions are flat around the true value of $\sigma
_{\phi }^{2}$ for $T=5$ and $n$ small. When $T=5$, large values of $n$ are
required to achieve reasonable power in the locality of the null hypothesis.
The power improves rapidly as $T$ and $n$ are increased and, in line with
the earlier results, the FDAC estimator performs better than the HetroGMM
estimator.

\vspace{-2ex}

\subsubsection{Robustness}

The FDAC estimators seem to be reasonably robust to departures from Gaussian
errors and the presence of GARCH effects. Table \ref{tab:MCm1e} of the
online supplement provides results for the four combinations of error
distributions, Gaussian and non-Gaussian, without and with GARCH effects for
estimation of $\mu _{\phi }$. This table reports the results for the
uniformly distributed AR coefficients with $\phi _{i} \in (-1,1]$ and $\mu
_{\phi }=0.5$. We obtain similar results when we generate $\phi _{i}$
following a categorical distribution. The RMSE and size distortions of the
FDAC estimator increase only slightly as we move from Gaussian to
non-Gaussian errors and as we allow for GARCH effects. In contrast, the
HetroGMM estimator is much more adversely affected by departures from
Gaussian errors. Its bias and RMSE are much higher, with large size
distortions, particularly with small $n$ ($n=100$). The performances of both
estimators are adversely affected when non-Gaussian errors are combined with
GARCH effects. Estimation of $\sigma _{\phi }^{2}$ is similarly adversely
affected when we allow for non-Gaussian errors as well as GARCH effects. The
related simulation results are summarized in Table \ref{tab:MCve} of the
online supplement.

Overall, the FDAC estimator outperforms the HetroGMM estimator and seems to
be reasonably robust to non-Gaussian errors and GARCH effects. It is also
simple to compute. In what follows we focus on the estimation of $\mu _{\phi
}$ and compare the FDAC estimator with the HomoGMM estimators as well as the
MSW estimator that allows for slope heterogeneity.

\vspace{-2ex}

\subsection{Comparison of FDAC and HomoGMM estimators\label{MChomo}}

Tables \ref{tab:hetro_u1_a} and \ref{tab:hetro_u2_a} in the online
supplement summarize the results comparing the FDAC estimator with FDLS, AH,
AAH, AB and BB estimators, where $\phi _{i}$ is uniformly distributed, $\phi
_{i}=\mu _{\phi }+v_{i}$ and $v_{i}\sim IIDU[-a,a]$ with $a=0.5$ and $\mu
_{\phi }=0.4$ ($|\phi _{i}|<1$) and $\mu _{\phi }=0.5$ ($\phi _{i} \in (-1,1]$%
). We use the sample size combinations, $T=4,6,10$, and $n=100,1000,5000,$
in the baseline case where the errors are Gaussian without GARCH effects.
The simulation results with the other error processes are available upon
request.

In line with our theoretical derivations, the HomoGMM estimators that
neglect heterogeneity are severely biased and show large size distortions,
whilst the bias of the FDAC estimator is close to zero and its size is
around the five per cent nominal level, irrespective of whether $|\phi
_{i}|<1$ or $\phi _{i} \in (-1,1]$. Also, with increases in $n$ and/or $T$, the
biases of the HomoGMM estimators do not shrink to zero and, as a result, the
size distortions of the HomoGMM estimators become even more pronounced. The
simulation results also confirm the magnitude of the asymptotic bias of the
AH estimator given by (\ref{AsyBiasAH}) in Section \ref{bias}, and those of
AB and BB estimators provided in Section \ref{NegBias} of the online
supplement.

Since it is not known if the heterogeneity bias is serious, it is natural to
ask if the FDAC estimator continues to perform equally well under
homogeneity ($\phi _{i}=\mu _{\phi }=0.5$ for all $i$), and if its
performance under homogeneity is comparable to those of HomoGMM estimators
of $\phi $. Accordingly, we also compute bias, RMSE, and size of the FDAC
and HomoGMM estimators under slope homogeneity $(a=0)$ with $\mu _{\phi
}=0.5 $. The results for Gaussian errors without GARCH effects are
summarized in Table \ref{tab:homo_a} of the online supplement. As can be
seen, the FDAC estimator continues to perform well even under slope
homogeneity. Its bias is close to zero and shows only a small degree of size
distortions when $n=100$. In terms of assumptions, the FDAC estimator is
closest to the FDLS estimator under homogeneity. Figure \ref%
{fig:pw_fdac_fdls_homo_a} in the online supplement compares the empirical
power functions of FDAC and FDLS estimators. Compared to the FDLS estimator,
the FDAC estimator makes use of higher order autocorrelation of first
differences that are not needed for identification of $\mu _{\phi }$ under
homogeneity. As a result, the FDLS estimator is marginally more powerful
than the FDAC for small $T=4$, while the opposite is the case for $T=10$.

When comparing the FDAC and the other HomoGMM estimators (such as AAH, BB,
or AB) one needs to be cautious however, since these estimators do allow for
the distribution of $y_{i0}$ to depart from the steady state distribution of 
$\{y_{it}\}$. With this in mind, we note that the FDAC estimator performs
well when compared to AH, AAH and AB estimators, although it is marginally
less efficient when compared to the BB estimator. Also, the FDAC estimator
has less size distortion and better power performance compared to all
HomoGMM estimators as $T$ is increased. In short, these results demonstrate
the FDAC estimator is reliable and has desirable small-sample performance
even in homogeneous panels with stationary outcome processes.

Figure \ref{fig:fdac_u25_errors_ab} in the online supplement shows the
empirical power functions for the FDAC estimator under homogeneity with $%
\phi_{i} = \mu_{\phi} = 0.5$ for all $i$ and heterogeneity with uniformly
distributed $\phi _{i} \in (-1,1]$ $(\mu_{\phi} = 0.5)$, in the cases
of Gaussian and non-Gaussian errors without GARCH effects. The empirical
power functions for the FDAC estimator in the cases of Gaussian errors
without and with GARCH effects are displayed in Figure \ref%
{fig:fdac_u25_errors_ac} of the online supplement. The power functions
become steeper as $n$ and $T$ increase. In general, the power of the FDAC
estimator is similar under heterogeneous and homogeneous $\phi _{i}$.
Consistent with the previous findings, with non-Gaussian errors and/or GARCH
effects, particularly for small $n=100$, the power functions become
noticeably flatter, and the size distortions become more pronounced.

\vspace{-2ex}

\subsection{Comparison of FDAC and MSW estimators\label{MCmsw}}

This section compares the small-sample performance of the FDAC estimator
with the MSW estimator by \cite{MavroeidisEtal2015}. Table \ref{tab:msw_m1}
in the online supplement reports bias, RMSE, and size of the FDAC and MSW
estimators for $\mu _{\phi } $ for $T=4,6,10$, and $n=100,1000$, with
uniformly distributed $\phi _{i}$ and Gaussian errors without GARCH effects.
The left and right panels of the table report results for $\mu _{\phi }=0.4$
($|\phi _{i}|<1$) and $\mu _{\phi }=0.5$ ($\phi _{i} \in (-1,1]$),
respectively. The performance of the FDAC estimator is in line with the ones
already discussed and as noted earlier is not affected by whether some $\phi
_{i}=1$ or not. In contrast, the MSW estimator performs rather poorly in the
presence of a high degree of heterogeneity in $\phi _{i}$ and shows large
biases and substantial size distortions across the examined sample sizes. In
the case of $\phi _{i} \in (-1,1]$, the MSW estimator shows greater bias, RMSE,
and size distortions.

\vspace{-2ex}

\subsection{Non-stationary initializations \label{Initial}}

Since the first differences of $y_{it}$ do not depend on the initial values
when $\phi _{i}=1$, non-stationary initialization matters only if $|\phi
_{i}|<1$. In this case, using (\ref{IniVal}) it is clear initial values
matter only when $M_{i}$ is small. Therefore, to investigate the robustness
of the FDAC estimator to different initializations we consider relatively
small values of $M_{i}=1$ and $3$ for all $i$, compared with the baseline
case where we set $M_{i}=100$ for all units with $\left\vert \phi
_{i}\right\vert <1$. The initial values are generated as $(y_{i,-M_{i}}-\mu
_{i})\sim IIDN(b,\kappa \sigma _{i}^{2})$ with $b=1$ and $\kappa =2$,
compared to their steady state values of $b=0$ and $\kappa_{i} =1/(1-\phi
_{i}^{2})$, respectively. When $\phi _{i}$ are generated from a categorical
distribution we set $M_{i}=1$ for all units with $\phi _{H}=1$.

We consider both uniformly and categorically distributed $\phi _{i}\,$. The
results for the uniformly distributed $\phi _{i}$ under the three
initializations $M_{i}\in \{100,3,1\}$ are summarized in Table \ref%
{tab:MCAu_mi} of the online supplement. Similar results when $\phi _{i}$
follow the categorical distribution are given in Table \ref{tab:MCAc_mi} of
the online supplement. It is clear that the FDAC estimator is adversely
affected when $M_{i}=1$ and displays bias and substantial size distortions.
As to be expected, the magnitude of the bias is not affected by $n$ but
falls sharply with $T$. As a result when $M_{i}=1$ we observe substantial
size distortions when $n$ is large. Comparing the upper and lower panels, as
the first differences of a unit root process are not affected by the initial
values, having some $\phi _{i}$ being close to one mitigates the negative
impact of non-stationary initializations on the FDAC estimator. These
impacts are more pronounced for categorically distributed $\phi _{i}$ where
the variances of $\phi _{i}$ are smaller, as shown in Table \ref{tab:MCAc_mi}
versus Table \ref{tab:MCAu_mi}. More importantly, as to be expected, the
bias and size distortion of the FDAC estimator disappear as $M_{i}$ is increased. When
moving from $M_{i}=1$ to $M_{i}=3$, the bias and size distortion shrink
fast, with only a slight size distortion observed when $M_{i}=3$.

We also consider the relative performance of the FDAC and HomoGMM estimators
under different initialization scenarios, for both cases of homogeneous and
heterogeneous panels. Results for the homogenous case when $\phi _{i}=\mu
_{\phi }=0.5$ are summarized in Table \ref{tab:homo_a_mi}, and results for
the heterogenous case are provided in Tables \ref{tab:hetro_u1_a_mi} and \ref%
{tab:hetro_u2_a_mi} for cases where $\mu _{\phi }=0.4$ ($|\phi _{i}|<1$) and 
$\mu _{\phi }=0.5$ ($\phi _{i} \in (-1.1]$), respectively. In the homogeneous
case, when $M_{i}=1$, the FDAC, FDLS, and BB estimators all show sizeable
bias and size distortions that do not vanish as $n$ increases. Also, as to
be expected, under homogeneity, the AH, AAH, and AB estimators are robust to
non-stationary initialization and have similar performances across different
values of $M_{i}$. In the case of heterogeneous panels, the performance of
the FDAC estimator is as discussed above. For the HomoGMM estimators, the
magnitude of neglected heterogeneity bias is smaller with less serious size
distortions when $M_{i}=1$ or $3$, as compared to $M_{i}=100$ (which
approximately corresponds to the stationary case). The AH estimator seems to
be an exception. Nonetheless, the HomoGMM estimators exhibit substantial
size distortions across most of the considered sample sizes, leading to
incorrect inference.

In short, for moderate values of $M_{i}$ (in the case of our experiments
when $M_{i}>3)$, the performance of the FDAC estimator is satisfactory even
when $y_{i0} - \mu_{i}$ are not drawn from the steady distribution of the
underlying processes, $\{y_{it}-\mu _{i}\}$. Comparisons of the FDAC and
HomoGMM estimators also highlight the trade-off that exists between the
\textquotedblleft non-stationary initialization" bias of the FDAC estimator
and the neglected heterogenous bias of the HomoGMM estimators. It remains a
challenge to simultaneously deal with heterogeneity of $\phi _{i}$ and the
non-stationarity of the initial values.

\vspace{-3ex}

\section{Empirical application: heterogeneity in earnings dynamics\label%
{application}}

\vspace{-2ex}

\subsection{Literature review of estimation of earnings dynamics}

Estimating earnings equations is crucial for answering some of the most
important economic

\noindent questions.\footnote{%
See p. 58 in \cite{Guvenen2009} for a brief summary of several economic
inquiries hinging on the estimation of earnings functions.} Variance of
earnings has been modeled and decomposed to measure income uncertainties in 
\cite{LillardWeiss1979}, \cite{MaCurdy1982}, \cite{CarrollSamwick1997}, \cite%
{MeghirPistaferri2004}, \cite{AltonjiEtal2013} and to quantify earnings
mobility in \cite{LillardWillis1978} and \cite{GewekeKeane2000}. The
covariance structures between earnings and other households'
characteristics, for example, work hours, consumptions and savings, have
been studied by \cite{AbowdCard1989}, \cite{HubbardEtal1995}, \cite%
{Guvenen2007}, and \cite{AlanEtal2018}.

Among these studies, a homogeneous AR or ARMA process is often used as a
component when modeling \textit{innovations} in earnings processes. Based on
the Restricted Income Profiles model that assumes homogenous linear trends
proposed in \cite{MaCurdy1982}, \cite{MaCurdy1982} and \cite{HubbardEtal1995} 
obtained close to unit root estimates for the AR(1) coefficient, ranging
from 0.946 to 0.998.\footnote{%
See Table 5 on p. 111 in \cite{MaCurdy1982} using an ARMA(1,1) process. See
Table 2 on p. 380 in \cite{HubbardEtal1995} based on an AR(1) process.}
Following this literature, a unit root assumption was imposed in \cite%
{CarrollSamwick1997} and \cite{MeghirPistaferri2004}. On the other hand,
using the Heterogeneous Income Profiles, by assuming unit-specific linear
trends, \cite{LillardWeiss1979} obtained estimates of the AR(1) coefficient
(assumed to be homogeneous) ranging from 0.153 to 0.860 for a sample with
PhD degrees. \cite{Guvenen2009} obtained estimates ranging
from 0.809 to 0.899 using PSID data.\footnote{%
See Tables 2, 4, 6 and 7 in \cite{LillardWeiss1979}, Table 1 on p. 64 in 
\cite{Guvenen2009}, and the abstract of \cite{GuKoenker2017}.}

There are also a number of studies that allow for heterogeneity in the AR(1)
coefficients. Prominent examples are \cite{BrowningEtal2010}, \cite%
{AlanEtal2018}, \cite{BrowningEtal2010}, and \cite{GuKoenker2017}. These
studies are typically based on panels with a moderate time dimension and
make parametric assumptions regarding the distribution of the AR(1)
coefficients; often using a Bayesian framework.\footnote{%
See pp. 227--232 in \cite{BrowningEjrnaes2013} for a comprehensive survey of
heterogeneity in parameters of earnings functions.} The application of the
FDAC estimator to earnings equation allows for heterogeneity in the AR(1)
coefficients without making any strong parametric assumptions, even when $T$
is as small as $5$. Also because of first differencing prior to estimation,
the FDAC estimator is robust to unobserved individual-specific
characteristics and is not subject to misspecification bias that could arise
when log real wages are filtered for individual-specific characteristics
before investigating the dynamics of the earnings process.

\subsection{A heterogeneous panel AR(1) model of earnings dynamics with
linear trends}

We consider estimating the earnings equation with fixed effects,
heterogeneous autoregressive coefficients, without imposing any restrictions
on the joint distributions of $\alpha _{i}$, $\phi _{i}$, and $y_{i0}$.
However, to accommodate growth in real earnings we extend our baseline model
in (\ref{Par1}) to allow for linear trends: 
\begin{equation}
y_{it}=\alpha _{i}+g_{i}(1-\phi _{i})t+\phi _{i}y_{i,t-1}+u_{it},
\label{earnings}
\end{equation}%
where $y_{it}=log(earnings_{it}/p_{t}),$ $earnings_{it}$ is the reported
earnings of individual $i$ in year $t$, $p_{t}$ is a general price, and $%
g_{i}$ is the growth rate of real earnings for individual $i$. (\ref%
{earnings}) can be written equivalently as 
\begin{equation*}
\tilde{y}_{it}(g_{i})=b_{i}+\phi _{i}\tilde{y}_{i,t-1}(g_{i})+u_{it},
\end{equation*}%
with $\tilde{y}_{it}(g_{i})=y_{it}-g_{i}t$ and $b_{i}=\alpha _{i}-g_{i}\phi
_{i}$. For $\vert \phi_{i}\vert < 1$, the steady state distribution of $%
y_{it}$ can now be derived using 
\begin{equation}
y_{it}=b_{i}+g_{i}t+\sum_{s=0}^{\infty }\phi _{i}^{s}u_{i,t-s}.  \label{Lyit}
\end{equation}%
When $T$ is sufficiently large, individual-specific growth rates, $g_{i}$,
can be estimated $\sqrt{T}$-consistently by running individual least squares
regressions of $y_{it}$ on an intercept and a linear trend, and then using
the residuals from these regressions to estimate the moments of $\phi _{i}$.
This approach requires $n$ and $T$ to be both large. In the case of the
present empirical application where $T$ is short ($5$ or $10)$, we provide
estimates of the moments of $\phi _{i}$ assuming that $g_{i}=g$ for
individuals within a given group, but allow $g$ to differ across groups,
classified by the educational attainment levels. $\sqrt{n}$-consistent
estimators of $g$ can be obtained either from the pooled regression of $%
y_{it}$ on fixed effects and a common linear trend, namely%
\begin{equation}
\hat{g}_{FE}=\left[\sum_{t=1}^{T}\left( t-\frac{(T+1)}{2}\right) ^{2}\right]%
^{-1}\left[\sum_{t=1}^{T}(\bar{y}_{\circ t}-\bar{y}_{\circ \circ })t\right],
\label{gFE}
\end{equation}%
with $\bar{y}_{\circ t}=n^{-1}\sum_{i=1}^{n}y_{it}$ and $\bar{y}_{\circ
\circ }=T^{-1}\sum_{t=1}^{T}\bar{y}_{\circ t}$, or after first differencing
of (\ref{Lyit}) by%
\begin{equation}
\hat{g}_{FD}=\frac{\sum_{t=2}^{T}\sum_{i=1}^{n}\Delta y_{it}}{n(T-1)}.
\label{gFD}
\end{equation}%
For small $T$ there is little to choose between these two estimators, and
they are identical when $T=2$. Given either of the above estimators,
generically denoted by $\hat{g}$, $\tilde{y}_{it}(\hat{g})=y_{it}-\hat{g} t$
can now be used to estimate the moments of $\phi _{i}$ using the FDAC or MSW
procedures.\footnote{%
Consistent estimation of $E(\phi _{i})$ in the presence of heterogeneity in
both $\phi _{i}$ and $g_{i}$ requires moderate to large values of $T$. The
approach used in the empirical literature whereby $y_{it}$ are first
de-meaned and de-trended for each $i$ prior to the estimation of $E(\phi
_{i})$ is subject to \cite{Nickell1981} bias in the case of short $T$
panels, even if $E(\phi _{i})=\phi $.}

In addition to the FDAC estimates, we also present estimates based on three 
estimation methods assuming homogeneous slope coefficients, namely AAH, AB,
and BB estimators proposed by \cite{ChudikPesaran2021}, \cite%
{ArellanoBond1991}, and \cite{BlundellBond1998}, and the MSW estimator of 
\cite{MavroeidisEtal2015}. Following \cite{MeghirPistaferri2004},
individuals in each time series sample are divided into three education
categories, where \textquotedblleft HSD" refers to high school dropouts with
less than 12 years of education, \textquotedblleft HSG" refers to high
school graduates with at least 12 but less than 16 years of education, and
\textquotedblleft CLG" refers to college graduates with at least 16 years of
education.\footnote{%
The sample for all individuals in both $5$ and $10$ yearly samples covered $%
3,113$ individuals with consecutive observations of nine years or more, and
36,325 individual-year observations.} To allow for possible time variations
in the estimates of mean earnings persistence we provide estimates for $five$
and $ten$ yearly non-overlapping sub-periods. The five yearly samples are
1976--1980, 1981--1985, 1986--1990 and 1991--1995. The ten yearly samples
are 1976--1985, 1981--1990 and 1991--1995. For each sub-period, we provide
estimates for all categories combined, as well as separate estimates for the
three educational sub-categories.\footnote{%
From 1997 PSID data are updated every two years. We confine our analysis to
the years 1976 to 1995 to construct panels with $5$ and $10$ consecutive
years.} To save space, the results for the last five and ten yearly samples
are given in the paper. The estimates for the earlier sub-periods are
provided in the online supplement.

Table \ref{tab:PSIDm1} gives the estimates of mean earnings persistence, $%
\mu_{\phi} = E(\phi _{i}),$ and the common linear trend coefficient, $g$,
for the sub-periods 1991--1995 ($T=5$) and 1986--1995 ($T=10$). The
estimates of $g$ are on average around $2$ per cent per annum with some
modest variations across the sub-samples and educational categories. The
HomoGMM estimates (AAH, AB and BB) differ a great deal, both over
sub-periods and across educational categories. The AAH estimates are all
around 0.50 and show little variations across the two sub-periods and the
educational categories. The AB estimates tend to be quite low and are not
statistically significant for two of the educational categories in the
shorter sub-period $(T=5)$. In contrast, the BB estimates are much larger
and in many instances are close to unity. For example, for the sub-period
1986--1995 $(T=10)$, the BB estimates of earnings persistence for the three
educational categories HSD, HSG and CLG are 0.923 (0.003), 0.914 (0.003) and
0.992 (0.004), respectively, with standard errors in brackets.

\begin{table}[h!]
\caption{Estimates of mean persistence ($\protect\mu_{\protect\phi} = E(%
\protect\phi_{i})$) of log real earnings in a panel AR(1) model with a
common linear trend using PSID data over 1991--1995 and 1986--1995}
\label{tab:PSIDm1}
\begin{center}
\scalebox{0.79}{
\renewcommand{\arraystretch}{1}
\begin{tabular}{cccccccccccc}
\hline \hline
 & \multicolumn{5}{c}{1991--1995, $T=5$} &  & \multicolumn{5}{c}{1986--1995, $T=10$} \\ \cline{2-6} \cline{8-12}
 & All &  & \multicolumn{3}{l}{Category by education} &  & All &  & \multicolumn{3}{l}{Category by education} \\ \cline{4-6} \cline{10-12}
 & categories &  & HSD & HSG & CLG &  & categories &  & HSD & HSG & CLG \\ \hline
\multicolumn{1}{l}{Homogeneous slopes} &  &  &  &  &  &  &  &  &  &  &  \\
AAH & 0.526 &  & 0.490 & 0.547 & 0.447 &  & 0.546 &  & 0.569 & 0.535 & 0.522 \\
 & (0.046) &  & (0.072) & (0.061) & (0.072) &  & (0.028) &  & (0.024) & (0.033) & (0.038) \\
AB & 0.278 &  & 0.105 & 0.320 & -0.013 &  & 0.311 &  & 0.310 & 0.335 & 0.232 \\
 & (0.081) &  & (0.147) & (0.097) & (0.133) &  & (0.039) &  & (0.045) & (0.044) & (0.070) \\
BB & 0.488 &  & 0.872 & 0.602 & 0.964 &  & 0.880 &  & 0.923 & 0.914 & 0.992 \\
 & (0.059) &  & (0.031) & (0.042) & (0.074) &  & (0.004) &  & (0.003) & (0.003) & (0.004) \\
 \multicolumn{1}{l}{Heterogeneous slopes} &  &  &  &  &  &  &  &  &  &  \\
FDAC & 0.586 &  & 0.582 & 0.567 & 0.635 &  & 0.636 &  & 0.580 & 0.611 & 0.734 \\
 & (0.042) &  & (0.132) & (0.056) & (0.065) &  & (0.023) &  & (0.071) & (0.028) & (0.040) \\
MSW & 0.437 &  & 0.431 & 0.436 & 0.452 &  & 0.458 &  & 0.459 & 0.452 & 0.460 \\
 & (0.040) &  & (0.044) & (0.043) & (0.045) &  & (0.054) &  & (0.038) & (0.046) & (0.063) \\
 &  &  &  &  &  &  &  &  &  &  &  \\
\multicolumn{1}{l}{Common linear trend} & 0.023 &  & 0.008 & 0.027 & 0.020 &  & 0.019 &  & 0.024 & 0.020 & 0.013\\ \hline
$n$ & 1,366 &  & 127 & 832 & 407 &  & 1,139 &  & 109 & 689 & 341\\
\hline\hline
\end{tabular}}
\end{center}
\par
{\footnotesize Notes: The estimates are based on $y_{it}=\alpha_{i}+g(1-%
\phi_{i})t+\phi_{i} y_{i,t-1}+u_{it}$, where $%
y_{it}=log(earnings_{it}/p_{t}) $ using the PSID data over the sub-periods
1991--1995 and 1986--1995. \textquotedblleft HSD" refers to high school
dropouts with less than 12 years of education, \textquotedblleft HSG" refers
to high school graduates with at least 12 but less than 16 years of
education, and \textquotedblleft CLG" refers to college graduates with at
least 16 years of education. $\hat{g}_{FD}$ is computed by (\ref{gFD}), then 
$\mu_{\phi}$ is estimated based on $\tilde{y}_{it} = y_{it} - \hat{g}_{FD} t$%
. \textquotedblleft AAH", \textquotedblleft AB", and \textquotedblleft BB"
denote the 2-step GMM estimators by \cite{ChudikPesaran2021}, \cite%
{ArellanoBond1991}, and \cite{BlundellBond1998}. The FDAC estimator is
calculated by (\ref{Estm1}). \textquotedblleft MSW" denotes the estimator by 
\cite{MavroeidisEtal2015}.}
\end{table}

We also find sizeable differences in the estimates of mean earnings
persistence when we consider the FDAC and MSW estimators. The MSW estimates
are all around 0.45 and do not vary with the level of educational
attainment. In contrast, the FDAC estimates are somewhat larger (lie in the
range of 0.570--0.734) and rise with the level of educational attainment.
This pattern can be seen in both sub-periods. For example, for the longer
sub-period (1986-1995), the mean persistence for HSD, HSG and CLG categories
are estimated to be 0.580 (0.071), 0.611 (0.028) and 0.735 (0.040),
respectively. Similar results are obtained for the other sub-periods. See
Tables \ref{tab:PSIDm1t5} and \ref{tab:PSIDm1t10} of the supplement.
Interestingly, the higher earnings persistence of the college graduate
category is a prominent feature of the FDAC estimates for all sub-periods.
This result is also in line with a number of theoretical arguments in the
literature in terms of higher mobility of college graduates and their
relative job stability, for example, \cite{CarrollSamwick1997} and \cite%
{CarneiroEtal2023}.

Although we have not developed a formal statistical test of the
heterogeneity $\phi _{i}$, the estimates of $\sigma _{\phi }^{2}$ provide a
good indication of the degree of within-group heterogeneity. Estimates of $%
\sigma _{\phi }^{2}$ based on MSW and FDAC procedures for the various
sub-periods are given in Tables \ref{tab:PSIDvar}--\ref{tab:PSIDvart5} of
the online supplement. The FDAC estimates are much larger than the MSW
estimates. For example, for the sub-period 1986--1995 the MSW estimates of $%
\sigma _{\phi }^{2}$ are all around 0.011 with standard errors in the range
of 0.005--0.011, whilst the FDAC estimates of $\sigma _{\phi }^{2}$ for the
same sub-period are 0.122 (0.06), 0.12 (0.031) and 0.141 (0.036) for the
three educational categories of HSD, HSG and CLG, respectively. The degree
of within-group heterogeneity also seems to vary over time. For example, for
the shorter sub-period (1991-1995), the FDAC estimates of $\sigma _{\phi
}^{2}$ are generally smaller with larger standard errors for the two
categories of HSG and CLG.

\vspace{-3ex}

\section{Conclusion\label{conclusion}}

\vspace{-1.5ex}

This paper considers the estimation of heterogeneous panel AR(1) models with
short $T$, as $n\rightarrow \infty $. It allows for individual fixed effects
and proposes estimating the moments of the AR(1) coefficients, $E(\phi
_{i}^{s})$, for $s=1,2,...,S$, using the autocorrelation functions of first
differences. It is shown that the standard GMM estimators proposed in the
literature for short $T$ homogeneous panels are inconsistent in the presence
of slope heterogeneity. Analytical expressions for the bias are derived and
shown to be very close to estimates obtained from stochastic simulations.

We propose two moment based estimators. A simple estimator based on
autocorrelations of first differences, denoted by FDAC, and a GMM estimator
based on autocovariances of first differences denoted by HetroGMM. Both
estimators allow for some of the cross section units to have unit roots.

The small sample properties of the proposed estimators are investigated
using Monte Carlo experiments. It is shown that the FDAC estimators of $\mu
_{\phi }$ and $\sigma _{\phi }^{2}$ perform much better than the
corresponding HetroGMM estimator. We also find that quite large samples
might be required for reliable estimation of $\sigma _{\phi }^{2}$, assuming
that the true value of $\sigma _{\phi }^{2}$ is not too close to zero.

The simulation results also show that the FDAC estimator of $\mu _{\phi }$
is robust to different distributions of autoregressive coefficients and
error processes. Further, we find that the FDAC estimator performs well even
under homogeneous AR(1) coefficients. The magnitudes of bias and RMSE of the
FDAC estimator are comparable to the HomoGMM estimators, and the size of the
tests based on the FDAC estimator is mostly around the 5 per cent nominal
level. But when initializations of the outcome processes deviate from their
associated steady state distributions, the FDAC estimator could suffer from
bias and size distortions. There is a trade-off between heterogeneity bias
and the bias due to the non-stationary initializations.

The utility of the FDAC estimators of $\mu _{\phi }$ and $\sigma _{\phi
}^{2} $ is illustrated by an empirical application using the 1976--1995 PSID
data to estimate heterogeneous AR(1) panels in log real earnings with a
common linear trend. We provide estimates of $\mu _{\phi }$ and $\sigma
_{\phi }^{2} $ over a number of $5$ and $10$ yearly sub-periods, with three 
educational groups. The estimates of $\mu _{\phi }$ differ systematically
across the education groups, with the mean persistence of real earnings
rising with the level of educational attainments (high school dropouts, high
school graduates, and college graduates). The estimates of $\sigma _{\phi
}^{2}$ differ across periods and levels of educational attainment but do not
display any particular patterns.

It is important to acknowledge that the scope of the present paper is
limited, with a number of remaining challenges: (a) allowing for
individual-specific time-varying covariates, and (b) simultaneously dealing
with heterogeneity and non-stationary initializations. It is not clear that
such extensions will be possible without relaxing the assumption that $T$ is
short and fixed, as $n\rightarrow \infty $. But these are clearly important
topics for future research.

{\small \setstretch{1.26} 
\bibliographystyle{chicago}
\bibliography{HPARref}
}



\newpage

\begin{center}
\thispagestyle{empty}

{\LARGE Online Supplement to }

{\LARGE \bigskip }

{\LARGE \textquotedblleft Heterogeneous Autoregressions in Short $T$ Panel
Data Models \textquotedblright\ }{\Large \bigskip \bigskip }

{\large {M. Hashem Pesaran } }

{\normalsize University of Southern California, and Trinity College,
Cambridge }

{\normalsize \vskip 0.5em }{\large Liying Yang }

{\normalsize Postdoctoral research fellow, University of British Columbia }

{\large \vskip 1.5em \today
}

{\normalsize \vskip 30pt }
\end{center}

\clearpage
\makeatletter
\setcounter{page}{1}\renewcommand{\thepage}{S\arabic{page}} %
\setcounter{table}{0} \renewcommand{\thetable}{S.\arabic{table}} %
\renewcommand{\theHtable}{S.\arabic{table}} \setcounter{section}{0} %
\renewcommand{\thesection}{S.\arabic{section}} \setcounter{figure}{0} %
\renewcommand{\thefigure}{S.\arabic{figure}} \setcounter{footnote}{0} %
\renewcommand{\thefootnote}{S\arabic{footnote}} \renewcommand{%
\thetheorem}{S.\arabic{theorem}}\setcounter{theorem}{0} \renewcommand{%
\theproposition}{S.\arabic{proposition}}\setcounter{proposition}{0} %
\renewcommand{\theassumption}{S.\arabic{assumption}}%
\setcounter{assumption}{0} \renewcommand{\thelemma}{S.\arabic{lemma}}%
\setcounter{lemma}{0} \renewcommand{\theremark}{S.\arabic{remark}}%
\setcounter{remark}{0} \makeatother

\renewcommand{\thesection}{S}

\subsection{Introduction}

This online supplement is organized as follows. Section \ref{ExiAutocov}
provides a proof of Lemma \ref{VarAuto} under the stationarity of the first
differences, $\Delta y_{it} = y_{it} - y_{i,t-1}$. Section \ref{ExU} further
illustrates the convergence property with uniformly distributed
autoregressive coefficients, $\phi _{i}$. Section \ref{NegBias} derives
expressions for the analytical bias of the AB and BB estimators under
heterogeneity of $\phi _{i}$ when $T=4$. Section \ref{Asyvar} derives the
asymptotic covariance matrix for the HetroGMM estimator of the first two
moments and its consistent estimator. Section \ref{EPF} provides formulae
for empirical power functions of the tests based on our proposed estimators
in the Monte Carlo simulations. Section \ref{MCevidence} provides additional
Monte Carlo evidence. Section \ref{subpsid} describes the sample
(1976--1995) of the Panel Study of Income Dynamics (PSID) data used in the
empirical application and provides estimation results for a number of
sub-periods in addition to the ones reported in the main paper.

\subsection{Proof of Lemma \protect\ref{VarAuto}: Existence of
autocovariances of first differences\label{ExiAutocov}}

We first establish conditions under which first differences, $\Delta y_{it},$
are covariance stationary for any given $t$ and $i$. Consider the result (%
\ref{DyitG}) in the main paper which we reproduce here for convenience: 
\begin{equation}
\Delta y_{it}=u_{it}-(1-\phi _{i})\sum_{\ell =1}^{M_{i}+t-1}\phi _{i}^{\ell
-1}u_{i,t-\ell }-\phi _{i}^{M_{i}+t-1}(1-\phi _{i})\left( y_{i,-M_{i}}-\mu
_{i}\right) ,\text{ }  \label{DyitApp}
\end{equation}%
for $t=2,3,...,T,$ where $R_{i}\left( y_{i,-M_{i}}\right) =-\phi
_{i}^{M_{i}-1}(1-\phi _{i})\left( y_{i,-M_{i}}-\mu _{i}\right) $. Assuming $%
\phi _{i}$ and $u_{it}$ are independently distributed and since the initial
values, $y_{i,-M_{i}}-\mu _{i},$ are given, we have 
\begin{equation*}
E\left\vert \Delta y_{it}\right\vert \leq E\left\vert u_{it}\right\vert +\sum_{\ell
=1}^{M_{i}+t-1}E\left\vert \phi _{i}^{\ell -1}(1-\phi _{i})\right\vert
E\left\vert u_{i,t-\ell }\right\vert +E\left[ \left\vert \phi
_{i}^{M_{i}+t-1}(1-\phi _{i})\right\vert \right] \left\vert y_{i,-M_{i}}-\mu
_{i}\right\vert .
\end{equation*}%
Also since $\phi _{i}\in (-1,1]$, then $E\left\vert \phi _{i}^{s}(1-\phi
_{i})\right\vert \leq c^{s},$ for some $c<1$, and we have 
\begin{equation*}
\sup_{i,t}E\left[ \left\vert \Delta y_{it} \right\vert \left\vert (y_{i,-M_{i}}-\mu
_{i})\right. \right] \leq \sup_{i,t}\left\vert u_{it}\right\vert \left[ 1+%
\frac{1-c^{M_{i}+t-1}}{1-c}\right] +c^{M_{i}+t-1}\left\vert y_{i,-M_{i}}-\mu
_{i}\right\vert <C<\infty .
\end{equation*}%
Hence, $E\left\vert y_{it}\right\vert $ exists for all values of $\phi
_{i}\in (-1,1]$ and is given by 
\begin{equation*}
E\left( \Delta y_{it}\left\vert y_{i,-M_{i}}-\mu _{i}\right. \right) =-E\left[ \phi
_{i}^{M_{i}+t-1}(1-\phi _{i})\right] \left( y_{i,-M_{i}}-\mu _{i}\right) .
\end{equation*}%
It is clear that, since $t=1,2,...,T$ and $T$ is finite, then $E\left(
\Delta y_{it} \right) $ varies with $t$ and in general depends on the initial
values, $y_{i,-M_{i}}$. $E\left( \Delta y_{it}\left\vert y_{i,-M_{i}}-\mu
_{i}\right. \right) $ is time-invariant if and only if $M_{i}\rightarrow
\infty $, and hence unconditionally we have $E(\Delta y_{it})=0$, for all $i$ and $%
t $, if $M_{i}\rightarrow \infty $.

By Cauchy-Schwarz inequality $\left\vert \gamma _{\Delta }(h)\right\vert
=\left\vert E\left( \Delta y_{it}\Delta y_{i,t-h}\right) \right\vert \leq %
\left[ E\left( \Delta y_{it}\right) ^{2}E\left( \Delta y_{i,t-h}\right) ^{2}%
\right] ^{\frac{1}{2}}$, thus for the existence of autocovariances of $%
\Delta y_{it}$, it is sufficient to show that $E\left( \Delta y_{it}\right)
^{2}<\infty $. Using (\ref{Sdy1}) in the main paper, it readily follows that%
\begin{equation}
E\left( \Delta y_{it}\right) ^{2}=E\left[ E\left( \left( \Delta
y_{it}\right) ^{2}\left\vert \phi _{i},\sigma _{i}^{2}\right. \right) \right]
=E\left( \sigma _{i}^{2}\right) +E\left[ (1-\phi _{i})^{2}\sum_{\ell
=1}^{\infty }\phi _{i}^{2\left( \ell -1\right) }\sigma _{i}^{2}\right] ,
\label{Dyit2}
\end{equation}%
and given the independence of $\sigma _{i}^{2}$ and $\phi _{i}$ (see
Assumption \ref{error variances} in the main paper) we have%
\begin{equation*}
E\left( \Delta y_{it}\right) ^{2}=\sigma ^{2}+\sigma ^{2}\sum_{\ell
=1}^{\infty }E\left[ (1-\phi _{i})^{2}\phi _{i}^{2\left( \ell -1\right) }%
\right] \leq \sigma ^{2}+\sigma ^{2}\sum_{s=0}^{\infty }E\left[ (1-\phi
_{i})^{2}\phi _{i}^{2s}\right] .
\end{equation*}%
We now show that $\sum_{s=0}^{\infty }E\left[ (1-\phi _{i})^{2}\phi _{i}^{2s}%
\right] $ is convergent for any probability distributions of $\phi _{i}$
defined over the interval $(-1,+1]$. Note that for any finite $M$ 
\begin{eqnarray*}
\sum_{s=0}^{M}E\left[ (1-\phi _{i})^{2}\phi _{i}^{2s}\right] &=&E\left[
\sum_{s=0}^{M}(1-\phi _{i})^{2}\phi _{i}^{2s}\right] =E\left[ \frac{(1-\phi
_{i})^{2}\left( 1-\phi _{i}^{2M+2}\right) }{1-\phi _{i}^{2}}\right] \\
&=&E\left[ \frac{(1-\phi _{i})\left( 1-\phi _{i}^{2M+2}\right) }{1+\phi _{i}}%
\right],
\end{eqnarray*}%
where $1+\phi _{i}>\epsilon >0$, and $-1 < \phi_{i} \leq 1$. 
\begin{equation*}
\frac{(1-\phi _{i})\left( 1-\phi _{i}^{2M+2}\right) }{1+\phi _{i}}\leq
(1/\epsilon )(1+\left\vert \phi _{i}\right\vert +\left\vert \phi
_{i}^{2M+2}\right\vert +\left\vert \phi _{i}^{2M+3}\right\vert ).
\end{equation*}%
Hence, 
\begin{equation*}
\sum_{s=0}^{M}E\left[ (1-\phi _{i})^{2}\phi _{i}^{2s}\right] \leq
(1/\epsilon )(1+\left\vert \phi _{i}\right\vert +\left\vert \phi
_{i}^{2M+2}\right\vert +\left\vert \phi _{i}^{2M+3}\right\vert ).
\end{equation*}%
But since $\phi_{i} \in (-1, 1]$, $E\left\vert \phi
_{i}^{\ell }\right\vert \leq 1$ for any $\ell =0,1,...,$ and it follows that 
$\sum_{s=0}^{M}E\left[ (1-\phi _{i})^{2}\phi _{i}^{2s}\right] $ $\leq
4/\epsilon $ for any finite $M$ and as $M\rightarrow \infty $. Therefore, it
follows that $\left\vert \gamma _{\Delta }(h)\right\vert <C$, as required.

Having established the existence of $\gamma _{\Delta }(h)$, using (\ref%
{Dyit2}) and recalling that under Assumption \ref{error variances} in the
main paper $\phi _{i}$ and $\sigma _{i}^{2}$ are independently distributed
we have 
\begin{eqnarray*}
Var(\Delta y_{it}) &=&\gamma _{\Delta }(0)=E\left[ \sigma _{i}^{2}+(1-\phi
_{i})^{2}\sum_{\ell =1}^{\infty }\phi _{i}^{2\left( \ell -1\right) }\sigma
_{i}^{2}\right] \\
&=&E\left[ \sigma _{i}^{2}+\frac{(1-\phi _{i})^{2}}{1-\phi _{i}^{2}}\sigma
_{i}^{2}\right] =2\sigma ^{2}E\left( \frac{1}{1+\phi _{i}}\right) .
\end{eqnarray*}%
Similarly, to derive $\gamma _{\Delta }(h)=E\left( \Delta y_{it}\Delta
y_{i,t-h}\right) $ we first note that $M_{i}\rightarrow \infty $, then using
(\ref{Dyit2}) we have 
\begin{equation*}
\Delta y_{it}=u_{it}-(1-\phi _{i})\sum_{\ell =1}^{\infty }\phi _{i}^{\ell
-1}u_{i,t-\ell },\text{ }
\end{equation*}%
\begin{equation*}
\Delta y_{i,t-h}=u_{i,t-h}-(1-\phi _{i})\sum_{\ell =1}^{\infty }\phi
_{i}^{\ell -1}u_{i,t-\ell -h},\text{ }
\end{equation*}%
and for $h=1,2,...,$ 
\begin{eqnarray*}
E\left( \Delta y_{it}\Delta y_{i,t-h}\right) &=&E\left[ (1-\phi
_{i})^{2}\left( \sum_{\ell =1}^{\infty }\phi _{i}^{\ell -1}u_{i,t-\ell
}\right) \left( \sum_{\ell =1}^{\infty }\phi _{i}^{\ell -1}u_{i,t-\ell
-h}\right) \right] \\
&&-E\left[ (1-\phi _{i})\left( \sum_{\ell =1}^{\infty }\phi _{i}^{\ell
-1}u_{i,t-\ell }u_{i,t-h}\right) \right].
\end{eqnarray*}%
First, we consider the second term, and note that 
\begin{equation*}
E\left[ (1-\phi _{i})\left( \sum_{\ell =1}^{\infty }\phi _{i}^{\ell
-1}u_{i,t-\ell }u_{i,t-h}\right) \right] =E\left[ \sigma _{i}^{2}(1-\phi
_{i})\phi _{i}^{h-1}\right] .
\end{equation*}%
Also%
\begin{eqnarray*}
E\left[ (1-\phi _{i})^{2}\left( \sum_{\ell =1}^{\infty }\phi _{i}^{\ell
-1}u_{i,t-\ell }\right) \left( \sum_{\ell =1}^{\infty }\phi _{i}^{\ell
-1}u_{i,t-\ell -h}\right) \right] = E\left[ \sigma _{i}^{2}(1-\phi
_{i})^{2}\left( \phi _{i}^{h}+\phi _{i}^{h+2}+\phi _{i}^{h+4}+...\right) %
\right].
\end{eqnarray*}%
Hence%
\begin{equation*}
E(\Delta y_{it}\Delta y_{i,t-h})=-E\left[ \sigma _{i}^{2}(1-\phi _{i})\phi
_{i}^{h-1}\right] +E\left[ \sigma _{i}^{2}(1-\phi _{i})^{2}\left( \phi
_{i}^{h}+\phi _{i}^{h+2}+\phi _{i}^{h+4}+...\right) \right] ,
\end{equation*}%
and since $\phi _{i}$ and $\sigma _{i}^{2}$ are independently distributed we
have%
\begin{equation*}
E(\Delta y_{it}\Delta y_{i,t-h})=-E\left( \sigma _{i}^{2}\right) E\left[
(1-\phi _{i})\phi _{i}^{h-1}-(1-\phi _{i})^{2}\left( \phi _{i}^{h}+\phi
_{i}^{h+2}+\phi _{i}^{h+4}+...\right) \right] ,
\end{equation*}%
As before, for all $\phi _{i}\in (-1,1],$ we have $E\left\vert (1-\phi
_{i})^{2}\phi _{i}^{h+s}\right\vert \leq E\left\vert (1-\phi
_{i})^{2}\left\vert \phi _{i}\right\vert ^{h+s}\right\vert \leq c^{h+s}$,
where $c<1$, and the series is convergent, and we have 
\begin{equation*}
E(\Delta y_{it}\Delta y_{i,t-h})=-E\left( \sigma _{i}^{2}\right) E\left[
(1-\phi _{i})\phi _{i}^{h-1}-\frac{(1-\phi _{i})\phi _{i}^{h}}{1+\phi _{i}}%
\right],
\end{equation*}%
or%
\begin{equation}
E(\Delta y_{it}\Delta y_{i,t-h})=-E\left( \sigma _{i}^{2}\right) E\left[ 
\frac{(1-\phi _{i})\phi _{i}^{h-1}}{1+\phi _{i}}\right] ,\text{ for }%
h=1,2,...,  \label{dym1}
\end{equation}%
as required. Similarly 
\begin{eqnarray*}
E\left( \phi _{i}\Delta y_{it}\Delta y_{i,t-h}\right) &=&-E\left( \sigma
_{i}^{2}\right) E\left[ (1-\phi _{i})\phi _{i}^{h}-\frac{(1-\phi _{i})\phi
_{i}^{h+1}}{1+\phi _{i}}\right] \\
&=&-E\left( \sigma _{i}^{2}\right) E\left[ \frac{(1-\phi _{i})\phi _{i}^{h}}{%
1+\phi _{i}}\right] ,\text{ for }h=1,2,....
\end{eqnarray*}%
The results of Lemma \ref{VarAuto} are now established noting that $E\left(
\sigma _{i}^{2}\right) =\sigma ^{2}$.

\subsection{Examples: uniform distributions\label{ExU}}

It is also instructive to consider the important case where $\phi _{i}$ is
uniformly distributed. First suppose that $\phi _{i}\thicksim Uniform(0,a] $
for $0<a\leq 1$, then \thinspace $E\left( \phi _{i}^{\ell }\right) =\frac{%
a^{\ell }}{\ell +1}$, and%
\begin{equation*}
E\left[ (1-\phi _{i})^{2}\phi _{i}^{2s}\right] =\frac{a^{2s}}{2s+1}-\frac{%
2a^{2s+1}}{2s+2}+\frac{a^{2s+2}}{2s+3}.
\end{equation*}%
Hence 
\begin{equation*}
\sum_{s=0}^{\infty }E\left[ (1-\phi _{i})^{2}\phi _{i}^{2s}\right]
=\sum_{s=0}^{\infty }\left( \frac{a^{2s}}{2s+1}-\frac{2a^{2s+1}}{2s+2}+\frac{%
a^{2s+2}}{2s+3}\right) .
\end{equation*}%
When $a<1$, all the three individual sums in the above expression are
bounded by $C/(1-a^{2})$. However, this does not follow when $a=1$, and the
series $\sum_{s=0}^{\infty }\frac{1}{2s+1}$, $\sum_{s=0}^{\infty }\frac{2}{%
2s+2}$, and $\sum_{s=0}^{\infty }\frac{1}{2s+3}$, diverge individually.
Hence, to investigate the convergence property of $\sum_{s=0}^{\infty }E%
\left[ (1-\phi _{i})^{2}\phi _{i}^{2s}\right] $ when $a=1$, we need to
consider all the terms together. For $a=1$, 
\begin{equation*}
\sum_{s=0}^{\infty }E\left[ (1-\phi _{i})^{2}\phi _{i}^{2s}\right]
=\sum_{s=0}^{\infty }\left( \frac{1}{2s+1}-\frac{2}{2s+2}+\frac{1}{2s+3}%
\right),
\end{equation*}%
and after some algebra we have 
\begin{equation*}
\frac{1}{2s+1}-\frac{2}{2s+2}+\frac{1}{2s+3}=\frac{2}{\left( 2s+1\right)
\left( 2s+2\right) \left( 2s+3\right) },
\end{equation*}%
\begin{equation*}
\sum_{s=0}^{\infty }E\left[ (1-\phi _{i})^{2}\phi _{i}^{2s}\right]
=\sum_{s=0}^{\infty }\frac{2}{\left( 2s+1\right) \left( 2s+2\right) \left(
2s+3\right) }<C<\infty .
\end{equation*}%
Similarly, for $\phi _{i}\thicksim Unifrom(-1+\epsilon ,0]$ we have $E\left(
\phi _{i}^{\ell }\right) =-\frac{(-1)^{\ell }\left( 1-\epsilon \right)
^{\ell }}{\ell +1}$, and we have 
\begin{equation*}
E\left[ (1-\phi _{i})^{2}\phi _{i}^{2s}\right] =\frac{\left( 1-\epsilon
\right) ^{2s}}{2s+1}+\frac{2\left( 1-\epsilon \right) ^{2s+1}}{2s+2}+\frac{%
\left( 1-\epsilon \right) ^{2s+2}}{2s+3},
\end{equation*}%
and $\sum_{s=0}^{\infty }E\left[ (1-\phi _{i})^{2}\phi _{i}^{2s}\right] $ is
convergent for $\epsilon >0$, and diverges if $\epsilon =0$. The latter case
is ruled out under Assumption \ref{hetro} in the main paper, which
establishes the necessity of ruling out the boundary value of $\phi _{i}=-1$.

\subsection{Proof of Proposition \protect\ref{NegBiasAH}: Neglected
heterogeneity bias of the AH estimator \label{BiasAH}}

Result (\ref{AH3}) follows directly from (\ref{AH2}), after subtracting $%
E\left( \phi _{i}\right) $ from\ both sides. Also, since $\phi _{i}\in
\lbrack -1+\epsilon ,1]$, for some $\epsilon >0$, then $1+E(\phi _{i})>0$,
and $E\left( \frac{1-\phi _{i}}{1+\phi _{i}}\right) >0$. Since $1/(1+\phi
_{i})$ is a convex function of $\phi _{i}$ on $[-1+\epsilon ,1]$, then by
Jensen inequality $E\left( \frac{1}{1+\phi _{i}}\right) \geq \frac{1}{%
1+E(\phi _{i})}$, and it follows that $plim_{n\rightarrow \infty }\hat{\phi}%
_{AH}\leq E(\phi _{i})=\mu _{\phi }$. Since $1+\mu _{\phi }=1+E(\phi _{i})>0$%
, the asymptotic bias is zero if and only if $\frac{1}{1+\mu _{\phi }}%
=E\left( \frac{1}{1+\phi _{i}}\right) $, and due to the convexity of $%
1/(1+\phi _{i})$, this condition is met only if $\phi _{i}=\mu _{\phi }$ for
all $i$.

\subsection{Neglected heterogeneity bias in AB and BB estimators\label%
{NegBias}}

The AB estimator proposed by \cite{ArellanoBond1991} is based on the
following moment conditions:\footnote{%
See equation (8) on p. 5 in \cite{ChudikPesaran2021}.} 
\begin{equation}
E(y_{is}\Delta u_{it})=0\text{, for }i=1,2,...,n,s=1,2,...,t-2\text{, and }%
t=3,4,...,T,  \label{sys1}
\end{equation}%
which can also be written as $E[y_{is}(\Delta y_{it}-\phi _{i}\Delta
y_{i,t-1})]=0,$ with $(T-1)(T-2)/2$ moment conditions in total. When $T=4$,
under homogeneity of $\phi_{i}$, the AB moment conditions are given by $%
E[y_{i1}(\Delta y_{i3}-\phi \Delta y_{i2})]=0$, $E[y_{i1}(\Delta y_{i4}-\phi
\Delta y_{i3})]=0$, and $E[y_{i2}(\Delta y_{i4}-\phi \Delta y_{i3})]=0$.
With a fixed weight matrix $\mathbf{W}_{AB}$, the AB estimator can be
written as 
\begin{equation}
\hat{\phi}_{AB}=\left( \boldsymbol{\bar{z}}_{na}^{\prime }\boldsymbol{W}_{AB}%
\boldsymbol{\bar{z}}_{na}\right) ^{-1}\left( \boldsymbol{\bar{z}}%
_{na}^{\prime }\boldsymbol{W}_{AB}\boldsymbol{\bar{z}}_{nb}\right) ,
\label{AB}
\end{equation}%
where $\boldsymbol{\bar{z}}_{na}=n^{-1}\left( \sum_{i=1}^{n}y_{i1}\Delta
y_{i2},\sum_{i=1}^{n}y_{i1}\Delta y_{i3},\sum_{i=1}^{n}y_{i2}\Delta
y_{i3}\right) ^{\prime },$ and

\noindent $\boldsymbol{\bar{z}}_{nb}=n^{-1}\left( \sum_{i=1}^{n}y_{i1}\Delta
y_{i3},\sum_{i=1}^{n}y_{i1}\Delta y_{i4},\sum_{i=1}^{n}y_{i2}\Delta
y_{i4}\right) ^{\prime }.$

Using (\ref{Par2}) in the main paper 
\begin{equation}
y_{it}=\mu _{i}+\phi _{i}^{M_{i}+t}(y_{i,-M_{i}}-\mu _{i})+\sum_{\ell
=0}^{M_{i}+t-1}\phi _{i}^{\ell }u_{i,t-\ell },  \label{y2}
\end{equation}%
and assuming that $\mu _{i}$ is distributed independently of $\{u_{it}\}$
(as assumed under AB) then using (\ref{DyitG}) in the main paper and (\ref%
{y2}) we have 
\begin{align*}
& E\left( y_{i,t-h}\Delta y_{it}|\alpha _{i},\phi _{i},\sigma _{i}^{2}\right)
\\
=& E\left[ \left( \mu _{i}+\phi _{i}^{M_{i}+t-h}(y_{i,-M_{i}}-\mu
_{i})+\sum_{\ell =0}^{M_{i}+t-h-1}\phi _{i}^{\ell }u_{i,t-h-\ell }\right)
\right. \\
& \left. \left. \times \left( u_{it}-(1-\phi _{i})\sum_{\ell
=1}^{M_{i}+t-1}\phi _{i}^{\ell -1}u_{i,t-\ell }-\phi _{i}^{M_{i}+t-1}(1-\phi
_{i})\left( y_{i,-M_{i}}-\mu _{i}\right) \right) \right\vert \alpha
_{i},\phi _{i},\sigma _{i}^{2}\right] \\
=& E\left[ \left. -(1-\phi _{i})\left( \sum_{\ell =0}^{M_{i}+t-h-1}\phi
_{i}^{h-1+2\ell }u_{i,t-h-\ell }^{2}\right) -(1-\phi _{i})\phi
_{i}^{2M_{i}+2t-h-1}\left( y_{i,-M_{i}}-\mu _{i}\right) \right\vert \phi
_{i},\sigma _{i}^{2}\right]
\end{align*}%
As $M_{i}\rightarrow \infty $ for $|\phi _{i}|<1$ (with finite $M_{i}$ for $%
\phi _{i}=1$), 
\begin{equation}
E\left( y_{i,t-h}\Delta y_{it}|\alpha _{i},\phi _{i},\sigma _{i}^{2}\right)
=\left\{ 
\begin{array}{ll}
0, & \text{for }\phi _{i}=1\text{ and }h=2,3,..., \\ 
-\frac{\sigma _{i}^{2}\phi _{i}^{h-1}}{1+\phi _{i}}, & \text{for }|\phi
_{i}|<1\text{ and }h=1,2,....%
\end{array}%
\right.  \label{Mhl}
\end{equation}%
Given (\ref{Mhl}), if $Pr(\phi _{i}=1)=0$ and $\phi _{i}\in (-1,1]$, we have 
\begin{align*}
\boldsymbol{z}_{a}=\underset{n\rightarrow \infty }{plim}\boldsymbol{\bar{z}}%
_{na}& =-\left( E\left( \frac{\sigma _{i}^{2}}{1+\phi _{i}}\right) ,E\left( 
\frac{\sigma _{i}^{2}\phi _{i}}{1+\phi _{i}}\right) ,E\left( \frac{\sigma
_{i}^{2}}{1+\phi _{i}}\right) \right) ^{\prime }, \\
\text{and}\quad \boldsymbol{z}_{b}=\underset{n\rightarrow \infty }{plim}%
\boldsymbol{\bar{z}}_{nb}& =-\left( E\left( \frac{\sigma _{i}^{2}\phi _{i}}{%
1+\phi _{i}}\right) ,E\left( \frac{\sigma _{i}^{2}\phi _{i}^{2}}{1+\phi _{i}}%
\right) ,E\left( \frac{\sigma _{i}^{2}\phi _{i}}{1+\phi _{i}}\right) \right)
^{\prime }.
\end{align*}%
Since $\phi _{i}$ is distributed independently of $\sigma _{i}^{2}$, for
uniformly distributed $\phi _{i}=\mu _{\phi }+v_{i}$ with $v_{i}$ $\thicksim
IIDU(-a,a)$, $a>0$ and $\phi _{i} \in (-1,1]$, 
\begin{equation}
\underset{n\rightarrow \infty }{plim}\left( \hat{\phi}_{AB}-E(\phi
_{i})\right) =\left( \boldsymbol{z}_{a}^{\prime }\boldsymbol{W}_{AB}%
\boldsymbol{z}_{a}\right) ^{-1}\left( \boldsymbol{z}_{a}^{\prime }%
\boldsymbol{W}_{AB}\boldsymbol{z}_{b}\right) -\mu _{\phi },  \label{biasAB}
\end{equation}%
where $\boldsymbol{z}_{a}=-\sigma ^{2}(c_{\phi },1-c_{\phi },c_{\phi
})^{\prime }$ and $\boldsymbol{z}_{a}=-\sigma ^{2}(1-c_{\phi },\mu _{\phi
}-1+c_{\phi },1-c_{\phi })^{\prime }$ with $\sigma ^{2}=E(\sigma _{i}^{2})$
and $c_{\phi }=E\left( \frac{1}{1+\phi _{i}}\right) =\frac{1}{2a}\ln \left( 
\frac{1+\mu _{\phi }+a}{1+\mu _{\phi }-a}\right) $.

In addition to (\ref{sys1}), consider the following moment conditions, used
in the system GMM estimator proposed by \cite{BlundellBond1998}, and note
that under homogeneity we have for $i=1,2,...,n$, and $t=3,4,...,T$,\footnote{%
See equation (9) on p. 5 in \cite{ChudikPesaran2021}.} 
\begin{equation}
E\left[ \Delta y_{i,t-1}(\mu _{i}(1-\phi _{i})+u_{it})\right] =E[\Delta
y_{i,t-1}(y_{it}-\phi y_{i,t-1})]=0.  
\label{sys2}
\end{equation}%
For $T=4$, with a given weight matrix $\boldsymbol{W}_{BB}$, the BB
estimator based on the moment conditions in (\ref{sys1}) and (\ref{sys2}) is
given by 
\begin{equation}
\hat{\phi}_{BB}=\left( \boldsymbol{\bar{z}}_{nc}^{\prime }\boldsymbol{W}_{BB}%
\boldsymbol{\bar{z}}_{nc}\right) ^{-1}\left( \boldsymbol{\bar{z}}%
_{nc}^{\prime }\boldsymbol{W}_{BB}\boldsymbol{\bar{z}}_{nd}\right) ,
\label{BB}
\end{equation}%
where 
\begin{align*}
\boldsymbol{\bar{z}}_{nc}& =n^{-1}\left( \sum_{i=1}^{n}y_{i1}\Delta
y_{i2},\,\sum_{i=1}^{n}y_{i1}\Delta y_{i3},\sum_{i=1}^{n}y_{i2}\Delta
y_{i3},\sum_{i=1}^{n}y_{i2}\Delta y_{i2},\sum_{i=1}^{n}y_{i3}\Delta
y_{i3}\right) ^{\prime }, \\
\text{and}\quad \boldsymbol{\bar{z}}_{nd}& =n^{-1}\left(
\sum_{i=1}^{n}y_{i1}\Delta y_{i3},\sum_{i=1}^{n}y_{i1}\Delta
y_{i4},\sum_{i=1}^{n}y_{i2}\Delta y_{i4},\sum_{i=1}^{n}y_{i3}\Delta
y_{i2},\sum_{i=1}^{n}y_{i4}\Delta y_{i3}\right) ^{\prime }.
\end{align*}%
Using (\ref{DyitG}) in the main paper and (\ref{y2}), similarly, we can
derive the following equations as $M_{i}\rightarrow \infty $ for $|\phi
_{i}|<1$, 
\begin{equation}
E\left( y_{it}\Delta y_{i,t-h}\right) =E\left( \frac{\sigma _{i}^{2}\phi
_{i}^{h}}{1+\phi _{i}}\right) \text{, for }h=0,1,2,...,  \label{Mhl2}
\end{equation}%
and for $\phi _{i}=1$ and finite $M_{i}$, $E(\Delta
y_{i,t-1}y_{it})=E(\Delta y_{i,t-1}y_{i,t-1})=\sigma _{i}^{2}$. In the case
of $Pr(\phi _{i}=1)=0$ with $\phi _{i}\in (-1,1]$, it follows that 
\begin{align*}
\boldsymbol{z}_{c}& =\underset{n\rightarrow \infty }{plim}\boldsymbol{\bar{z}%
}_{nc}=\left( -E\left( \frac{\sigma _{i}^{2}}{1+\phi _{i}}\right) ,-E\left( 
\frac{\sigma _{i}^{2}\phi _{i}}{1+\phi _{i}}\right) ,-E\left( \frac{\sigma
_{i}^{2}}{1+\phi _{i}}\right) ,E\left( \frac{\sigma _{i}^{2}}{1+\phi _{i}}%
\right) ,E\left( \frac{\sigma _{i}^{2}}{1+\phi _{i}}\right) \right) ^{\prime
}, \\
\boldsymbol{z}_{d}& =\underset{n\rightarrow \infty }{plim}\boldsymbol{\bar{z}%
}_{nd} \\
& =\left( -E\left( \frac{\sigma _{i}^{2}\phi _{i}}{1+\phi _{i}}\right)
,-E\left( \frac{\sigma _{i}^{2}\phi _{i}^{2}}{1+\phi _{i}}\right) ,-E\left( 
\frac{\sigma _{i}^{2}\phi _{i}}{1+\phi _{i}}\right) ,E\left( \sigma _{i}^{2}-%
\frac{\sigma _{i}^{2}}{1+\phi _{i}}\right) ,E\left( \sigma _{i}^{2}-\frac{%
\sigma _{i}^{2}}{1+\phi _{i}}\right) \right) ^{\prime }.
\end{align*}%
Since $\phi _{i}$ is distributed independently of $\sigma _{i}^{2}$, for
uniformly distributed $\phi _{i}=\mu _{\phi }+v_{i}$ with $v_{i}$ $\thicksim
IIDU(-a,a)$, $a>0$ and $\phi _{i}\in (-1,1]$, 
\begin{equation}
plim_{n\rightarrow \infty }\left[ \hat{\phi}_{BB}-E(\phi _{i})\right]
=\left( \boldsymbol{z}_{c}^{\prime }\boldsymbol{W}_{BB}\boldsymbol{z}%
_{c}\right) ^{-1}\left( \boldsymbol{z}_{c}^{\prime }\boldsymbol{W}_{BB}%
\boldsymbol{z}_{d}\right) ,  \label{biasBB}
\end{equation}%
where $\boldsymbol{z}_{c}=\sigma ^{2}(-c_{\phi },-1+c_{\phi },-c_{\phi
},c_{\phi },c_{\phi })^{\prime }$ and $\boldsymbol{z}_{d}=\sigma
^{2}(c_{\phi }-1,1-\mu _{\phi }-c_{\phi },-1+c_{\phi },1-c_{\phi },1-c_{\phi
})^{\prime },$ with $E(\sigma _{i}^{2})=\sigma ^{2}$, and $c_{\phi }=E\left( 
\frac{1}{1+\phi _{i}}\right) =\frac{1}{2a}\ln \left( \frac{1+\mu _{\phi }+a}{%
1+\mu _{\phi }-a}\right) $.

To approximate the values of the asymptotic bias of AB and BB estimators
corresponding to our Monte Carlo experiments, we replace $\boldsymbol{W}%
_{AB} $ and $\boldsymbol{W}_{BB}$ by the simulated weight matrices\footnote{%
The simulated weight matrices are calculated as the average of the weight
matrices used in calculating the two-step AB and BB estimators across 2,000
replications.} with $a=0.5$, $\mu _{\phi }\in \{0.4,0.5\}$, and Gaussian
errors without GARCH effects for $T=4$, and $n=5,000$. In this case, the
biases of AB and BB estimators are around -0.055 and -0.045 for $\mu _{\phi
}=0.4$, and -0.062 and -0.044 for $\mu _{\phi }=0.5$, respectively. These
results are close to the simulated bias of these estimators reported in
Tables \ref{tab:hetro_u1_a} ($\mu _{\phi }=0.4$) and \ref{tab:hetro_u2_a} ($%
\mu _{\phi }=0.5$) for $T=4$ and $n=5,000$.

\subsection{Asymptotic variances of the first two moments \label{Asyvar}}

Suppose that Assumptions \ref{fe}--\ref{initial} in the main paper hold, $%
T\geq 5$, and $M_{i}\rightarrow \infty $. The asymptotic distribution of $%
\boldsymbol{\hat{\theta}}_{HetroGMM}=(\hat{\theta}_{1,HetroGMM},\hat{\theta}%
_{2,HetroGMM})^{\prime }$ is given by 
\begin{equation*}
\sqrt{n}(\boldsymbol{\hat{\theta}}_{HetroGMM}-\boldsymbol{\theta }%
_{0})\rightarrow _{d}N(0,\mathbf{V}_{\boldsymbol{\theta }}),
\end{equation*}%
with $\boldsymbol{\theta }_{0}=(\theta _{1,0},\theta _{2,0})^{\prime }$, and 
\begin{equation*}
\mathbf{V}_{\boldsymbol{\theta }}^{-1}=plim_{n\rightarrow \infty }(\mathbf{H}%
_{\boldsymbol{\theta },nT}^{\prime }\mathbf{S}_{\boldsymbol{\theta },T}^{-1}(%
\boldsymbol{\theta }_{0})\mathbf{H}_{\boldsymbol{\theta },nT}),
\end{equation*}%
where $\mathbf{H}_{\boldsymbol{\theta },nT}=\frac{1}{n}\sum_{i=1}^{n}\mathbf{%
H}_{\boldsymbol{\theta },iT}$, 
\begin{equation*}
\mathbf{H}_{\boldsymbol{\theta },iT}=\left( 
\begin{array}{cc}
\mathbf{h}_{iT} & \boldsymbol{0}_{(T-3)\times 1} \\ 
\boldsymbol{0}_{(T-4)\times 1} & \mathbf{h}_{2,iT}%
\end{array}%
\right),
\end{equation*}%
$\mathbf{S}_{\boldsymbol{\theta },T}(\boldsymbol{\theta }_{0})=\frac{1}{n}%
\sum_{i=1}^{n}(\mathbf{g}_{\boldsymbol{\theta },iT}-\mathbf{H}_{\boldsymbol{%
\theta },iT}\boldsymbol{\theta }_{0})(\mathbf{g}_{\boldsymbol{\theta },iT}-%
\mathbf{H}_{\theta ,iT}\boldsymbol{\theta }_{0})^{\prime }$, $\mathbf{g}_{%
\boldsymbol{\boldsymbol{\theta }},iT}=(\mathbf{g}_{iT}^{\prime },\mathbf{g}%
_{2,iT}^{\prime })^{\prime }$, and $\mathbf{h}_{iT}$, $\mathbf{h}_{2,iT}$, $%
\mathbf{g}_{iT}$ and $\mathbf{g}_{2,iT}$ are given by (\ref{h_iT}), (\ref%
{h_2iT}), (\ref{g_iT}) and (\ref{g_2iT}) in the main paper, respectively. $%
\mathbf{V}_{\boldsymbol{\theta }}$ can be consistently estimated by 
\begin{equation}
\mathbf{\hat{V}}_{\boldsymbol{\theta }}=\left( \mathbf{H}_{\boldsymbol{%
\theta },nT}^{\prime }\mathbf{\hat{S}}_{\boldsymbol{\theta },T}^{-1}(%
\boldsymbol{\hat{\theta}}_{HetroGMM})\mathbf{H}_{\boldsymbol{\theta }%
,nT}\right) ^{-1},  \label{Vartheta}
\end{equation}%
with 
\begin{equation*}
\mathbf{\hat{S}}_{\boldsymbol{\theta },T}(\boldsymbol{\hat{\theta}}%
_{HetroGMM})=\frac{1}{n}\sum_{i=1}^{n}(\mathbf{g}_{\boldsymbol{\theta },iT}-%
\mathbf{H}_{\boldsymbol{\theta },iT}\boldsymbol{\hat{\theta}}_{HetroGMM})(%
\mathbf{g}_{\boldsymbol{\theta },iT}-\mathbf{H}_{\theta ,iT}\boldsymbol{\hat{%
\theta}}_{HetroGMM})^{\prime }.
\end{equation*}

\subsection{Empirical power functions\label{EPF}}

The test statistics for $\mu _{\phi } =E(\phi _{i})$ and $\sigma _{\phi
}^{2} = Var(\phi _{i})$ are given by 
\begin{equation*}
S_{N,\mu }\left( \mu _{\phi }\right) =\frac{\hat{\mu}_{\phi }-\mu _{\phi }}{%
\left[ \widehat{Var(\hat{\mu}_{\phi })}\right] ^{1/2}}\quad \text{and}\quad
S_{N,\sigma }\left( \sigma _{\phi }^{2}\right) =\frac{\hat{\sigma}_{\phi
}^{2}-\sigma _{\phi }^{2}}{\left[ \widehat{Var\left( \hat{\sigma}_{\phi
}^{2}\right) }\right] ^{1/2}},
\end{equation*}%
respectively, where FDAC and HetroGMM estimators of $\hat{\mu}_{\phi }=\hat{%
\theta}_{1}$ are given by (\ref{Estm1}) and (\ref{theta1GMM}) in the main
paper, respectively. $\hat{\sigma}_{\phi }^{2}$ is computed as the plug-in
estimator given by (\ref{Estvar}) in the main paper. In the Monte Carlo
experiments, the empirical power functions (EPF) are computed as the
simulated rejection frequencies for replications $r=1,2,...,R$: 
\begin{equation*}
EPF_{R}(\mu _{\phi })=R^{-1}\sum_{r=1}^{R}I\left[ \left\vert \frac{\hat{\mu}%
_{\phi }^{(r)}-\mu _{\phi }}{\left[ \widehat{Var\left( \hat{\mu}_{\phi
}\right) }^{(r)}\right] ^{1/2}}\right\vert >1.96\right],
\end{equation*}
and 
\begin{equation*}
EPF_{R}(\sigma _{\phi }^{2})=R^{-1}\sum_{r=1}^{R}I\left[ \left\vert \frac{%
\left( \hat{\sigma}_{\phi }^{2}\right) ^{(r)}-\sigma _{\phi }^{2}}{\left[ 
\widehat{Var\left( \hat{\sigma}_{\phi }^{2}\right) }^{(r)}\right] ^{1/2}}%
\right\vert >1.96\right].
\end{equation*}

\subsection{Monte Carlo evidence \label{MCevidence}}

\subsubsection{Comparison of FDAC and HetroGMM estimators \label{MCmm}}

Tables \ref{tab:MCAm1u} and \ref{tab:MCAm1c} summarize bias, RMSE, and size
of FDAC and HetroGMM estimators of $\mu _{\phi }=E(\phi _{i})$ with
uniformly and categorically distributed $\phi _{i}$, respectively, in the
case of Gaussian errors without GARCH effects for the sample size
combinations $n=100,1000,5000$ and $T=4,5,6,10$. The empirical power
functions of FDAC and HetroGMM estimators of $\mu _{\phi }$ with uniformly
distributed $\phi_{i} \in [-1+\epsilon, 1]$ for some $\epsilon>0$ are shown in Figure \ref%
{fig:pw_fdac_hetrogmm_u2_a_m1}.

\begin{figure}[tbp]
\caption{Empirical power functions for FDAC and HetroGMM estimators of $%
\protect\sigma_{\protect\phi}^{2} = Var(\protect\phi_{i})$ $(\protect\sigma_{%
\protect\phi, 0}^{2} = 0.083)$ in a heterogeneous AR(1) panel with uniformly
distributed $\phi_{i} \in [-1+\epsilon, 1]$ for some $\epsilon>0$ and Gaussian errors
without GARCH effects}
\label{fig:pw_fdac_hetrogmm_u2_a_var}
\begin{center}
\includegraphics[scale=0.23]{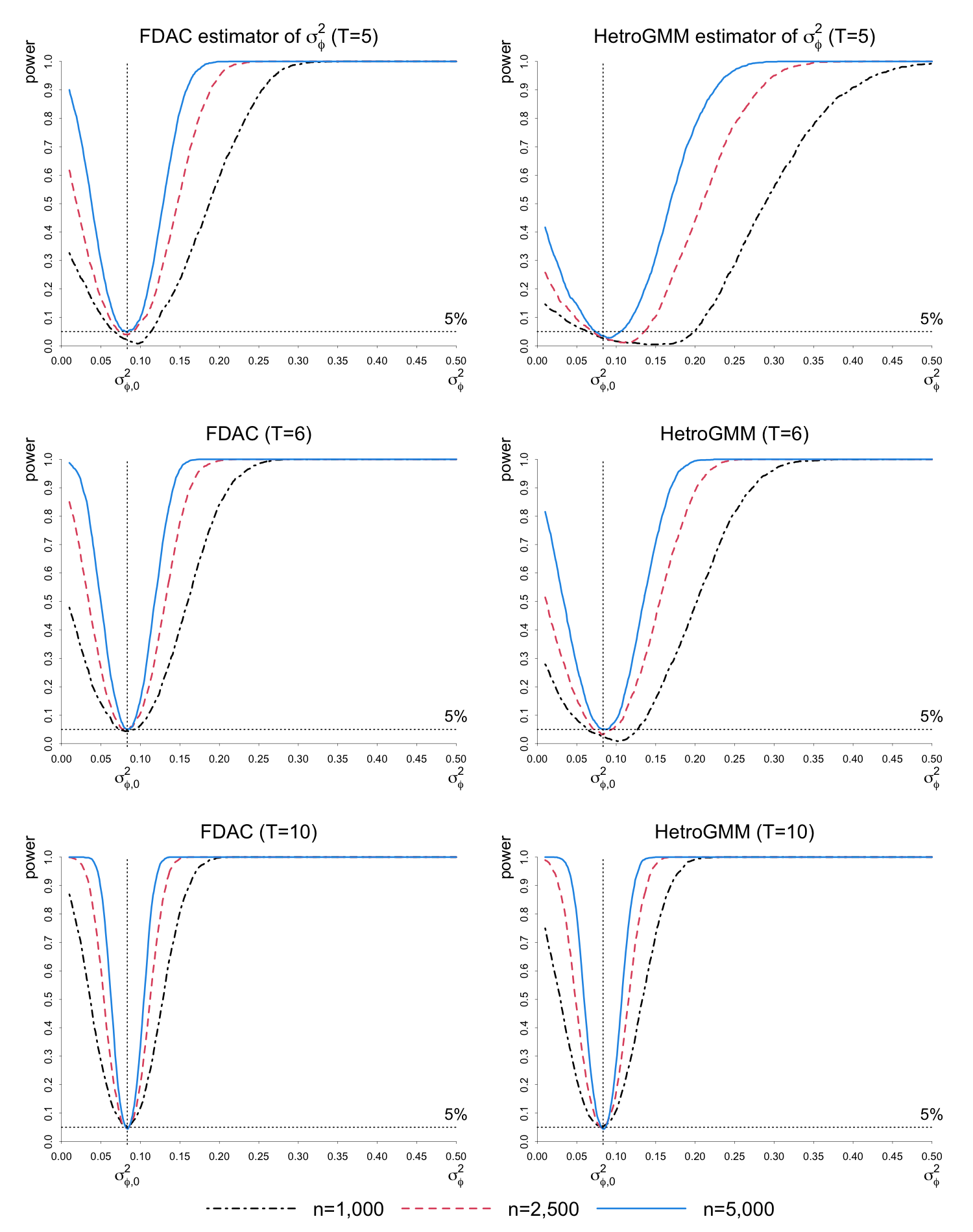}
\end{center}
\end{figure}

Table \ref{tab:npzig} reports the frequency where FDAC and HetroGMM
estimates of $\sigma _{\phi }^{2}$ are either negative or very close to
zero, using the threshold $\left( \hat{\sigma}_{\phi }^{2}\right)
^{(r)}<0.0001$, for replication $r=1,2,...,2000$, respectively, with
uniformly distributed $\phi _{i}$ and Gaussian errors without GARCH effects
for $n=100,1000,2500,5000$ and $T=5,6,10$. Table \ref{tab:MCAvu} summarizes
simulated outcomes with positive estimates of $\sigma_{\phi }^{2}=Var(\phi
_{i})$ with uniformly distributed $\phi _{i}$ in the case of Gaussian errors
without GARCH effects for the sample size combinations $n=100,1000,5000$ and 
$T=5,6,10$. The empirical power functions of FDAC and HetroGMM estimators of 
$\sigma _{\phi }^{2}$ (for simulated outcomes of positive estimates) are
shown in Figure \ref{fig:pw_fdac_hetrogmm_u2_a_var} with $n=1000,2500,5000$
and $T=5,6,10$.

For the four combinations of error distributions, Gaussian and non-Gaussian,
without and with GARCH effects, Tables \ref{tab:MCm1e} and \ref{tab:MCve}
summarize simulation results of the estimation of $\mu _{\phi }$ and $%
\sigma_{\phi}^{2}$ (for simulated outcomes of positive estimates),
respectively, for uniformly distributed $\phi_{i} \in [-1+\epsilon, 1]$ for some $\epsilon>0$ with $\mu_{\phi}
= 0.5$. Table \ref{tab:npzige} reports the frequency where estimates of $%
\sigma_{\phi}^{2}$ are not positive.

\begin{table}[h!]
\caption{Bias, RMSE, and size of FDAC and HetroGMM estimators of $\protect\mu%
_{\protect\phi}= E(\protect\phi _{i})$ in a heterogeneous panel AR(1) model
with uniformly distributed $\protect\phi_{i}$ and Gaussian errors without
GARCH effects}
\label{tab:MCAm1u}
\begin{center}
\scalebox{0.7}{
\renewcommand{\arraystretch}{1.1}
\begin{tabular}{rrcrrcccccrcrrrccccr}
\hline\hline
 & & & \multicolumn{2}{c}{Bias} &  & \multicolumn{2}{c}{RMSE} &  & \multicolumn{2}{c}{Size ($\times 100$)} &  & \multicolumn{2}{c}{Bias } &  & \multicolumn{2}{c}{RMSE} &  & \multicolumn{2}{c}{Size ($\times 100$)} \\  \cline{4-5} \cline{7-8} \cline{10-11} \cline{13-14} \cline{16-17} \cline{19-20}
& &&  & Hetro &&  & Hetro &&  & Hetro &&  & Hetro &&  & Hetro &  &  & Hetro \\ 
$T$ & $n$ && \multicolumn{1}{c}{FDAC} & GMM &  &  \multicolumn{1}{c}{FDAC} & GMM &  &  \multicolumn{1}{c}{FDAC} & GMM &  &  \multicolumn{1}{c}{FDAC} & GMM &  &  \multicolumn{1}{c}{FDAC} & GMM &  & FDAC & GMM \\ \hline
&  &  & \multicolumn{8}{c}{$\mu_{\phi} = 0.4$ with $\vert \phi_{i} \vert < 1$} &  & \multicolumn{8}{c}{$\mu_{\phi} = 0.5$ with $\phi_{i} \in [-1+\epsilon, 1]$ for some $\epsilon>0$} \\ \cline{4-11} \cline{13-20}
4 & 100 &  & -0.008 & -0.043 &  & 0.174 & 0.297 &  & 7.0 & 5.2 &  & 0.003 & -0.013 &  & 0.176 & 0.264 &  & 7.6 & 5.7 \\
4 & 1,000 &  & 0.000 & -0.004 &  & 0.057 & 0.086 &  & 5.1 & 5.8 &  & 0.000 & 0.003 &  & 0.057 & 0.080 &  & 5.1 & 5.2 \\
4 & 5,000 &  & 0.000 & -0.002 &  & 0.025 & 0.038 &  & 3.8 & 4.5 &  & 0.001 & 0.002 &  & 0.026 & 0.036 &  & 5.4 & 5.4 \\
 &  &  &  &  &  &  &  &  &  &  &  &  &  &  &  &  &  &  &  \\
5 & 100 &  & -0.003 & 0.006 &  & 0.134 & 0.163 &  & 6.3 & 7.3 &  & -0.003 & 0.008 &  & 0.134 & 0.157 &  & 7.1 & 8.5 \\
5 & 1,000 &  & 0.000 & 0.001 &  & 0.043 & 0.052 &  & 5.1 & 5.1 &  & -0.001 & 0.002 &  & 0.042 & 0.051 &  & 5.1 & 5.8 \\
5 & 5,000 &  & 0.000 & 0.000 &  & 0.019 & 0.023 &  & 4.4 & 4.9 &  & 0.001 & 0.002 &  & 0.019 & 0.022 &  & 4.3 & 4.3 \\
 &  &  &  &  &  &  &  &  &  &  &  &  &  &  &  &  &  &  &  \\
6 & 100 &  & -0.004 & 0.010 &  & 0.111 & 0.120 &  & 6.3 & 7.4 &  & -0.004 & 0.009 &  & 0.112 & 0.119 &  & 7.1 & 8.6 \\
6 & 1,000 &  & -0.001 & 0.001 &  & 0.037 & 0.040 &  & 5.8 & 5.2 &  & -0.002 & 0.001 &  & 0.035 & 0.039 &  & 4.5 & 6.2 \\
6 & 5,000 &  & 0.000 & 0.000 &  & 0.016 & 0.018 &  & 4.4 & 5.2 &  & 0.000 & 0.001 &  & 0.016 & 0.017 &  & 4.7 & 4.9 \\
 &  &  &  &  &  &  &  &  &  &  &  &  &  &  &  &  &  &  &  \\
10 & 100 &  & 0.000 & 0.013 &  & 0.078 & 0.077 &  & 6.5 & 10.8 &  & -0.003 & 0.009 &  & 0.079 & 0.077 &  & 6.3 & 10.5 \\
10 & 1,000 &  & 0.000 & 0.001 &  & 0.026 & 0.026 &  & 5.8 & 6.5 &  & -0.001 & 0.001 &  & 0.025 & 0.026 &  & 4.8 & 5.7 \\
10 & 5,000 &  & 0.000 & 0.000 &  & 0.011 & 0.012 &  & 5.3 & 5.4 &  & 0.000 & 0.001 &  & 0.011 & 0.012 &  & 5.3 & 5.8 \\ 
\hline\hline
\end{tabular}}
\end{center}
\par
{\footnotesize Notes: The DGP is given by $y_{it}=\mu _{i}(1-\phi _{i})+\phi
_{i}y_{i,t-1}+h_{it}\varepsilon_{it}$ for $i=1,2,...,n$, and $t=-99,
-98,...,T$, with $\varepsilon_{it} \sim IIDN(0,1)$ and cross-sectional
heteroskedasticity, $h_{it} = \sigma_{i}$, where $\sigma_{i}^{2} \sim
IID(0.5+0.5z_{i}^{2})$ and $z_{i} \sim IIDN(0,1)$. The heterogeneous AR(1)
coefficients are generated by uniform distributions: $\phi_{i} = \mu_{\phi}
+ v_{i}$, with $v_{i} \sim IIDU[-a,a]$, $a=0.5$ and $\mu_{\phi} \in \{0.4,
0.5\}$. The initial values are generated as $(y_{i,-100} - \mu_{i}) \sim
IIDN(b, \kappa \sigma_{i}^{2})$ with $b=1$ and $\kappa=2$ for all $i$. For
each experiment, $(\alpha _{i},\phi _{i},\sigma _{i})^{\prime } $ are
generated differently across replications. FDAC and HetroGMM estimators of $%
\mu_{\phi}$ are computed based on (\ref{Estm1}) and (\ref{theta1GMM}) in the
main paper, respectively. The asymptotic variances are estimated by the
Delta method. The estimation is based on $\{y_{i1},y_{i2},...,y_{iT}\}$ for $%
i=1,2,...,n$. The nominal size of the tests is set to 5 per cent. The number
of replications is $2,000$. }
\end{table}

\begin{table}[h!]
\caption{Bias, RMSE, and size of FDAC and HetroGMM estimators of $\protect\mu%
_{\protect\phi} = E(\protect\phi_{i}) $ in a heterogeneous panel AR(1) model
with categorically distributed $\protect\phi_{i}$ and Gaussian errors
without GARCH effects}
\label{tab:MCAm1c}
\begin{center}
\scalebox{0.7}{
\renewcommand{\arraystretch}{1}
\begin{tabular}{rrrrrccrrrccrrrrrrrr}
\hline\hline 
 & & & \multicolumn{2}{c}{Bias} &  & \multicolumn{2}{c}{RMSE} &  & \multicolumn{2}{c}{Size ($\times 100$)} &  & \multicolumn{2}{c}{Bias } &  & \multicolumn{2}{c}{RMSE} &  & \multicolumn{2}{c}{Size ($\times 100$)} \\  \cline{4-5} \cline{7-8} \cline{10-11} \cline{13-14} \cline{16-17} \cline{19-20}
& &&  & Hetro &&  & Hetro &&  & Hetro &&  & Hetro &&  & Hetro &  &  & Hetro \\ 
$T$ & $n$ && \multicolumn{1}{c}{FDAC} & GMM &  &  \multicolumn{1}{c}{FDAC} & GMM &  &  \multicolumn{1}{c}{FDAC} & GMM &  &  \multicolumn{1}{c}{FDAC} & GMM &  &  \multicolumn{1}{c}{FDAC} & GMM &  & FDAC & GMM \\ \hline
 &  &  & \multicolumn{8}{c}{$\mu_{\phi} = 0.545$ with $\vert \phi_{i} \vert < 1$} &  & \multicolumn{8}{c}{$\mu_{\phi} = 0.525$ with $\phi_{i} \in [-1+\epsilon, 1]$ for some $\epsilon>0$} \\ \cline{4-11} \cline{13-20}
4 & 100 &  & 0.000 & -0.034 &  & 0.164 & 0.264 &  & 7.5 & 6.8 &  & -0.002 & -0.028 &  & 0.169 & 0.267 &  & 7.4 & 6.7 \\
4 & 1,000 &  & -0.001 & -0.005 &  & 0.053 & 0.077 &  & 5.2 & 5.0 &  & 0.001 & -0.002 &  & 0.054 & 0.074 &  & 6.2 & 4.5 \\
4 & 5,000 &  & 0.000 & 0.000 &  & 0.025 & 0.035 &  & 5.1 & 5.4 &  & 0.001 & 0.002 &  & 0.025 & 0.033 &  & 5.8 & 4.9 \\
 &  &  &  &  &  &  &  &  &  &  &  &  &  &  &  &  &  &  &  \\
5 & 100 &  & 0.001 & 0.004 &  & 0.118 & 0.144 &  & 7.0 & 7.0 &  & -0.003 & 0.003 &  & 0.118 & 0.142 &  & 5.9 & 7.0 \\
5 & 1,000 &  & -0.001 & -0.002 &  & 0.037 & 0.045 &  & 4.4 & 4.7 &  & 0.001 & 0.001 &  & 0.039 & 0.046 &  & 6.4 & 5.1 \\
5 & 5,000 &  & -0.001 & 0.000 &  & 0.017 & 0.021 &  & 4.9 & 5.2 &  & 0.001 & 0.002 &  & 0.017 & 0.021 &  & 5.8 & 5.2 \\
 &  &  &  &  &  &  &  &  &  &  &  &  &  &  &  &  &  &  &  \\
6 & 100 &  & -0.001 & 0.006 &  & 0.099 & 0.108 &  & 6.9 & 8.6 &  & -0.001 & 0.007 &  & 0.097 & 0.107 &  & 6.4 & 7.8 \\
6 & 1,000 &  & 0.000 & -0.001 &  & 0.031 & 0.034 &  & 5.1 & 5.1 &  & 0.001 & 0.002 &  & 0.032 & 0.035 &  & 5.3 & 5.9 \\
6 & 5,000 &  & 0.000 & 0.000 &  & 0.014 & 0.016 &  & 4.8 & 5.7 &  & 0.001 & 0.001 &  & 0.014 & 0.016 &  & 6.0 & 5.5 \\
 &  &  &  &  &  &  &  &  &  &  &  &  &  &  &  &  &  &  &  \\
10 & 100 &  & -0.001 & 0.004 &  & 0.064 & 0.064 &  & 5.9 & 8.8 &  & 0.000 & 0.006 &  & 0.066 & 0.067 &  & 5.4 & 10.8 \\
10 & 1,000 &  & 0.000 & 0.000 &  & 0.021 & 0.021 &  & 4.2 & 5.3 &  & 0.001 & 0.002 &  & 0.021 & 0.021 &  & 4.6 & 5.4 \\
10 & 5,000 &  & 0.000 & 0.000 &  & 0.009 & 0.010 &  & 5.0 & 5.7 &  & 0.000 & 0.001 &  & 0.010 & 0.010 &  & 5.4 & 4.9 \\ 
\hline\hline
\end{tabular}}
\end{center}
\par
{\footnotesize Notes: The DGP is given by $y_{it}=\mu _{i}(1-\phi _{i})+\phi
_{i}y_{i,t-1}+h_{it}\varepsilon_{it}$, for $i=1,2,...,n$, and $t=-M_{i}+1,
-M_{i}+2,...,T$, featuring Gaussian standardized errors with cross-sectional
heteroskedasticity. The heterogeneous AR(1) coefficients are generated by
categorical distributions: $\func{Pr}(\phi _{i}=\phi _{L})=\pi $ and $\func{%
Pr}(\phi _{i}=\phi _{H})=1-\pi $, where $(\phi_{H}, \phi_{L}, \pi) = (0.8,
0.5, 0.85)$ with $\vert \phi_{i} \vert < 1$ for all $i$ and $(1, 0.5, 0.95)$
with $\phi_{i} \in [-1+\epsilon, 1]$ for some $\epsilon>0$ and all $i$. The initial values are given
by $(y_{i,-M_{i}} - \mu_{i}) \sim IIDN(b, \kappa \sigma_{i}^{2})$ with $b=1$
and $\kappa=2$, where $M_{i} = 100$ for units with $|\phi_{i}|<1$, and $%
M_{i}=1$ for units with $\phi_{i} = 1$. For each experiment, $(\alpha_{i},
\phi_{i}, \sigma_{i})^{\prime}$ are generated differently across
replications. FDAC and HetroGMM estimators of $\mu_{\phi}$ are computed
based on (\ref{Estm1}) and (\ref{theta1GMM}) in the main paper,
respectively. The asymptotic variances are estimated by the Delta method.
The estimation is based on $\{y_{i1}, y_{i2},...,y_{iT}\}$ for $i=1,2,...,n$%
. The nominal size of the tests is set to 5 per cent. The number of
replications is $2,000$. }
\end{table}

\newpage\clearpage
\begin{figure}[tbp]
\caption{Empirical power functions for FDAC and HetroGMM estimators of $%
\protect\mu_{\protect\phi} = E(\protect\phi_{i})$ $(\protect\mu_{\protect\phi%
,0} = 0.5)$ in a heterogeneous AR(1) panel with uniformly distributed $\phi_{i} \in [-1+\epsilon, 1]$ for some $\epsilon>0$ and Gaussian errors without GARCH effects}
\label{fig:pw_fdac_hetrogmm_u2_a_m1}
\begin{center}
\includegraphics[scale=0.23]{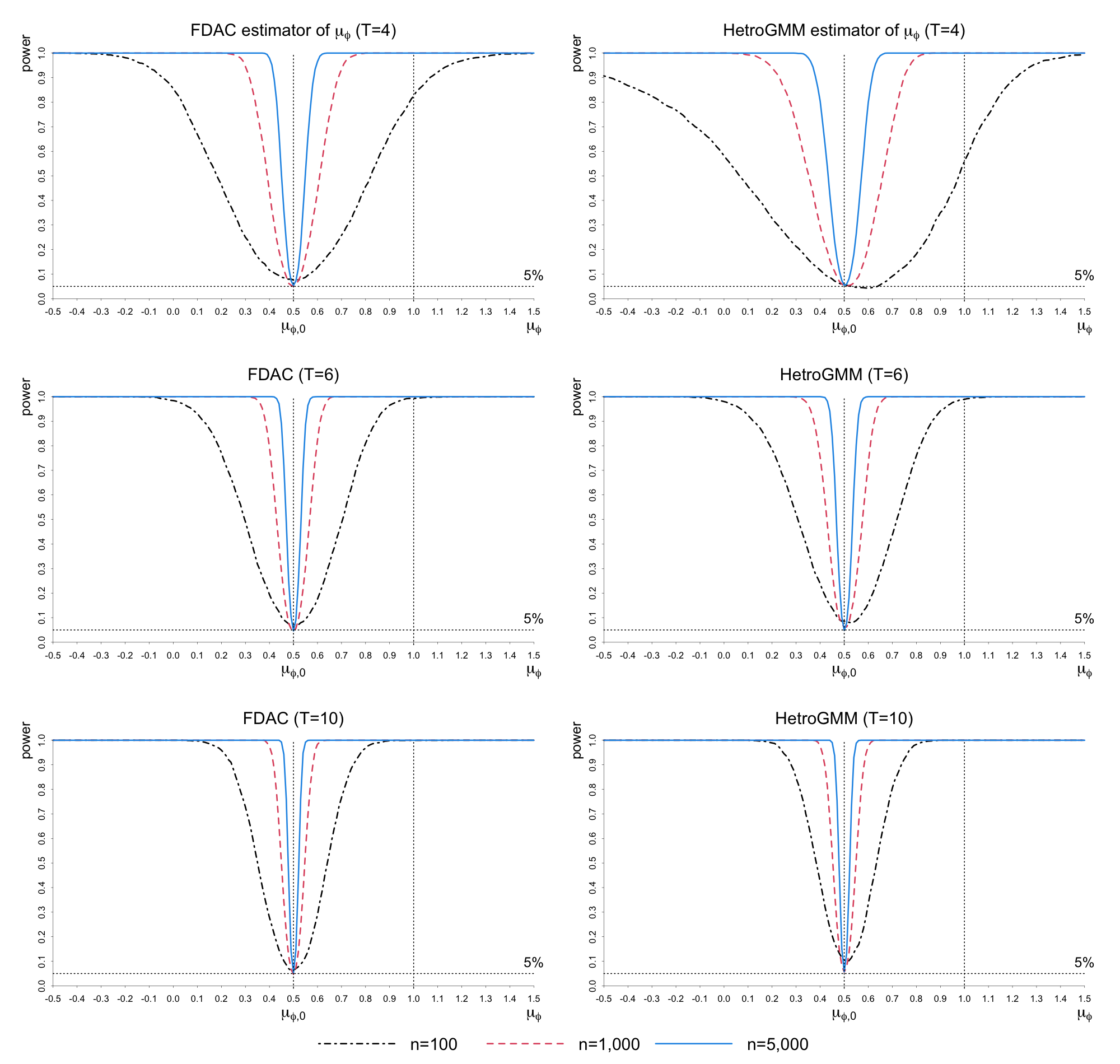}
\end{center}
\end{figure}

\newpage\clearpage
\begin{table}[tbp]
\caption{Frequency of FDAC and HetroGMM estimators of $\protect\sigma_{%
\protect\phi}^{2} = Var(\protect\phi_{i})$ being negative with uniformly
distributed $\protect\phi_i$ and Gaussian errors without GARCH effects}
\label{tab:npzig}
\begin{center}
\scalebox{0.8}{
\renewcommand{\arraystretch}{1}
\begin{tabular}{rrrrrcrr}
\hline\hline
   & &  & \multicolumn{2}{c}{ $\sigma_{\phi}^{2} = 0.083$} &  & \multicolumn{2}{c}{$\sigma_{\phi}^{2} = 0.083$ with } \\ 
   & &  & \multicolumn{2}{c}{ $\quad\,\,\,$ with $\vert \phi_{i}\vert<1$ $\quad\,\,\,$ } &  & \multicolumn{2}{c}{$\phi_{i} \in [-1+\epsilon, 1]$ ($\epsilon>0$)} \\ \cline{4-5} \cline{7-8}
  &&& & Hetro &&& Hetro \\
$T$ & $n$ &  & FDAC & GMM &  & FDAC & GMM \\ \hline
5 & 100 &  & 34.4 & 42.1 &  & 33.4 & 45.2 \\ 
5 & 1,000 &  & 5.9 & 21.3 &  & 6.4 & 20.6 \\
5 & 2,500 &  & 0.8 & 9.9 &  & 0.7 & 9.8 \\
5 & 5,000 &  & 0.0 & 2.8 &  & 0.0 & 2.0 \\
 &  &  &  &  &  &  &  \\
6 & 100 &  & 28.2 & 34.3 &  & 28.1 & 36.3 \\
6 & 1,000 &  & 1.9 & 8.4 &  & 1.7 & 8.2 \\
6 & 2,500 &  & 0.1 & 1.8 &  & 0.2 & 1.4 \\
6 & 5,000 &  & 0.0 & 0.2 &  & 0.0 & 0.1 \\
 &  &  &  &  &  &  &  \\
10 & 100 &  & 15.6 & 20.6 &  & 15.4 & 19.9 \\
10 & 1,000 &  & 0.1 & 0.0 &  & 0.1 & 0.2 \\
10 & 2,500 &  & 0.0 & 0.0 &  & 0.0 & 0.0 \\
10 & 5,000 &  & 0.0 & 0.0 &  & 0.0 & 0.0
\\ \hline\hline  
\end{tabular}}
\end{center}
\par
{\footnotesize Notes: The DGP is given by $y_{it}=\mu _{i}(1-\phi _{i})+\phi
_{i}y_{i,t-1}+h_{it}\varepsilon_{it}$, for $i=1,2,...,n$, and $t=-99,
-M_{i}+2,...,T$, featuring Gaussian standardized errors with cross-sectional
heteroskedasticity. The heterogeneous AR(1) coefficients are generated by
uniform distributions: $\phi_{i} = \mu_{\phi} + v_{i}$, with $v_{i} \sim
IIDU[-a,a]$, $a=0.5$ and $\mu_{\phi} \in \{0.4, 0.5\}$. The initial values
are given by $(y_{i,-100} - \mu_{i}) \sim IIDN(b, \kappa \sigma_{i}^{2})$
with $b=1$ and $\kappa=2$ for all $i$. For each experiment, $(\alpha_{i},
\phi_{i},\sigma_{i})^{\prime}$ are generated differently across
replications. The FDAC estimator of $\sigma_{\phi}^{2}$ is computed by
plugging (\ref{Estm1}) and (\ref{Estm2}) into (\ref{Estvar}) in the main
paper, and the HetroGMM estimator of $\sigma_{\phi}^{2}$ is computed by
plugging (\ref{theta1GMM}) and (\ref{theta2GMM}) into (\ref{Estvar}) in the
main paper. The asymptotic variances are estimated by the Delta method. The
estimation is based on $\{y_{i1}, y_{i2},...,y_{iT}\}$ for $i=1,2,...,n$.
The figure in the cell denotes the frequency (multiplied by 100) of
occurrences where the estimate of $\sigma_{\phi}^{2}$ is negative or close
to zero, $\left(\hat{\sigma}_{\phi}^{2}\right)^{(r)} < 0.0001$, for
replication $r$ over 2,000 replications.}
\end{table}

\begin{table}[h]
\caption{Bias, RMSE, and size of FDAC and HetroGMM estimators of $\protect%
\sigma _{\protect\phi }^{2} = Var(\protect\phi_{i})$ in a heterogeneous
panel AR(1) model with uniformly distributed $\protect\phi _{i}$ and
Gaussian errors without GARCH effects}
\label{tab:MCAvu}
\begin{center}
\scalebox{0.69}{
\renewcommand{\arraystretch}{1.1}
\begin{tabular}{rrcrrcccccccrrrccccc}
\hline\hline
 & & & \multicolumn{2}{c}{Bias} &  & \multicolumn{2}{c}{RMSE} &  & \multicolumn{2}{c}{Size ($\times 100$)} &  & \multicolumn{2}{c}{Bias } &  & \multicolumn{2}{c}{RMSE} &  & \multicolumn{2}{c}{Size ($\times 100$)} \\  \cline{4-5} \cline{7-8} \cline{10-11} \cline{13-14} \cline{16-17} \cline{19-20}
& &&  & Hetro &&  & Hetro &&  & Hetro &&  & Hetro &&  & Hetro &  &  & Hetro \\ 
$T$ & $n$ && \multicolumn{1}{c}{FDAC} & GMM &  &  \multicolumn{1}{c}{FDAC} & GMM &  &  \multicolumn{1}{c}{FDAC} & GMM &  &  \multicolumn{1}{c}{FDAC} & GMM &  &  \multicolumn{1}{c}{FDAC} & GMM &  & FDAC & GMM \\ \hline
&  &  & \multicolumn{8}{c}{$\sigma_{\phi}^{2} = 0.083$ with $\vert \phi_{i} \vert < 1$ and $\mu_{\phi} = 0.4$} &  & \multicolumn{8}{c}{$\sigma_{\phi}^{2} = 0.083$ with $\phi_{i} \in [-1+\epsilon, 1]$ for $\epsilon>0$ and $\mu_{\phi} = 0.5$} \\ \cline{4-11} \cline{13-20}
5 & 1,000 &  & 0.005 & 0.032 &  & 0.047 & 0.079 &  & 3.0 & 3.2 &  & 0.006 & 0.029 &  & 0.047 & 0.076 &  & 2.0 & 2.6 \\
5 & 2,500 &  & 0.000 & 0.012 &  & 0.033 & 0.053 &  & 4.6 & 2.9 &  & 0.000 & 0.010 &  & 0.031 & 0.053 &  & 4.0 & 2.8 \\
5 & 5,000 &  & -0.001 & 0.002 &  & 0.024 & 0.041 &  & 4.9 & 2.8 &  & 0.000 & 0.003 &  & 0.023 & 0.041 &  & 4.9 & 3.7 \\
 &  &  &  &  &  &  &  &  &  &  &  &  &  &  &  &  &  &  &  \\
6 & 1,000 &  & 0.001 & 0.007 &  & 0.038 & 0.052 &  & 3.8 & 3.0 &  & 0.002 & 0.008 &  & 0.038 & 0.050 &  & 4.5 & 2.4 \\
6 & 2,500 &  & 0.000 & 0.001 &  & 0.026 & 0.036 &  & 5.5 & 3.7 &  & 0.000 & 0.001 &  & 0.025 & 0.035 &  & 5.1 & 3.3 \\
6 & 5,000 &  & 0.000 & 0.000 &  & 0.018 & 0.027 &  & 5.8 & 5.0 &  & 0.001 & 0.001 &  & 0.018 & 0.026 &  & 5.1 & 5.1 \\
 &  &  &  &  &  &  &  &  &  &  &  &  &  &  &  &  &  &  &  \\
10 & 1,000 &  & -0.001 & -0.002 &  & 0.023 & 0.027 &  & 4.0 & 5.0 &  & 0.000 & -0.002 &  & 0.024 & 0.027 &  & 5.3 & 5.4 \\
10 & 2,500 &  & 0.000 & -0.001 &  & 0.015 & 0.017 &  & 5.2 & 6.0 &  & 0.000 & -0.001 &  & 0.015 & 0.017 &  & 4.6 & 5.2 \\
10 & 5,000 &  & 0.000 & -0.001 &  & 0.011 & 0.012 &  & 5.4 & 5.8 &  & 0.000 & 0.000 &  & 0.010 & 0.012 &  & 4.4 & 4.5 \\
\hline\hline
\end{tabular}}
\end{center}
\par
{\footnotesize Notes: The DGP is given by $y_{it}=\mu _{i}(1-\phi _{i})+\phi
_{i}y_{i,t-1}+h_{it}\varepsilon _{it}$ for $i=1,2,...,n$, and $%
t=-99,-98,...,T$, featuring Gaussian standardized errors with
cross-sectional heteroskedasticity without GARCH effects, where the
heterogeneous AR(1) coefficients are generated by uniform distributions. The
FDAC estimator of $\sigma _{\phi }^{2}$ is computed by plugging (\ref{Estm1}%
) and (\ref{Estm2}) into (\ref{Estvar}), and the HetroGMM estimator of $%
\sigma _{\phi }^{2}$ is computed by plugging (\ref{theta1GMM}) and (\ref%
{theta2GMM}) into (\ref{Estvar}) in the main paper. The asymptotic variances
are estimated by the Delta method. The estimation is based on $%
\{y_{i1},y_{i2},...,y_{iT}\}$ for $i=1,2,...,n$. The nominal size of the
tests is set to 5 per cent. The total number of replications is $2,000$. But
the reported results are based on simulated outcomes with $\left( \hat{\sigma%
}_{\phi }^{2}\right) ^{(r)} \geq 0.0001$. The frequencies with negative
outcomes, by sample sizes and estimation method, are reported in Table \ref%
{tab:npzig} of the online supplement. See also the footnotes to Table \ref%
{tab:MCAm1u} for further details of the DGP used.}
\end{table}

\newpage\clearpage
\begin{table}[h!]
\caption{Bias, RMSE, and size of FDAC and HetroGMM estimators of $\protect\mu%
_{\protect\phi} =E(\protect\phi _{i}) = 0.5$ in a heterogeneous panel AR(1)
model with uniformly distributed $\phi_{i} \in [-1+\epsilon, 1]$ for some $\epsilon>0$ under
different error processes}
\label{tab:MCm1e}
\begin{center}
\scalebox{0.7}{
\renewcommand{\arraystretch}{1.05}
\begin{tabular}{rrrrrrcccrrcrrrrrrrr}
\hline\hline
 & & & \multicolumn{2}{c}{Bias} &  & \multicolumn{2}{c}{RMSE} &  & \multicolumn{2}{c}{Size ($\times 100$)} &  & \multicolumn{2}{c}{Bias } &  & \multicolumn{2}{c}{RMSE} &  & \multicolumn{2}{c}{Size ($\times 100$)} \\  \cline{4-5} \cline{7-8} \cline{10-11} \cline{13-14} \cline{16-17} \cline{19-20}
& &&  & Hetro &&  & Hetro &&  & Hetro &&  & Hetro &&  & Hetro &  &  & Hetro \\ 
$T$ & $n$ && \multicolumn{1}{c}{FDAC} & GMM &  &  \multicolumn{1}{c}{FDAC} & GMM &  &  \multicolumn{1}{c}{FDAC} & GMM &  &  \multicolumn{1}{c}{FDAC} & GMM &  &  \multicolumn{1}{c}{FDAC} & GMM &  & FDAC & GMM \\ \hline
&  &  & \multicolumn{8}{c}{Gaussian errors without GARCH effects} &  & \multicolumn{8}{c}{Non-Gaussian errors without GARCH effects} \\  \cline{4-11} \cline{13-20}
4 & 100 &  & 0.003 & -0.013 &  & 0.176 & 0.264 &  & 7.6 & 5.7 &  & -0.008 & -0.113 &  & 0.214 & 0.607 &  & 9.5 & 8.3 \\
4 & 1,000 &  & 0.000 & 0.003 &  & 0.057 & 0.080 &  & 5.1 & 5.2 &  & 0.002 & -0.008 &  & 0.071 & 0.116 &  & 5.8 & 5.6 \\
4 & 5,000 &  & 0.001 & 0.002 &  & 0.026 & 0.036 &  & 5.4 & 5.4 &  & 0.000 & -0.002 &  & 0.031 & 0.050 &  & 5.2 & 5.2 \\
 &  &  &  &  &  &  &  &  &  &  &  &  &  &  &  &  &  &  &  \\
5 & 100 &  & -0.003 & 0.008 &  & 0.134 & 0.157 &  & 7.1 & 8.5 &  & -0.005 & 0.019 &  & 0.151 & 0.189 &  & 7.4 & 11.1 \\
5 & 1,000 &  & -0.001 & 0.002 &  & 0.042 & 0.051 &  & 5.1 & 5.8 &  & 0.001 & 0.004 &  & 0.051 & 0.064 &  & 5.4 & 5.8 \\
5 & 5,000 &  & 0.001 & 0.002 &  & 0.019 & 0.022 &  & 4.3 & 4.3 &  & 0.000 & 0.002 &  & 0.023 & 0.029 &  & 5.0 & 5.1 \\
 &  &  &  &  &  &  &  &  &  &  &  &  &  &  &  &  &  &  &  \\
6 & 100 &  & -0.004 & 0.009 &  & 0.112 & 0.119 &  & 7.1 & 8.6 &  & -0.002 & 0.027 &  & 0.125 & 0.134 &  & 6.5 & 11.5 \\
6 & 1,000 &  & -0.002 & 0.001 &  & 0.035 & 0.039 &  & 4.5 & 6.2 &  & 0.001 & 0.006 &  & 0.042 & 0.047 &  & 5.5 & 6.4 \\
6 & 5,000 &  & 0.000 & 0.001 &  & 0.016 & 0.017 &  & 4.7 & 4.9 &  & 0.000 & 0.001 &  & 0.019 & 0.021 &  & 5.6 & 5.1 \\
 &  &  &  &  &  &  &  &  &  &  &  &  &  &  &  &  &  &  &  \\
10 & 100 &  & -0.003 & 0.009 &  & 0.079 & 0.077 &  & 6.3 & 10.5 &  & -0.001 & 0.023 &  & 0.085 & 0.081 &  & 6.4 & 14.4 \\
10 & 1,000 &  & -0.001 & 0.001 &  & 0.025 & 0.026 &  & 4.8 & 5.7 &  & 0.002 & 0.006 &  & 0.028 & 0.028 &  & 5.9 & 7.1 \\
10 & 5,000 &  & 0.000 & 0.001 &  & 0.011 & 0.012 &  & 5.3 & 5.8 &  & 0.000 & 0.001 &  & 0.013 & 0.013 &  & 4.9 & 5.0 \\
 &  &  &  &  &  &  &  &  &  &  &  &  &  &  &  &  &  &  &  \\
 &  &  & \multicolumn{8}{c}{Gaussian errors with GARCH effects} &  & \multicolumn{8}{c}{Non-Gaussian errors with GARCH effects} \\  \cline{4-11} \cline{13-20}
4 & 100 &  & 0.000 & -0.017 &  & 0.205 & 0.306 &  & 9.0 & 6.4 &  & -0.028 & -0.082 &  & 0.302 & 1.029 &  & 13.5 & 8.3 \\
4 & 1,000 &  & 0.000 & 0.002 &  & 0.069 & 0.095 &  & 5.7 & 5.2 &  & -0.002 & -0.016 &  & 0.117 & 0.181 &  & 6.1 & 5.7 \\
4 & 5,000 &  & 0.000 & 0.001 &  & 0.031 & 0.043 &  & 6.2 & 5.4 &  & -0.001 & -0.003 &  & 0.059 & 0.087 &  & 5.1 & 4.3 \\
 &  &  &  &  &  &  &  &  &  &  &  &  &  &  &  &  &  &  &  \\
5 & 100 &  & -0.004 & 0.010 &  & 0.159 & 0.178 &  & 9.1 & 8.9 &  & -0.015 & 0.008 &  & 0.215 & 0.251 &  & 11.1 & 12.3 \\
5 & 1,000 &  & -0.001 & 0.003 &  & 0.051 & 0.061 &  & 5.3 & 6.2 &  & -0.001 & 0.004 &  & 0.085 & 0.092 &  & 6.7 & 5.9 \\
5 & 5,000 &  & 0.000 & 0.002 &  & 0.023 & 0.027 &  & 4.7 & 4.9 &  & 0.000 & 0.001 &  & 0.045 & 0.045 &  & 5.3 & 4.9 \\
 &  &  &  &  &  &  &  &  &  &  &  &  &  &  &  &  &  &  &  \\
6 & 100 &  & -0.006 & 0.009 &  & 0.133 & 0.137 &  & 8.1 & 10.5 &  & -0.007 & 0.017 &  & 0.181 & 0.166 &  & 10.5 & 14.0 \\
6 & 1,000 &  & -0.002 & 0.002 &  & 0.042 & 0.046 &  & 4.6 & 6.0 &  & 0.000 & 0.002 &  & 0.072 & 0.065 &  & 5.7 & 6.9 \\
6 & 5,000 &  & 0.000 & 0.002 &  & 0.019 & 0.021 &  & 5.0 & 4.9 &  & -0.001 & -0.001 &  & 0.038 & 0.034 &  & 5.4 & 6.0 \\
 &  &  &  &  &  &  &  &  &  &  &  &  &  &  &  &  &  &  &  \\
10 & 100 &  & -0.005 & 0.008 &  & 0.095 & 0.087 &  & 6.9 & 13.2 &  & -0.005 & 0.008 &  & 0.130 & 0.104 &  & 9.4 & 19.4 \\
10 & 1,000 &  & -0.001 & 0.001 &  & 0.031 & 0.030 &  & 4.9 & 6.3 &  & 0.001 & -0.001 &  & 0.053 & 0.040 &  & 6.0 & 9.7 \\
10 & 5,000 &  & 0.000 & 0.001 &  & 0.014 & 0.014 &  & 5.7 & 6.1 &  & 0.000 & -0.002 &  & 0.027 & 0.020 &  & 5.4 & 5.9 \\
\hline\hline
\end{tabular}}
\end{center}
\par
{\footnotesize Notes: The DGP is given by $y_{it}=\mu _{i}(1-\phi _{i})+\phi
_{i}y_{i,t-1}+h_{it}\varepsilon_{it}$, for $i=1,2,...,n$, and $t=-99,
-98,...,T$, where the heterogeneous AR(1) coefficients are generated by the
uniform distribution: $\phi_{i} = \mu_{\phi} + v_{i}$, with $v_{i} \sim
IIDU[-a,a]$, $a=0.5$ and $\mu_{\phi} = 0.5$. The standardized errors, $%
\varepsilon _{it}$, are generated as Gaussian, $\varepsilon _{it}\sim
IIDN(0,1)$, or non-Gaussian, $\varepsilon _{it}=\left( e_{it}-2\right) /2$
with $e_{it}\sim IID\chi _{2}^{2}$. The GARCH effect is generated as $%
h_{it}^{2}=\sigma _{i}^{2}(1-\psi _{0}-\psi _{1})+\psi
_{0}h_{i,t-1}^{2}+\psi _{1}(h_{i,t-1}\varepsilon_{i,t-1})^{2}$, with $\sigma
_{i}^{2}\sim IID\left( 0.5+0.5z_{i}^{2}\right) $ and $z_{i}\sim IIDN(0,1)$,
where $\psi _{0}=0.6$ and $\psi _{1}=0.2$, with $h_{i,-M_{i}}=\sigma _{i}$.
In the case of no GARCH effects, $\psi _{0}=\psi _{1}=0$. The initial values
are given by $(y_{i,-100} - \mu_{i}) \sim IIDN(b, \kappa \sigma_{i}^{2})$
with $b=1$ and $\kappa=2$ for all $i$. For each experiment, $(\alpha
_{i},\phi _{i},\sigma _{i})^{\prime }$ are generated differently across
replications. The FDAC estimator is calculated based on (\ref{Estm1}) in the
main paper, and its asymptotic variance is estimated by the Delta method.
The HetroGMM estimator and its asymptotic variance are calculated by (\ref%
{theta1GMM}) and (\ref{Shathat}) in the main paper. The estimation is based
on $\{y_{i1},y_{i2},...,y_{iT}\}$ for $i=1,2,...,n$. The nominal size of the
tests is set to 5 per cent. The number of replications is $2,000$. }
\end{table}

\begin{table}[tbp]
\caption{Bias, RMSE, and size of FDAC and HetroGMM estimators of $\protect%
\sigma_{\protect\phi}^{2} = Var(\protect\phi_{i}) = 0.083$ in a
heterogeneous panel AR(1) model with uniformly distributed $\phi_{i} \in [-1+\epsilon, 1]$ for some $\epsilon>0$ under different error processes}
\label{tab:MCve}
\begin{center}
\scalebox{0.7}{
\renewcommand{\arraystretch}{1}
\begin{tabular}{rrrrrccrrrccrrrrrrrr}
\hline\hline 
 & & & \multicolumn{2}{c}{Bias} &  & \multicolumn{2}{c}{RMSE} &  & \multicolumn{2}{c}{Size ($\times 100$)} &  & \multicolumn{2}{c}{Bias } &  & \multicolumn{2}{c}{RMSE} &  & \multicolumn{2}{c}{Size ($\times 100$)} \\  \cline{4-5} \cline{7-8} \cline{10-11} \cline{13-14} \cline{16-17} \cline{19-20}
& &&  & Hetro &&  & Hetro &&  & Hetro &&  & Hetro &&  & Hetro &  &  & Hetro \\ 
$T$ & $n$ && \multicolumn{1}{c}{FDAC} & GMM &  &  \multicolumn{1}{c}{FDAC} & GMM &  &  \multicolumn{1}{c}{FDAC} & GMM &  &  \multicolumn{1}{c}{FDAC} & GMM &  &  \multicolumn{1}{c}{FDAC} & GMM &  & FDAC & GMM \\ \hline
&  &  & \multicolumn{8}{c}{Gaussian errors without GARCH effects} &  & \multicolumn{8}{c}{Non-Gaussian errors without GARCH effects} \\ \cline{4-11} \cline{13-20}
5 & 1,000 &  & 0.006 & 0.029 &  & 0.047 & 0.076 &  & 2.0 & 2.6 &  & 0.005 & 0.039 &  & 0.047 & 0.088 &  & 2.3 & 4.4 \\
5 & 2,500 &  & 0.000 & 0.010 &  & 0.031 & 0.053 &  & 4.0 & 2.8 &  & 0.000 & 0.013 &  & 0.033 & 0.059 &  & 5.0 & 4.1 \\
5 & 5,000 &  & 0.000 & 0.003 &  & 0.023 & 0.041 &  & 4.9 & 3.7 &  & 0.001 & 0.004 &  & 0.024 & 0.046 &  & 4.4 & 3.0 \\
 &  &  &  &  &  &  &  &  &  &  &  &  &  &  &  &  &  &  &  \\
6 & 1,000 &  & 0.002 & 0.008 &  & 0.038 & 0.050 &  & 4.5 & 2.4 &  & -0.001 & 0.011 &  & 0.038 & 0.055 &  & 3.3 & 3.4 \\
6 & 2,500 &  & 0.000 & 0.001 &  & 0.025 & 0.035 &  & 5.1 & 3.3 &  & -0.002 & -0.001 &  & 0.026 & 0.039 &  & 5.5 & 2.6 \\
6 & 5,000 &  & 0.001 & 0.001 &  & 0.018 & 0.026 &  & 5.1 & 5.1 &  & 0.000 & -0.002 &  & 0.018 & 0.030 &  & 5.1 & 4.2 \\
 &  &  &  &  &  &  &  &  &  &  &  &  &  &  &  &  &  &  &  \\
10 & 1,000 &  & 0.000 & -0.002 &  & 0.024 & 0.027 &  & 5.3 & 5.4 &  & -0.002 & -0.003 &  & 0.024 & 0.029 &  & 5.2 & 6.3 \\
10 & 2,500 &  & 0.000 & -0.001 &  & 0.015 & 0.017 &  & 4.6 & 5.2 &  & -0.001 & -0.002 &  & 0.015 & 0.018 &  & 5.1 & 5.8 \\
10 & 5,000 &  & 0.000 & 0.000 &  & 0.010 & 0.012 &  & 4.4 & 4.5 &  & 0.000 & -0.001 &  & 0.011 & 0.013 &  & 5.6 & 5.9 \\ 
 &  &  &  &  &  &  &  &  &  &  &  &  &  &  &  &  &  &  &  \\
 &  &  & \multicolumn{8}{c}{Gaussian errors with GARCH effects} &  & \multicolumn{8}{c}{Non-Gaussian errors with GARCH effects} \\ \cline{4-11} \cline{13-20}
5 & 1,000 &  & 0.012 & 0.043 &  & 0.054 & 0.093 &  & 2.2 & 3.0 &  & 0.024 & 0.080 &  & 0.076 & 0.151 &  & 3.4 & 5.1 \\
5 & 2,500 &  & 0.002 & 0.016 &  & 0.037 & 0.061 &  & 2.9 & 3.3 &  & 0.010 & 0.046 &  & 0.056 & 0.101 &  & 3.4 & 4.8 \\
5 & 5,000 &  & 0.001 & 0.008 &  & 0.028 & 0.047 &  & 4.3 & 3.5 &  & 0.004 & 0.025 &  & 0.044 & 0.076 &  & 4.3 & 3.5 \\
 &  &  &  &  &  &  &  &  &  &  &  &  &  &  &  &  &  &  &  \\
6 & 1,000 &  & 0.004 & 0.016 &  & 0.044 & 0.059 &  & 3.3 & 2.4 &  & 0.011 & 0.032 &  & 0.058 & 0.087 &  & 2.8 & 3.8 \\
6 & 2,500 &  & 0.000 & 0.004 &  & 0.029 & 0.041 &  & 4.6 & 2.8 &  & 0.003 & 0.014 &  & 0.044 & 0.059 &  & 4.1 & 3.9 \\
6 & 5,000 &  & 0.001 & 0.003 &  & 0.022 & 0.031 &  & 4.8 & 4.6 &  & 0.002 & 0.006 &  & 0.034 & 0.046 &  & 4.9 & 3.8 \\
 &  &  &  &  &  &  &  &  &  &  &  &  &  &  &  &  &  &  &  \\
10 & 1,000 &  & 0.000 & -0.002 &  & 0.029 & 0.031 &  & 6.1 & 5.1 &  & 0.001 & -0.002 &  & 0.039 & 0.039 &  & 4.4 & 6.0 \\
10 & 2,500 &  & 0.000 & -0.001 &  & 0.018 & 0.021 &  & 4.8 & 5.3 &  & 0.000 & -0.002 &  & 0.028 & 0.028 &  & 6.0 & 5.9 \\
10 & 5,000 &  & 0.000 & 0.000 &  & 0.013 & 0.014 &  & 5.0 & 4.6 &  & 0.001 & -0.001 &  & 0.021 & 0.021 &  & 5.5 & 5.2 \\
\hline\hline
\end{tabular}}
\end{center}
\par
{\footnotesize Notes: The DGP is given by $y_{it}=\mu _{i}(1-\phi _{i})+\phi
_{i}y_{i,t-1}+h_{it}\varepsilon_{it}$, for $i=1,2,...,n$, and $t=-99,
-98,...,T$, where the heterogeneous AR(1) coefficients are generated by the
uniform distribution: $\phi_{i} = \mu_{\phi} + v_{i}$, with $v_{i} \sim
IIDU[-a,a]$, $a=0.5$ and $\mu_{\phi} = 0.5$. The standardized errors, $%
\varepsilon _{it}$, are generated as Gaussian, $\varepsilon _{it}\sim
IIDN(0,1)$, or non-Gaussian, $\varepsilon _{it}=\left( e_{it}-2\right) /2$
with $e_{it}\sim IID\chi _{2}^{2}$. The GARCH effect is generated as $%
h_{it}^{2}=\sigma _{i}^{2}(1-\psi _{0}-\psi _{1})+\psi
_{0}h_{i,t-1}^{2}+\psi _{1}(h_{i,t-1}\varepsilon_{i,t-1})^{2}$, with $\sigma
_{i}^{2}\sim IID\left( 0.5+0.5z_{i}^{2}\right) $ and $z_{i}\sim IIDN(0,1)$,
where $\psi _{0}=0.6$ and $\psi _{1}=0.2$, with $h_{i,-M_{i}}=\sigma _{i}$.
In the case of no GARCH effects, $\psi _{0}=\psi _{1}=0$. The initial values
are given by $(y_{i,-100} - \mu_{i}) \sim IIDN(b, \kappa \sigma_{i}^{2})$
with $b=1$ and $\kappa=2$ for all $i$. For each experiment, $(\alpha_{i},
\phi_{i}, \sigma_{i})^{\prime}$ are generated differently across
replications. The FDAC estimator of $\sigma_{\phi}^{2}$ is calculated by
plugging (\ref{Estm1}) and (\ref{Estm2}) into (\ref{Estvar}) in the main
paper. The HetroGMM estimator of $\sigma_{\phi}^{2}$ is calculated by
plugging (\ref{theta1GMM}) and (\ref{theta2GMM}) into (\ref{Estvar}) in the
main paper. The asymptotic variances are estimated by the Delta method. The
estimation is based on $\{y_{i1}, y_{i2},...,y_{iT}\}$ for $i=1,2,...,n$.
The nominal size of the tests is set to 5 per cent. The number of
replications is $2,000$. But the reported results are based on simulated
outcomes with $\left( \hat{\sigma}_{\phi }^{2}\right) ^{(r)} \geq 0.0001$.
The frequencies with negative outcomes, by sample sizes and estimation
method, are reported in Table \ref{tab:npzige}.}
\end{table}

\begin{table}[tbp]
\caption{Frequency of FDAC and HetroGMM estimators of $\protect\sigma_{%
\protect\phi}^{2} = Var(\protect\phi_{i}) = 0.083$ being negative with
uniformly distributed $\phi_{i} \in [-1+\epsilon, 1]$ for some $\epsilon>0$ under different
error processes}
\label{tab:npzige}
\begin{center}
\scalebox{0.8}{
\renewcommand{\arraystretch}{1}
\begin{tabular}{ccccrccrcrrcrr}
\hline\hline
 &  &  & \multicolumn{5}{c}{Without GARCH effects} &  & \multicolumn{5}{c}{With GARCH effects} \\ \cline{4-8} \cline{10-14}
 &  &  & \multicolumn{2}{c}{Gaussian} &  & \multicolumn{2}{c}{Non-Gaussian} &  & \multicolumn{2}{c}{Gaussian} &  & \multicolumn{2}{c}{Non-Gaussian} \\ \cline{4-5} \cline{7-8} \cline{10-11} \cline{13-14}
 &  &  &  & Hetro &  &  & Hetro &  &  & Hetro &  &  & Hetro \\
$T$ & $n$ &  & FDAC & GMM &  & FDAC & GMM &  & FDAC & GMM &  & FDAC & GMM \\ \hline
5  & 1,000 &  & 6.3 & 20.6 &  & 6.7 & 27.2 &  & 10.4 & 24.6 &  & 18.6 & 32.9 \\
5  & 2,500 &  & 0.7 & 9.8  &  & 1.4 & 15.6 &  & 2.2  & 12.9 &  & 8.8  & 26.0 \\
5  & 5,000 &  & 0.0 & 1.9  &  & 0.0 & 7.2  &  & 0.1  & 4.7  &  & 3.4  & 17.8 \\
   &       &  &     &      &  &     &      &  &      &      &  &      &      \\
6  & 1,000 &  & 1.7 & 8.2  &  & 2.4 & 13.4 &  & 3.6  & 11.1 &  & 11.5 & 19.4 \\
6  & 2,500 &  & 0.1 & 1.3  &  & 0.1 & 3.6  &  & 0.4  & 2.8  &  & 3.9  & 10.6 \\
6  & 5,000 &  & 0.0 & 0.0  &  & 0.0 & 0.4  &  & 0.0  & 0.3  &  & 1.7  & 4.8  \\
   &       &  &     &      &  &     &      &  &      &      &  &      &      \\
10 & 1,000 &  & 0.1 & 0.2  &  & 0.0 & 0.4  &  & 0.3  & 0.6  &  & 2.8  & 2.7  \\
10 & 2,500 &  & 0.0 & 0.0  &  & 0.0 & 0.0  &  & 0.0  & 0.0  &  & 0.5  & 0.4  \\
10 & 5,000 &  & 0.0 & 0.0  &  & 0.0 & 0.0  &  & 0.0  & 0.0  &  & 0.2  & 0.0 
\\ \hline\hline  
\end{tabular}}
\end{center}
\par
{\footnotesize Notes: The DGP is given by $y_{it}=\mu _{i}(1-\phi _{i})+\phi
_{i}y_{i,t-1}+h_{it}\varepsilon_{it}$, for $i=1,2,...,n$, and $t=-99,
-M_{i}+2,...,T$ where the heterogeneous AR(1) coefficients are generated by
uniform distributions with $\phi_{i} \in [-1+\epsilon, 1]$ for some $\epsilon>0$ and all $i$. The
estimation is based on $\{y_{i1}, y_{i2},...,y_{iT}\}$ for $i=1,2,...,n$.
The figure denotes the frequency (multiplied by 100) of occurrences where
the estimate of $\sigma_{\phi}^{2}$ is negative or close to zero, $\left(%
\hat{\sigma}_{\phi}^{2}\right)^{(r)} < 0.0001$, for replication $r$ over
2,000 replications. See also the footnotes to Table \ref{tab:MCve}.}
\end{table}

\newpage

\subsubsection{Comparison of the FDAC estimator with FDLS, AH, AAH, AB, and
BB estimators\label{homo_gmm}}

Tables \ref{tab:hetro_u1_a}--\ref{tab:hetro_u2_a} report bias, RMSE, and
size of the FDAC, FDLS, AH, AAH, AB, and BB estimators with $\phi _{i}=\mu
_{\phi }+v_{i}$, $v_{i}\sim IIDU(-a,a)$, $\mu _{\phi } \in \{0.4, 0.5\}$, $a
= 0.5$, and Gaussian errors without GARCH effects. Table \ref{tab:homo_a}
summarizes simulation results of FDAC and the above HomoGMM estimators with
homogeneous $\phi_{i} = \mu_{\phi} = 0.5$ and Gassuain errors without GARCH
effects.

\begin{figure}[h!]
\caption{Empirical power functions for FDAC and FDLS estimators of $\protect%
\phi_{0}=0.5$ in a homogeneous AR(1) panel with Gaussian errors without
GARCH effects}
\label{fig:pw_fdac_fdls_homo_a}
\begin{center}
\includegraphics[scale=0.25]{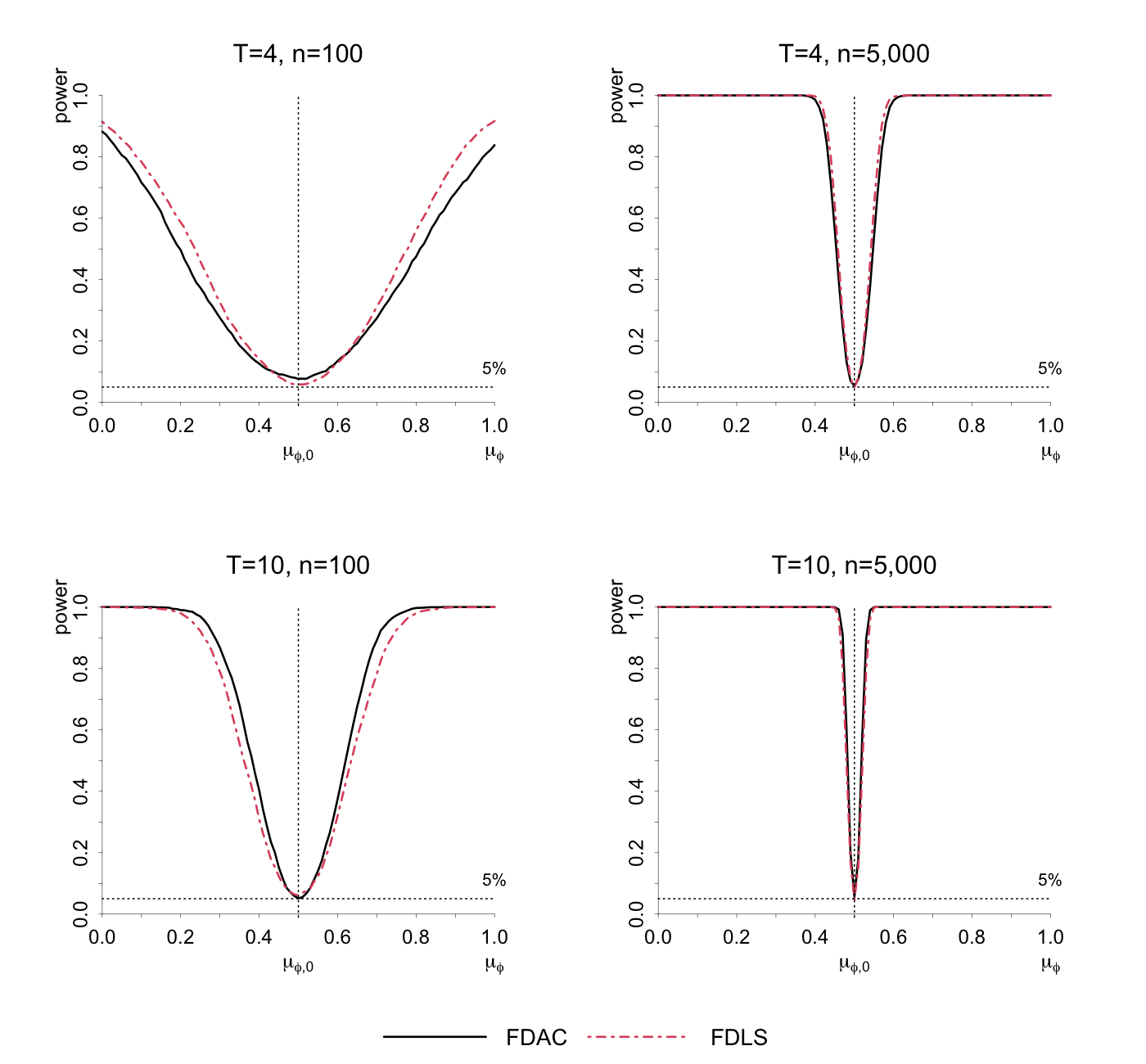}
\end{center}
\end{figure}

\begin{figure}[tbp]
\caption{Empirical power functions for the FDAC estimator of $\protect\mu_{%
\protect\phi}=E(\protect\phi_{i})$ $(\protect\mu_{\protect\phi,0}=0.5)$ in
homogeneous and heterogeneous AR(1) panels with Gaussian and non-Gaussian
error processes without GARCH effects}
\label{fig:fdac_u25_errors_ab}
\begin{center}
\includegraphics[scale=0.25]{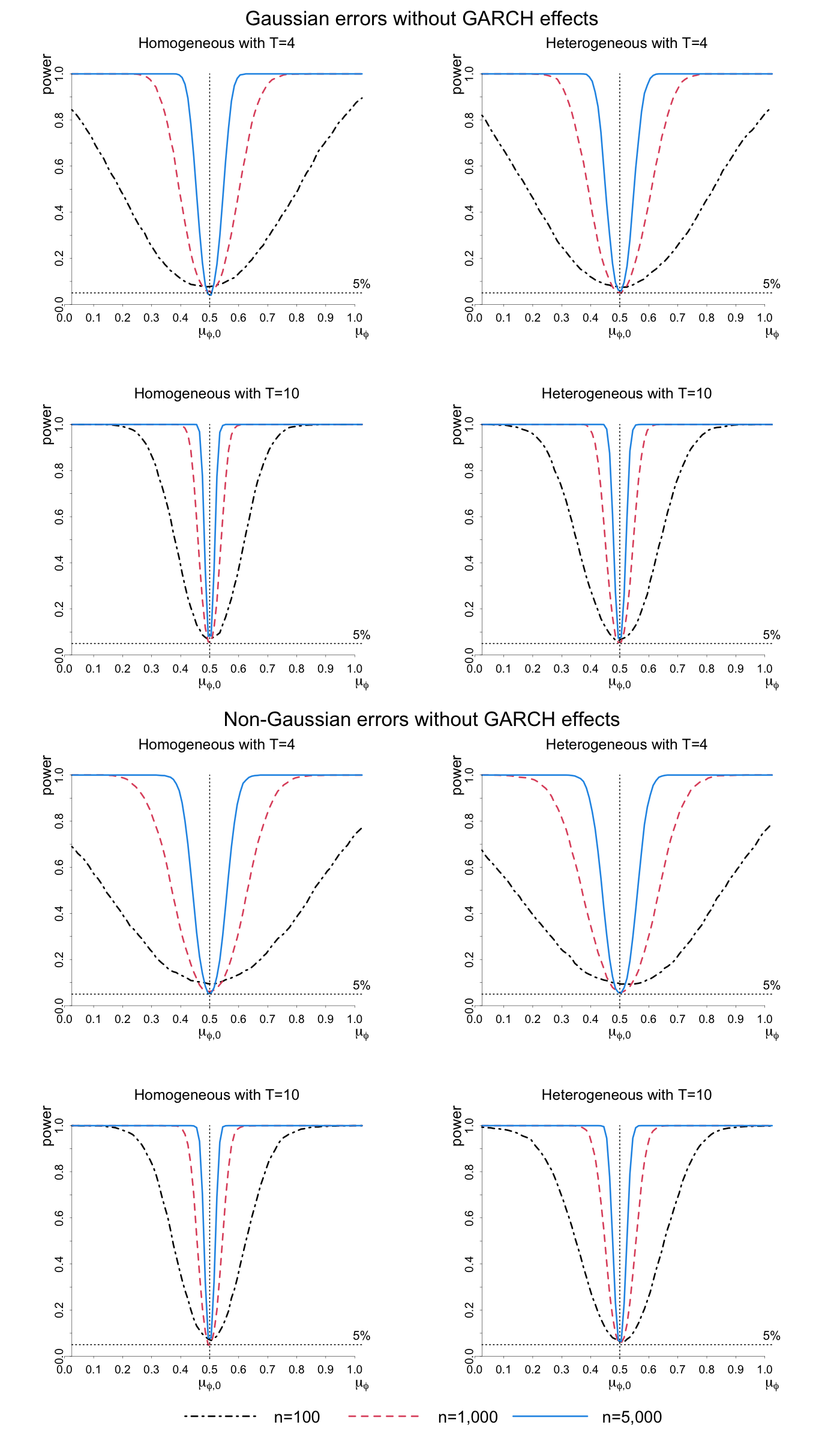}
\end{center}
\end{figure}

Figure \ref{fig:pw_fdac_fdls_homo_a} compares the empirical power functions
of FDAC and FDLS estimators under homogeneity of $\phi_{i}$ for $T=4,10$,
and $n=5,000$. Figures \ref{fig:fdac_u25_errors_ab} and \ref%
{fig:fdac_u25_errors_ac} plot the empirical power functions of the FDAC
estimator in homogeneous ($\phi_{i} = \mu_{\phi} = 0.5$ for all $i$) and
heterogeneous panel AR(1) panels, where the heterogeneous AR(1) coefficients
are generated by the above uniform distribution with $\phi_{i} \in (-1,1]$ and $\mu_{\phi} = 0.5$, under different error processes for $T=4, 10$
and $n=100, 1000, 5000$.

\begin{sidewaystable}
\begin{center}
\caption{Bias, RMSE, and size of FDAC, FDLS, AH, AAH, AB, and BB estimators of $\mu_{\phi} = E(\phi_{i})= 0.4$ in a heterogeneous panel AR(1) model with uniformly distributed $\vert \phi_{i}\vert < 1$ and Gaussian errors without GARCH effects} 
\label{tab:hetro_u1_a}
\scalebox{0.8}{
\begin{tabular}{rrrrrrrrrrrrrrcrrcrrrrr}
\\ \hline\hline & & & \multicolumn{6}{c}{Bias} && \multicolumn{6}{c}{RMSE} && \multicolumn{6}{c}{Size ($\times 100$)} \\ \cline{4-9} \cline{11-16} \cline{18-23} 
 $T$ & $n$ & & \multicolumn{1}{c}{FDAC} &  \multicolumn{1}{c}{FDLS} & \multicolumn{1}{c}{AH} & \multicolumn{1}{c}{AAH} & \multicolumn{1}{c}{AB} & \multicolumn{1}{c}{BB}   &&  \multicolumn{1}{c}{FDAC} &  \multicolumn{1}{c}{FDLS} & \multicolumn{1}{c}{AH} & \multicolumn{1}{c}{AAH} & \multicolumn{1}{c}{AB} & \multicolumn{1}{c}{BB}  &&  \multicolumn{1}{c}{FDAC} &  \multicolumn{1}{c}{FDLS} & \multicolumn{1}{c}{AH} & \multicolumn{1}{c}{AAH} & \multicolumn{1}{c}{AB} & \multicolumn{1}{c}{BB}   \\ \hline
4 & 100 &  & 0.005 & -0.057 & -0.124 & -0.007 & -0.081 & -0.041 &  & 0.182 & 0.168 & 1.930 & 0.261 & 0.236 & 0.157 &  & 8.2 & 8.2 & 15.5 & 18.2 & 14.9 & 16.2 \\
4 & 1,000 &  & -0.001 & -0.061 & -0.176 & -0.066 & -0.061 & -0.046 &  & 0.058 & 0.079 & 0.223 & 0.109 & 0.092 & 0.066 &  & 6.7 & 23.2 & 33.6 & 30.8 & 19.4 & 19.8 \\
4 & 5,000 &  & 0.000 & -0.062 & -0.185 & -0.073 & -0.055 & -0.044 &  & 0.026 & 0.065 & 0.195 & 0.077 & 0.063 & 0.049 &  & 5.4 & 78.1 & 85.8 & 81.4 & 46.8 & 56.4 \\
 &  &  &  &  &  &  &  &  &  &  &  &  &  &  &  &  &  &  &  &  &  &  \\
6 & 100 &  & 0.001 & -0.060 & -0.178 & -0.025 & -0.086 & -0.026 &  & 0.114 & 0.127 & 0.258 & 0.123 & 0.163 & 0.107 &  & 7.4 & 10.3 & 32.8 & 18.4 & 27.7 & 24.4 \\
6 & 1,000 &  & 0.000 & -0.061 & -0.143 & -0.033 & -0.052 & -0.022 &  & 0.036 & 0.072 & 0.155 & 0.046 & 0.068 & 0.039 &  & 4.1 & 40.8 & 71.5 & 18.9 & 30.0 & 15.5 \\
6 & 5,000 &  & 0.000 & -0.062 & -0.140 & -0.033 & -0.046 & -0.020 &  & 0.016 & 0.064 & 0.142 & 0.036 & 0.050 & 0.024 &  & 5.0 & 96.8 & 100.0 & 60.1 & 70.6 & 31.8 \\
 &  &  &  &  &  &  &  &  &  &  &  &  &  &  &  &  &  &  &  &  &  &  \\
10 & 100 &  & 0.000 & -0.061 & -0.121 & -0.031 & -0.060 & -0.018 &  & 0.078 & 0.102 & 0.159 & 0.084 & 0.110 & 0.080 &  & 5.2 & 13.1 & 57.5 & 41.5 & 48.1 & 48.4 \\
10 & 1,000 &  & 0.000 & -0.061 & -0.092 & -0.022 & -0.033 & -0.002 &  & 0.025 & 0.067 & 0.098 & 0.033 & 0.044 & 0.024 &  & 4.8 & 63.6 & 85.2 & 17.7 & 32.9 & 11.5 \\
10 & 5,000 &  & 0.000 & -0.061 & -0.090 & -0.021 & -0.029 & 0.000 &  & 0.011 & 0.063 & 0.091 & 0.024 & 0.032 & 0.011 &  & 5.2 & 100.0 & 100.0 & 44.8 & 70.9 & 9.6\\
\hline\hline
\end{tabular}
}
\end{center}
{\footnotesize 
Notes: The DGP is given by $y_{it}=\mu _{i}(1-\phi _{i})+\phi
_{i}y_{i,t-1}+h_{it}\varepsilon_{it}$, for $i=1,2,...,n$, and $t=-99, -98,...,T$, 
featuring Gaussian standardized errors with cross-sectional heteroskedasticity without GARCH effects. 
The heterogeneous AR(1) coefficients are generated by the uniform distribution: $\phi_{i} = \mu_{\phi} + v_{i}$, with $v_{i} \sim IIDU[-a,a]$, $a=0.5$ and $\mu_{\phi} = 0.4$. 
The initial values are given by $(y_{i,-100} - \mu_{i}) \sim IIDN(b, \kappa \sigma_{i}^{2})$ 
with $b=1$ and $\kappa=2$ for all $i$. For each experiment, $(\alpha_{i}, \phi_{i}, \sigma_{i})^{\prime}$ are generated differently across replications. The FDAC estimator is calculated by (\ref{Estm1}) in the main paper, and its asymptotic variance is estimated by the Delta method. \textquotedblleft FDLS" denotes the first difference least square estimator proposed by \cite{HanPhillips2010}. \textquotedblleft AH", \textquotedblleft AAH", \textquotedblleft AB", and \textquotedblleft BB" denote the 2-step GMM estimators proposed by \cite{AndersonHsiao1981,AndersonHsiao1982}, \cite{ChudikPesaran2021}, \cite{ArellanoBond1991}, and \cite{BlundellBond1998}.
The estimation is based on $\{y_{i1}, y_{i2},...,y_{iT}\}$ for $i=1,2,...,n$. The nominal size of the tests is set to 5 per cent. 
The number of replications is $2,000$. }
\end{sidewaystable}

\begin{sidewaystable}
\begin{center}
\caption{Bias, RMSE, and size of FDAC, FDLS, AH, AAH, AB, and BB estimators of $\mu_{\phi} = E(\phi_{i}) = 0.5$ in a heterogeneous panel AR(1) model with uniformly distributed $\phi_{i} \in [-1+\epsilon, 1]$ for some $\epsilon>0$ and Gaussian errors without GARCH effects} 
\label{tab:hetro_u2_a}
\scalebox{0.8}{
\begin{tabular}{rrrrrrrrrrrrrrcrrccrrrr}
\\ \hline\hline & & & \multicolumn{6}{c}{Bias} && \multicolumn{6}{c}{RMSE} && \multicolumn{6}{c}{Size ($\times 100$)} \\ \cline{4-9} \cline{11-16} \cline{18-23} 
 $T$ & $n$ & & \multicolumn{1}{c}{FDAC} &  \multicolumn{1}{c}{FDLS} & \multicolumn{1}{c}{AH} & \multicolumn{1}{c}{AAH} & \multicolumn{1}{c}{AB} & \multicolumn{1}{c}{BB}   &&  \multicolumn{1}{c}{FDAC} &  \multicolumn{1}{c}{FDLS} & \multicolumn{1}{c}{AH} & \multicolumn{1}{c}{AAH} & \multicolumn{1}{c}{AB} & \multicolumn{1}{c}{BB}  &&  \multicolumn{1}{c}{FDAC} &  \multicolumn{1}{c}{FDLS} & \multicolumn{1}{c}{AH} & \multicolumn{1}{c}{AAH} & \multicolumn{1}{c}{AB} & \multicolumn{1}{c}{BB}   \\ \hline
4 & 100 &  & 0.003 & -0.053 & -0.070 & -0.012 & -0.109 & -0.034 &  & 0.181 & 0.171 & 2.255 & 0.263 & 0.305 & 0.166 &  & 8.3 & 8.2 & 15.2 & 16.1 & 16.3 & 17.2 \\
4 & 1,000 &  & -0.001 & -0.057 & -0.190 & -0.062 & -0.072 & -0.043 &  & 0.057 & 0.076 & 0.256 & 0.125 & 0.113 & 0.068 &  & 6.2 & 19.8 & 30.0 & 29.5 & 18.5 & 16.4 \\
4 & 5,000 &  & 0.000 & -0.057 & -0.204 & -0.075 & -0.065 & -0.043 &  & 0.025 & 0.062 & 0.216 & 0.080 & 0.076 & 0.050 &  & 5.1 & 70.1 & 78.6 & 74.1 & 40.6 & 45.2 \\
 &  &  &  &  &  &  &  &  &  &  &  &  &  &  &  &  &  &  &  &  &  &  \\
6 & 100 &  & 0.001 & -0.056 & -0.224 & -0.021 & -0.119 & -0.022 &  & 0.111 & 0.127 & 0.314 & 0.133 & 0.202 & 0.114 &  & 7.2 & 9.6 & 35.8 & 21.0 & 30.8 & 27.8 \\
6 & 1,000 &  & 0.000 & -0.057 & -0.171 & -0.034 & -0.074 & -0.021 &  & 0.035 & 0.069 & 0.184 & 0.047 & 0.092 & 0.040 &  & 4.4 & 35.0 & 73.7 & 20.5 & 38.0 & 15.1 \\
6 & 5,000 &  & 0.000 & -0.057 & -0.166 & -0.034 & -0.068 & -0.020 &  & 0.016 & 0.060 & 0.169 & 0.037 & 0.073 & 0.025 &  & 5.1 & 93.3 & 100.0 & 61.2 & 84.3 & 29.1 \\
 &  &  &  &  &  &  &  &  &  &  &  &  &  &  &  &  &  &  &  &  &  &  \\
10 & 100 &  & 0.000 & -0.057 & -0.164 & -0.024 & -0.086 & -0.016 &  & 0.077 & 0.101 & 0.201 & 0.087 & 0.136 & 0.085 &  & 5.2 & 12.0 & 66.4 & 41.0 & 54.5 & 51.5 \\
10 & 1,000 &  & 0.000 & -0.057 & -0.123 & -0.018 & -0.059 & -0.003 &  & 0.025 & 0.063 & 0.129 & 0.031 & 0.069 & 0.025 &  & 4.6 & 55.3 & 94.6 & 13.0 & 58.1 & 11.9 \\
10 & 5,000 &  & 0.000 & -0.057 & -0.120 & -0.017 & -0.058 & 0.000 &  & 0.011 & 0.058 & 0.121 & 0.021 & 0.061 & 0.012 &  & 4.9 & 99.8 & 100.0 & 29.5 & 97.3 & 10.2 \\
\hline\hline
\end{tabular}
}
\end{center}
{\footnotesize 
Notes: The DGP is given by $y_{it}=\mu _{i}(1-\phi _{i})+\phi
_{i}y_{i,t-1}+h_{it}\varepsilon_{it}$, for $i=1,2,...,n$, and $t=-99,-98,...,T$, 
featuring Gaussian standardized errors with cross-sectional heteroskedasticity without GARCH effects. 
The heterogeneous AR(1) coefficients are generated by the uniform distribution: $\phi_{i} = \mu_{\phi} + v_{i}$, with $v_{i} \sim IIDU[-a,a]$, $a=0.5$ and $\mu_{\phi} = 0.5$. 
The initial values are given by $(y_{i,-100} - \mu_{i}) \sim IIDN(b, \kappa \sigma_{i}^{2})$ 
with $b=1$ and $\kappa=2$ for all $i$. For each experiment, $(\alpha_{i}, \phi_{i}, \sigma_{i})^{\prime}$ are generated differently across replications. The FDAC estimator is calculated by (\ref{Estm1}) in the main paper, and its asymptotic variance is estimated by the Delta method. 
\textquotedblleft FDLS" denotes the first difference least square estimator proposed by \cite{HanPhillips2010}. \textquotedblleft AH", \textquotedblleft AAH", \textquotedblleft AB", and \textquotedblleft BB" denote the 2-step GMM estimators proposed by \cite{AndersonHsiao1981,AndersonHsiao1982}, \cite{ChudikPesaran2021}, \cite{ArellanoBond1991}, and \cite{BlundellBond1998}.
The estimation is based on $\{y_{i1}, y_{i2},...,y_{iT}\}$ for $i=1,2,...,n$. The nominal size of the tests is set to 5 per cent. 
The number of replications is $2,000$. }
\end{sidewaystable}

\begin{sidewaystable}
\begin{center}
\caption{Bias, RMSE, and size of FDAC, FDLS, AH, AAH, AB, and BB estimators of $\phi$ $(\phi_{0}=0.5)$ in a homogeneous panel AR(1) model with Gaussian errors without GARCH effects} 
\label{tab:homo_a}
\scalebox{0.8}{
\begin{tabular}{rrrrrrrrrrrrrrcrrccrrrr}
\\ \hline\hline & & & \multicolumn{6}{c}{Bias} && \multicolumn{6}{c}{RMSE} && \multicolumn{6}{c}{Size ($\times 100$)} \\ \cline{4-9} \cline{11-16} \cline{18-23} 
 $T$ & $n$ & & \multicolumn{1}{c}{FDAC} &  \multicolumn{1}{c}{FDLS} & \multicolumn{1}{c}{AH} & \multicolumn{1}{c}{AAH} & \multicolumn{1}{c}{AB} & \multicolumn{1}{c}{BB}   &&  \multicolumn{1}{c}{FDAC} &  \multicolumn{1}{c}{FDLS} & \multicolumn{1}{c}{AH} & \multicolumn{1}{c}{AAH} & \multicolumn{1}{c}{AB} & \multicolumn{1}{c}{BB}  &&  \multicolumn{1}{c}{FDAC} &  \multicolumn{1}{c}{FDLS} & \multicolumn{1}{c}{AH} & \multicolumn{1}{c}{AAH} & \multicolumn{1}{c}{AB} & \multicolumn{1}{c}{BB}   \\ \hline
4 & 100 &  & 0.004 & 0.002 & 0.564 & 0.084 & -0.023 & -0.006 &  & 0.169 & 0.150 & 9.211 & 0.298 & 0.226 & 0.141 &  & 7.7 & 5.9 & 8.3 & 12.7 & 10.0 & 13.2 \\
4 & 1,000 &  & 0.000 & 0.001 & 0.021 & 0.072 & -0.004 & -0.002 &  & 0.054 & 0.049 & 0.206 & 0.209 & 0.070 & 0.044 &  & 6.2 & 5.6 & 4.2 & 18.4 & 5.9 & 5.2 \\
4 & 5,000 &  & 0.000 & 0.000 & 0.002 & 0.023 & 0.000 & 0.000 &  & 0.024 & 0.022 & 0.086 & 0.116 & 0.031 & 0.020 &  & 5.4 & 4.9 & 5.2 & 9.2 & 5.0 & 5.0 \\
 &  &  &  &  &  &  &  &  &  &  &  &  &  &  &  &  &  &  &  &  &  &  \\
6 & 100 &  & 0.001 & -0.001 & -0.072 & 0.023 & -0.041 & -0.001 &  & 0.098 & 0.104 & 0.220 & 0.152 & 0.130 & 0.088 &  & 8.1 & 6.6 & 15.9 & 22.6 & 16.9 & 21.1 \\
6 & 1,000 &  & 0.000 & 0.000 & -0.007 & 0.001 & -0.006 & 0.000 &  & 0.031 & 0.034 & 0.066 & 0.040 & 0.038 & 0.026 &  & 4.5 & 5.9 & 6.4 & 6.7 & 6.4 & 6.3 \\
6 & 5,000 &  & 0.000 & 0.000 & -0.001 & 0.000 & -0.001 & 0.000 &  & 0.014 & 0.015 & 0.030 & 0.014 & 0.017 & 0.012 &  & 4.9 & 5.3 & 5.4 & 5.2 & 6.5 & 5.8 \\
 &  &  &  &  &  &  &  &  &  &  &  &  &  &  &  &  &  &  &  &  &  &  \\
10 & 100 &  & 0.000 & -0.001 & -0.049 & 0.006 & -0.035 & -0.004 &  & 0.063 & 0.074 & 0.103 & 0.073 & 0.080 & 0.057 &  & 5.1 & 6.0 & 34.0 & 44.3 & 39.6 & 43.6 \\
10 & 1,000 &  & 0.000 & 0.000 & -0.004 & 0.000 & -0.004 & 0.000 &  & 0.020 & 0.024 & 0.028 & 0.016 & 0.021 & 0.016 &  & 5.0 & 5.1 & 8.5 & 8.8 & 8.7 & 8.6 \\
10 & 5,000 &  & 0.000 & 0.000 & -0.001 & 0.000 & 0.000 & 0.000 &  & 0.009 & 0.011 & 0.013 & 0.007 & 0.009 & 0.007 &  & 5.6 & 4.7 & 5.2 & 5.9 & 6.2 & 5.3 \\ 
\hline
\hline
\end{tabular}
}
\end{center}
{\footnotesize 
Notes: The DGP is given by $y_{it}=\mu _{i}(1-\phi _{i})+\phi
_{i}y_{i,t-1}+h_{it}\varepsilon_{it}$, for $i=1,2,...,n$, and $t=-99, -98,...,T$, 
featuring Gaussian standardized errors with cross-sectional heteroskedasticity without GARCH effects. 
The AR(1) coefficients are generated to be homogeneous: $\phi_{i} = \mu_{\phi}= 0.5$ for all $i$. 
The initial values are given by $(y_{i,-100} - \mu_{i}) \sim IIDN(b, \kappa \sigma_{i}^{2})$ 
with $b=1$ and $\kappa=2$ for all $i$. For each experiment, $(\alpha_{i}, \sigma_{i})^{\prime}$ are generated differently across replications. The FDAC estimator is calculated by (\ref{Estm1}) in the main paper, and its asymptotic variance is estimated by the Delta method. \textquotedblleft FDLS" denotes the first difference least square estimator proposed by \cite{HanPhillips2010}. \textquotedblleft AH", \textquotedblleft AAH", \textquotedblleft AB", and \textquotedblleft BB" denote the 2-step GMM estimators proposed by \cite{AndersonHsiao1981,AndersonHsiao1982}, \cite{ChudikPesaran2021}, \cite{ArellanoBond1991}, and \cite{BlundellBond1998}.
The estimation is based on $\{y_{i1}, y_{i2},...,y_{iT}\}$ for $i=1,2,...,n$. The nominal size of the tests is set to 5 per cent. 
The number of replications is $2,000$. }
\end{sidewaystable}

\newpage\clearpage

\newpage\clearpage

\begin{figure}[tbp]
\caption{Empirical power functions for the FDAC estimator of $\protect\mu_{%
\protect\phi} = E(\protect\phi_{i}) $ $(\protect\mu_{\protect\phi,0}=0.5)$
in homogeneous and heterogeneous AR(1) panels with Gaussian errors without
and with GARCH effects}
\label{fig:fdac_u25_errors_ac}
\begin{center}
\includegraphics[scale=0.25]{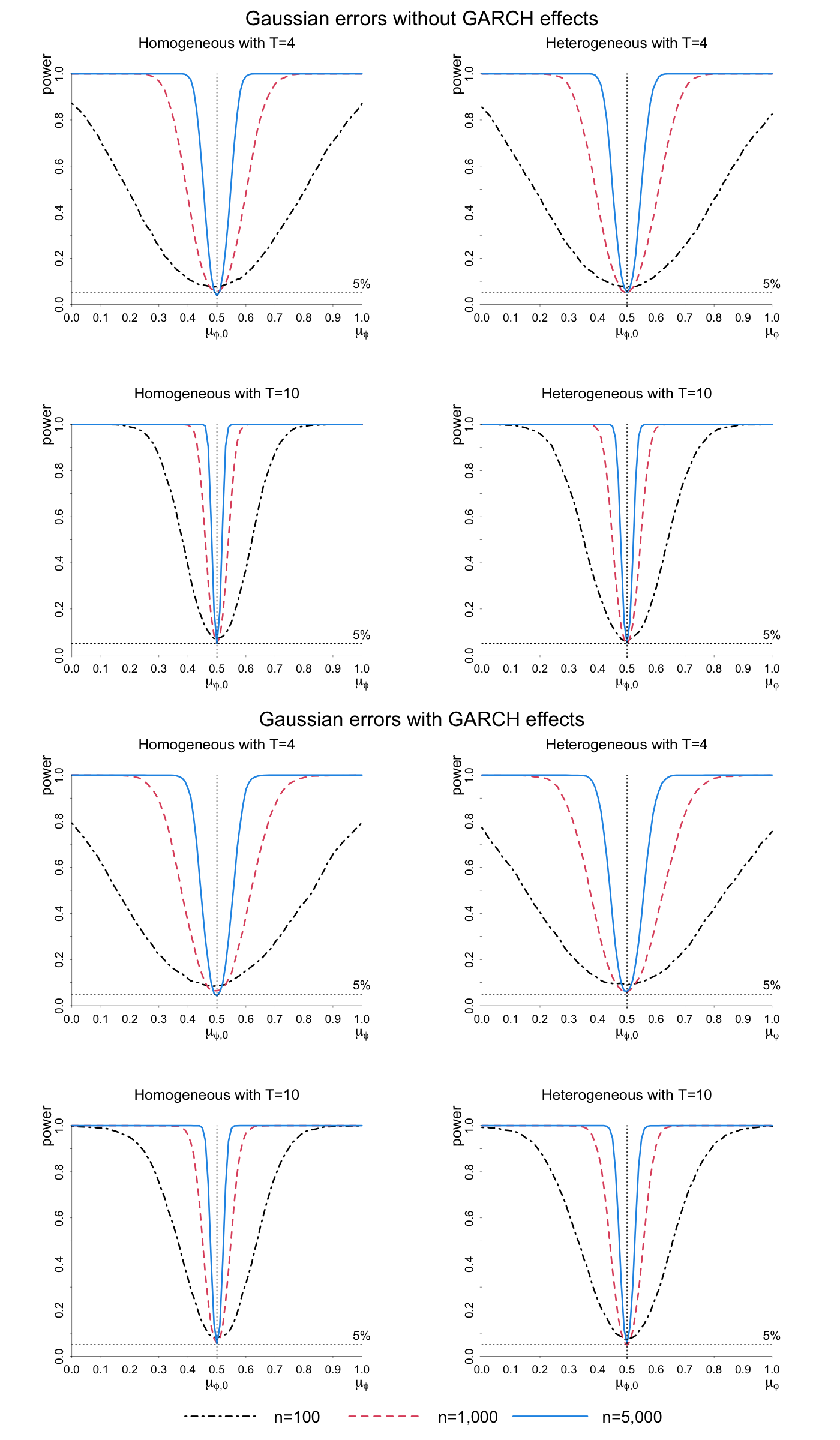}
\end{center}
\end{figure}

\clearpage \newpage
\subsubsection{Comparison of FDAC and MSW estimator}

Table \ref{tab:msw_m1} summarizes bias, RMSE, and size of FDAC and MSW
estimators of $\mu _{\phi }=E(\phi _{i})$ with uniformly distributed $\phi
_{i}$ in the case of Gaussian errors without GARCH effects for the sample
size combinations $n=100,1000$ and $T=4,6,10$.

\begin{table}[h!]
\caption{Bias, RMSE, and size of FDAC and MSW estimators of $\protect\mu_{%
\protect\phi} = E(\protect\phi_{i})$ in heterogeneous panel AR(1) models
with uniformly distributed $\protect\phi_{i}$ and Gaussian errors without
GARCH effects}
\label{tab:msw_m1}
\begin{center}
\scalebox{0.7}{
\renewcommand{\arraystretch}{1.05}
\begin{tabular}{rrrrrrccccrcrrrrrrcr}
\hline\hline
&  &  & \multicolumn{8}{c}{$\mu_{\phi} = 0.4$ with $\vert \phi_{i} \vert < 1$} &  & \multicolumn{8}{c}{$\mu_{\phi} = 0.5$ with $\phi_{i} \in [-1+\epsilon, 1]$ for some $\epsilon>0$} \\ \cline{4-11} \cline{13-20}
 & & & \multicolumn{2}{c}{Bias} &  & \multicolumn{2}{c}{RMSE} &  & \multicolumn{2}{c}{Size ($\times 100$)} &  & \multicolumn{2}{c}{Bias } &  & \multicolumn{2}{c}{RMSE} &  & \multicolumn{2}{c}{Size ($\times 100$)} \\  \cline{4-5} \cline{7-8} \cline{10-11} \cline{13-14} \cline{16-17} \cline{19-20}
$T$ & $n$ &  & FDAC & MSW &  & FDAC & MSW &  & FDAC & MSW &  & FDAC & MSW &  & FDAC & MSW &  & FDAC & MSW \\  \hline
4 & 100 &  & -0.005 & -0.145 &  & 0.177 & 0.157 &  & 8.0 & 82.0 &  & -0.005 & -0.207 &  & 0.175 & 0.221 &  & 8.7 & 84.3 \\
4 & 1,000 &  & 0.000 & -0.128 &  & 0.056 & 0.130 &  & 4.7 & 100.0 &  & 0.000 & -0.194 &  & 0.056 & 0.196 &  & 5.0 & 100.0 \\
 &  &  &  &  &  &  &  &  &  &  &  &  &  &  &  &  &  &  &  \\
6 & 100 &  & -0.004 & -0.144 &  & 0.113 & 0.155 &  & 5.7 & 79.3 &  & -0.004 & -0.202 &  & 0.111 & 0.215 &  & 5.5 & 81.2 \\
6 & 1,000 &  & -0.001 & -0.129 &  & 0.037 & 0.130 &  & 6.3 & 100.0 &  & -0.001 & -0.187 &  & 0.036 & 0.189 &  & 5.2 & 100.0 \\
 &  &  &  &  &  &  &  &  &  &  &  &  &  &  &  &  &  &  &  \\
10 & 100 &  & -0.001 & -0.146 &  & 0.079 & 0.158 &  & 6.4 & 71.2 &  & -0.001 & -0.198 &  & 0.079 & 0.213 &  & 6.7 & 74.3 \\
10 & 1,000 &  & 0.000 & -0.141 &  & 0.026 & 0.143 &  & 5.7 & 100.0 &  & -0.001 & -0.192 &  & 0.025 & 0.194 &  & 5.9 & 100.0 \\
\hline\hline
\end{tabular}}
\end{center}
\par
{\footnotesize Notes: The DGP is given by $y_{it}=\mu _{i}(1-\phi _{i})+\phi
_{i}y_{i,t-1}+h_{it}\varepsilon_{it}$, for $i=1,2,...,n$, and $t=-99,
-98,...,T$, featuring Gaussian standardized errors with cross-sectional
heteroskedasticity without GARCH effects, where the heterogeneous AR(1)
coefficients are generated by uniform distributions. The FDAC estimator is
calculated by (\ref{Estm1}) in the main paper. The asymptotic variance is
estimated by the Delta method. \textquotedblleft MSW" denotes the estimator
proposed by \cite{MavroeidisEtal2015}. The estimation is based on $\{y_{i1},
y_{i2},...,y_{iT}\}$ for $i=1,2,...,n$. The nominal size of the tests is set
to 5 per cent. Due to the extensive computations required for the
implementation of the MSW estimator, the number of replications is $1,000$.
See also footnotes to Table \ref{tab:MCAm1u}.}
\end{table}

\newpage\clearpage

\subsubsection{Simulation results with different initializations\label{InMC}}

Tables \ref{tab:MCAu_mi} and \ref{tab:MCAc_mi} summarize the bias, RMSE, and
size of the FDAC estimator of $E(\phi _{i})$ with uniformly and
categorically distributed $\phi_{i}$, respectively, under different
initializations $M_{i}=100, 3, 1$ for all $i$, (except a case of
categorically distributed $\phi_{i}$ where $M_{i}=100$ for units with $%
\phi_{i}=\phi_{L} = 0.5$ and $M_{i}=1$ for units with $\phi_{i} = \phi_{H}=1$%
). Table \ref{tab:homo_a_mi} reports the bias, RMSE, and sizes of the FDAC,
FDLS, AH, AAH, AB, and BB estimators in homogeneous panels for $%
M_{i}=100,3,1 $ for all $i$. The simulation results for heterogeneous panels
with uniformly distributed $\phi_{i}$ are shown in Table \ref%
{tab:hetro_u1_a_mi} for $\mu_{\phi}=0.4$, and Table \ref{tab:hetro_u2_a_mi}
for $\mu_{\phi}=0.5$. Table \ref{tab:msw_m1_mi} summarizes results of FDAC
and MSW estimators in both homogeneous and heterogeneous panels for
different initializations with $M_{i}=100,1$ for all $i$. .

\begin{table}[h!]
\caption{Bias, RMSE, and size of the FDAC estimator of $\protect\mu_{\protect%
\phi} = E(\protect\phi _{i})$ in a heterogeneous panel AR(1) model with
uniformly distributed $\protect\phi_{i}$, Gaussian errors without GARCH
effects, and different initializations}
\label{tab:MCAu_mi}
\begin{center}
\scalebox{0.75}{
\renewcommand{\arraystretch}{1.05}
\begin{tabular}{rrcccccccccccr}
\hline\hline
&  &  & \multicolumn{3}{c}{Bias} &  & \multicolumn{3}{c}{RMSE} &  & \multicolumn{3}{c}{Size $(\times 100)$} \\ \cline{4-6} \cline{8-10} \cline{12-14}
$T$ & $n$/$M_{i}$ &  & 100 & 3 & 1 &  & 100 & 3 & 1 &  & 100 & 3 & \multicolumn{1}{c}{1} \\ \hline
\multicolumn{14}{l}{$\mu_{\phi} = 0.4$ with $\vert \phi_{i} \vert < 1$}  \\ \hline
4 & 100 &  & -0.008 & 0.004 & 0.052 &  & 0.174 & 0.180 & 0.194 &  & 7.0 & 7.9 & 8.5 \\
4 & 1,000 &  & 0.000 & 0.016 & 0.063 &  & 0.057 & 0.060 & 0.087 &  & 5.1 & 5.6 & 18.2 \\
4 & 5,000 &  & 0.000 & 0.014 & 0.062 &  & 0.025 & 0.029 & 0.068 &  & 3.8 & 8.5 & 64.8 \\
 &  &  &  &  &  &  &  &  &  &  &  &  &  \\
5 & 100 &  & -0.003 & 0.008 & 0.041 &  & 0.134 & 0.136 & 0.147 &  & 6.3 & 7.2 & 8.8 \\
5 & 1,000 &  & 0.000 & 0.014 & 0.048 &  & 0.043 & 0.046 & 0.066 &  & 5.1 & 6.5 & 19.0 \\
5 & 5,000 &  & 0.000 & 0.012 & 0.048 &  & 0.019 & 0.023 & 0.052 &  & 4.4 & 9.2 & 65.2 \\
 &  &  &  &  &  &  &  &  &  &  &  &  &  \\
6 & 100 &  & -0.004 & 0.007 & 0.034 &  & 0.111 & 0.113 & 0.124 &  & 6.3 & 6.2 & 8.8 \\
6 & 1,000 &  & -0.001 & 0.011 & 0.038 &  & 0.037 & 0.038 & 0.054 &  & 5.8 & 5.1 & 16.4 \\
6 & 5,000 &  & 0.000 & 0.010 & 0.038 &  & 0.016 & 0.019 & 0.042 &  & 4.4 & 9.0 & 60.9 \\
 &  &  &  &  &  &  &  &  &  &  &  &  &  \\
10 & 100 &  & 0.000 & 0.005 & 0.015 &  & 0.078 & 0.082 & 0.084 &  & 6.5 & 7.2 & 7.0 \\
10 & 1,000 &  & 0.000 & 0.007 & 0.020 &  & 0.026 & 0.026 & 0.033 &  & 5.8 & 6.6 & 12.8 \\
10 & 5,000 &  & 0.000 & 0.006 & 0.021 &  & 0.011 & 0.013 & 0.024 &  & 5.3 & 7.5 & 41.9 \\
 &  &  &  &  &  &  &  &  &  &  &  &  &  \\
\multicolumn{14}{l}{$\mu_{\phi} = 0.5$ with $\phi_{i} \in [-1+\epsilon, 1]$ for some $\epsilon>0$}  \\ \hline
4 & 100 &  & 0.003 & 0.008 & 0.050 &  & 0.176 & 0.174 & 0.191 &  & 7.6 & 7.8 & 9.0 \\
4 & 1,000 &  & 0.000 & 0.012 & 0.058 &  & 0.057 & 0.057 & 0.084 &  & 5.1 & 5.2 & 17.0 \\
4 & 5,000 &  & 0.001 & 0.013 & 0.057 &  & 0.026 & 0.029 & 0.063 &  & 5.4 & 7.8 & 57.9 \\
 &  &  &  &  &  &  &  &  &  &  &  &  &  \\
5 & 100 &  & -0.003 & 0.007 & 0.039 &  & 0.134 & 0.136 & 0.145 &  & 7.1 & 6.7 & 8.5 \\
5 & 1,000 &  & -0.001 & 0.010 & 0.044 &  & 0.042 & 0.044 & 0.063 &  & 5.1 & 6.2 & 17.1 \\
5 & 5,000 &  & 0.001 & 0.010 & 0.044 &  & 0.019 & 0.022 & 0.048 &  & 4.3 & 9.8 & 59.2 \\
 &  &  &  &  &  &  &  &  &  &  &  &  &  \\
6 & 100 &  & -0.004 & 0.003 & 0.030 &  & 0.112 & 0.114 & 0.121 &  & 7.1 & 7.3 & 9.2 \\
6 & 1,000 &  & -0.002 & 0.007 & 0.035 &  & 0.035 & 0.037 & 0.051 &  & 4.5 & 6.2 & 15.9 \\
6 & 5,000 &  & 0.000 & 0.008 & 0.035 &  & 0.016 & 0.018 & 0.039 &  & 4.7 & 8.8 & 56.6 \\
 &  &  &  &  &  &  &  &  &  &  &  &  &  \\
10 & 100 &  & -0.003 & 0.001 & 0.016 &  & 0.079 & 0.078 & 0.083 &  & 6.3 & 5.9 & 8.1 \\
10 & 1,000 &  & -0.001 & 0.004 & 0.019 &  & 0.025 & 0.026 & 0.032 &  & 4.8 & 5.9 & 11.9 \\
10 & 5,000 &  & 0.000 & 0.005 & 0.019 &  & 0.011 & 0.012 & 0.022 &  & 5.3 & 7.3 & 36.6 \\
\hline\hline
\end{tabular}}
\end{center}
\par
{\footnotesize Notes: The DGP is given by $y_{it}=\mu _{i}(1-\phi _{i})+\phi
_{i}y_{i,t-1}+h_{it}\varepsilon_{it}$, for $i=1,2,...,n$, and $t=-M_{i}+1,
-M_{i}+2,...,T$, featuring Gaussian standardized errors with cross-sectional
heteroskedasticity without GARCH effects. The heterogeneous AR(1)
coefficients are generated by the uniform distribution: $\phi_{i} =
\mu_{\phi} + v_{i}$, with $v_{i} \sim IIDU[-a,a]$, $a=0.5$ and $\mu_{\phi}
\in \{0.4, 0.5\}$ The initial values are given by $(y_{i,-M_{i}} - \mu_{i})
\sim IIDN(b, \kappa \sigma_{i}^{2})$ with $b=1$ and $\kappa=2$, where $M_{i}
\in \{ 100, 3, 1\}$ for all $i$. For each experiment, $(\alpha_{i},
\phi_{i}, \sigma_{i})^{\prime}$ are generated differently across
replications. The FDAC estimator is calculated based on (\ref{Estm1}) in the
main paper, and its asymptotic variance is estimated by the Delta method.
The HetroGMM estimator and its asymptotic variance are calculated by (\ref%
{theta1GMM}) and (\ref{Shathat}) in the main paper. The estimation is based
on $\{y_{i1}, y_{i2},...,y_{iT}\}$ for $i=1,2,...,n$. The nominal size of
the tests is set to 5 per cent. The number of replications is $2,000$. }
\end{table}

\begin{table}[h!]
\caption{Bias, RMSE, and size of the FDAC estimator of $\protect\mu_{\protect%
\phi} = E(\protect\phi _{i})$ in a heterogeneous panel AR(1) model with
categorically distributed $\protect\phi_{i}$, Gaussian errors without GARCH
effects, and different initializations}
\label{tab:MCAc_mi}
\begin{center}
\scalebox{0.75}{
\renewcommand{\arraystretch}{1.05}
\begin{tabular}{rrcccccccccccr}
\hline\hline
&  &  & \multicolumn{3}{c}{Bias} &  & \multicolumn{3}{c}{RMSE} &  & \multicolumn{3}{c}{Size $(\times 100)$} \\ \cline{4-6} \cline{8-10} \cline{12-14}
$T$ & $n$/$M_{i}$ &  & 100 & 3 & 1 &  & 100 & 3 & 1 &  & 100 & 3 & \multicolumn{1}{c}{1} \\ \hline
\multicolumn{14}{l}{$\mu_{\phi} = 0.545$ with $\vert \phi_{i} \vert < 1$}   \\ \hline
4 & 100 &  & 0.000 & 0.010 & 0.079 &  & 0.164 & 0.168 & 0.193 &  & 7.5 & 8.0 & 11.1 \\
4 & 1,000 &  & -0.001 & 0.012 & 0.081 &  & 0.053 & 0.055 & 0.099 &  & 5.2 & 5.9 & 30.1 \\
4 & 5,000 &  & 0.000 & 0.012 & 0.082 &  & 0.025 & 0.027 & 0.085 &  & 5.1 & 7.8 & 91.1 \\
 &  &  &  &  &  &  &  &  &  &  &  &  &  \\
5 & 100 &  & 0.001 & 0.008 & 0.058 &  & 0.118 & 0.121 & 0.138 &  & 7.0 & 6.6 & 10.5 \\
5 & 1,000 &  & -0.001 & 0.009 & 0.061 &  & 0.037 & 0.040 & 0.073 &  & 4.4 & 5.8 & 33.0 \\
5 & 5,000 &  & -0.001 & 0.009 & 0.060 &  & 0.017 & 0.020 & 0.063 &  & 4.9 & 9.0 & 91.7 \\
 &  &  &  &  &  &  &  &  &  &  &  &  &  \\
6 & 100 &  & -0.001 & 0.008 & 0.046 &  & 0.099 & 0.101 & 0.112 &  & 6.9 & 6.8 & 10.2 \\
6 & 1,000 &  & 0.000 & 0.007 & 0.048 &  & 0.031 & 0.033 & 0.058 &  & 5.1 & 6.3 & 32.4 \\
6 & 5,000 &  & 0.000 & 0.008 & 0.047 &  & 0.014 & 0.016 & 0.049 &  & 4.8 & 8.1 & 89.2 \\
 &  &  &  &  &  &  &  &  &  &  &  &  &  \\
10 & 100 &  & -0.001 & 0.004 & 0.024 &  & 0.064 & 0.065 & 0.072 &  & 5.9 & 5.2 & 7.4 \\
10 & 1,000 &  & 0.000 & 0.004 & 0.025 &  & 0.021 & 0.021 & 0.034 &  & 4.2 & 5.6 & 22.4 \\
10 & 5,000 &  & 0.000 & 0.004 & 0.025 &  & 0.009 & 0.010 & 0.027 &  & 5.0 & 7.0 & 75.0 \\
 &  &  &  &  &  &  &  &  &  &  &  &  &  \\
\multicolumn{14}{l}{$\mu_{\phi} = 0.525$ with $\phi_{i} \in [-1+\epsilon, 1]$ for some $\epsilon>0$}   \\ \hline
4 & 100 &  & -0.002 & 0.003 & 0.068 &  & 0.169 & 0.165 & 0.182 &  & 7.4 & 7.6 & 8.7 \\
4 & 1,000 &  & 0.001 & 0.003 & 0.072 &  & 0.054 & 0.055 & 0.091 &  & 6.2 & 6.2 & 25.2 \\
4 & 5,000 &  & 0.001 & 0.005 & 0.073 &  & 0.025 & 0.024 & 0.077 &  & 5.8 & 5.5 & 83.9 \\
 &  &  &  &  &  &  &  &  &  &  &  &  &  \\
5 & 100 &  & -0.003 & 0.001 & 0.051 &  & 0.118 & 0.117 & 0.132 &  & 5.9 & 6.6 & 9.6 \\
5 & 1,000 &  & 0.001 & 0.002 & 0.050 &  & 0.039 & 0.039 & 0.064 &  & 6.4 & 5.4 & 25.3 \\
5 & 5,000 &  & 0.001 & 0.003 & 0.051 &  & 0.017 & 0.017 & 0.054 &  & 5.8 & 5.7 & 82.6 \\
 &  &  &  &  &  &  &  &  &  &  &  &  &  \\
6 & 100 &  & -0.001 & 0.001 & 0.037 &  & 0.097 & 0.100 & 0.106 &  & 6.4 & 7.2 & 8.2 \\
6 & 1,000 &  & 0.001 & 0.001 & 0.038 &  & 0.032 & 0.032 & 0.050 &  & 5.3 & 6.0 & 23.2 \\
6 & 5,000 &  & 0.001 & 0.002 & 0.039 &  & 0.014 & 0.014 & 0.041 &  & 6.0 & 5.5 & 76.4 \\
 &  &  &  &  &  &  &  &  &  &  &  &  &  \\
10 & 100 &  & 0.000 & 0.002 & 0.016 &  & 0.066 & 0.066 & 0.066 &  & 5.4 & 5.9 & 5.7 \\
10 & 1,000 &  & 0.001 & 0.001 & 0.020 &  & 0.021 & 0.021 & 0.029 &  & 4.6 & 5.7 & 14.9 \\
10 & 5,000 &  & 0.000 & 0.001 & 0.020 &  & 0.010 & 0.010 & 0.022 &  & 5.4 & 5.8 & 55.8 \\
\hline\hline
\end{tabular}}
\end{center}
\par
{\footnotesize Notes: The DGP is given by $y_{it}=\mu _{i}(1-\phi _{i})+\phi
_{i}y_{i,t-1}+h_{it}\varepsilon_{it}$, for $i=1,2,...,n$, and $t=-M_{i}+1,
-M_{i}+2,...,T$, featuring Gaussian standardized errors with cross-sectional
heteroskedasticity without GARCH effects. The heterogeneous AR(1)
coefficients are generated by the categorical distribution: $\func{Pr}(\phi
_{i}=\phi _{L})=\pi $ and $\func{Pr}(\phi _{i}=\phi _{H})=1-\pi $, where $%
(\phi_{H}, \phi_{L}, \pi)^{\prime} = (0.8, 0.5, 0.85)^{\prime}$ with $\vert
\phi_{i} \vert < 1$ for all $i$ and $(1, 0.5, 0.95)^{\prime}$ with $\phi_{i} \in [-1+\epsilon, 1]$ for some 
$\epsilon>0$ and all $i$. The initial values are given by $%
(y_{i,-M_{i}} - \mu_{i}) \sim IIDN(b, \kappa \sigma_{i}^{2})$ with $b=1$ and 
$\kappa=2$, where $M_{i} \in \{ 100, 3, 1\}$ for all $i$, except a case with 
$M_{i} = 100 $ for units with $\phi_{i} = \phi_{L}= 0.5$ and $M_{i} = 1$ for
units with $\phi_{i} = \phi_{H} = 1$. For each experiment, $(\alpha_{i},
\phi_{i}, \sigma_{i})^{\prime}$ are generated differently across
replications. The FDAC estimator is calculated based on (\ref{Estm1}) in the
main paper, and its asymptotic variance is estimated by the Delta method.
The HetroGMM estimator and its asymptotic variance are calculated by (\ref%
{theta1GMM}) and (\ref{Shathat}) in the main paper. The estimation is based
on $\{y_{i1}, y_{i2},...,y_{iT}\}$ for $i=1,2,...,n$. The nominal size of
the tests is set to 5 per cent. The number of replications is $2,000$. }
\end{table}

\begin{sidewaystable}
\caption{FDAC, FDLS, AH, AAH, AB, and BB estimators of $\phi$ $(\phi_{0} = 0.5)$ in a homogeneous panel AR(1) model with Gaussian errors without GARCH effects, and different initializations} 
\label{tab:homo_a_mi}
\begin{center}
\scalebox{0.6}{
{\large\begin{tabular}{l|rrrrrrrrrrrrrrrrrrrrrrrrrr}
   \hline\hline  
& & & &\multicolumn{3}{c}{FDAC} && \multicolumn{3}{c}{FDLS} &&  \multicolumn{3}{c}{AH} && \multicolumn{3}{c}{AAH} 
&& \multicolumn{3}{c}{AB} && \multicolumn{3}{c}{BB} \\  \cline{5-7} \cline{9-11} \cline{13-15} \cline{17-19} \cline{21-23} \cline{25-27}    & $T$ & $n/M_i$ &  & 100 & 3 & 1 &  & 100 & 3 & 1 &  & 100 & 3 & 1 &  & 100 & 3 & 1 &  & 100 & 3 & 1 &  & 100 & 3 & 1 \\ \hline
 & 4 & 100 &  & 0.004 & 0.007 & 0.086 &  & 0.002 & 0.009 & 0.072 &  & 0.564 & 0.412 & 0.255 &  & 0.084 & 0.090 & 0.093 &  & -0.023 & -0.026 & -0.017 &  & -0.006 & -0.016 & -0.016 \\
 & 4 & 1,000 &  & 0.000 & 0.006 & 0.090 &  & 0.001 & 0.005 & 0.069 &  & 0.021 & 0.016 & 0.027 &  & 0.072 & 0.065 & 0.073 &  & -0.004 & -0.001 & -0.003 &  & -0.002 & -0.014 & -0.023 \\
 & 4 & 5,000 &  & 0.000 & 0.006 & 0.090 &  & 0.000 & 0.005 & 0.069 &  & 0.002 & 0.004 & 0.006 &  & 0.023 & 0.023 & 0.017 &  & 0.000 & 0.001 & 0.000 &  & 0.000 & -0.014 & -0.023 \\
[2mm]
 & 6 & 100 &  & 0.001 & 0.003 & 0.046 &  & -0.001 & 0.005 & 0.038 &  & -0.072 & -0.070 & -0.086 &  & 0.023 & 0.026 & 0.028 &  & -0.041 & -0.039 & -0.032 &  & -0.001 & -0.008 & -0.011 \\
Bias & 6 & 1,000 &  & 0.000 & 0.003 & 0.049 &  & 0.000 & 0.002 & 0.037 &  & -0.007 & -0.008 & -0.013 &  & 0.001 & 0.001 & -0.001 &  & -0.006 & -0.003 & -0.003 &  & 0.000 & -0.008 & -0.011 \\
 & 6 & 5,000 &  & 0.000 & 0.003 & 0.049 &  & 0.000 & 0.003 & 0.037 &  & -0.001 & -0.001 & -0.004 &  & 0.000 & 0.000 & 0.000 &  & -0.001 & 0.000 & -0.001 &  & 0.000 & -0.008 & -0.011 \\
[2mm]
 & 10 & 100 &  & 0.000 & 0.002 & 0.024 &  & -0.001 & 0.003 & 0.019 &  & -0.049 & -0.044 & -0.053 &  & 0.006 & 0.008 & 0.005 &  & -0.035 & -0.031 & -0.030 &  & -0.004 & -0.005 & -0.009 \\
 & 10 & 1,000 &  & 0.000 & 0.002 & 0.025 &  & 0.000 & 0.001 & 0.018 &  & -0.004 & -0.005 & -0.006 &  & 0.000 & 0.000 & 0.000 &  & -0.004 & -0.003 & -0.003 &  & 0.000 & -0.003 & -0.003 \\
 & 10 & 5,000 &  & 0.000 & 0.002 & 0.025 &  & 0.000 & 0.001 & 0.019 &  & -0.001 & -0.001 & -0.002 &  & 0.000 & 0.000 & 0.000 &  & 0.000 & -0.001 & -0.001 &  & 0.000 & -0.003 & -0.003 \\\hline 
 &  &  &  &  &  &  &  &  &  &  &  &  &  &  &  &  &  &  &  &  &  &  &  &  &  &  \\ \hline
 & 4 & 100 &  & 0.169 & 0.164 & 0.185 &  & 0.150 & 0.148 & 0.171 &  & 9.211 & 9.900 & 12.229 &  & 0.298 & 0.301 & 0.292 &  & 0.226 & 0.216 & 0.185 &  & 0.141 & 0.136 & 0.135 \\
 & 4 & 1,000 &  & 0.054 & 0.053 & 0.105 &  & 0.049 & 0.048 & 0.085 &  & 0.206 & 0.201 & 0.262 &  & 0.209 & 0.198 & 0.216 &  & 0.070 & 0.066 & 0.055 &  & 0.044 & 0.047 & 0.047 \\
 & 4 & 5,000 &  & 0.024 & 0.025 & 0.093 &  & 0.022 & 0.022 & 0.072 &  & 0.086 & 0.087 & 0.109 &  & 0.116 & 0.116 & 0.106 &  & 0.031 & 0.030 & 0.025 &  & 0.020 & 0.024 & 0.029 \\
[2mm]
 & 6 & 100 &  & 0.098 & 0.095 & 0.107 &  & 0.104 & 0.103 & 0.115 &  & 0.220 & 0.210 & 0.242 &  & 0.152 & 0.156 & 0.155 &  & 0.130 & 0.124 & 0.112 &  & 0.088 & 0.085 & 0.084 \\
RMSE & 6 & 1,000 &  & 0.031 & 0.031 & 0.058 &  & 0.034 & 0.034 & 0.051 &  & 0.066 & 0.068 & 0.077 &  & 0.040 & 0.038 & 0.036 &  & 0.038 & 0.037 & 0.033 &  & 0.026 & 0.027 & 0.027 \\
 & 6 & 5,000 &  & 0.014 & 0.014 & 0.051 &  & 0.015 & 0.015 & 0.040 &  & 0.030 & 0.029 & 0.034 &  & 0.014 & 0.014 & 0.013 &  & 0.017 & 0.017 & 0.015 &  & 0.012 & 0.014 & 0.016 \\
[2mm]
 & 10 & 100 &  & 0.063 & 0.062 & 0.069 &  & 0.074 & 0.073 & 0.080 &  & 0.103 & 0.098 & 0.108 &  & 0.073 & 0.074 & 0.072 &  & 0.080 & 0.077 & 0.073 &  & 0.057 & 0.057 & 0.057 \\
 & 10 & 1,000 &  & 0.020 & 0.020 & 0.032 &  & 0.024 & 0.024 & 0.030 &  & 0.028 & 0.028 & 0.030 &  & 0.016 & 0.017 & 0.016 &  & 0.021 & 0.021 & 0.019 &  & 0.016 & 0.016 & 0.016 \\
 & 10 & 5,000 &  & 0.009 & 0.009 & 0.026 &  & 0.011 & 0.011 & 0.021 &  & 0.013 & 0.013 & 0.013 &  & 0.007 & 0.007 & 0.007 &  & 0.009 & 0.009 & 0.009 &  & 0.007 & 0.008 & 0.007 \\ \hline 
 &  &  &  &  &  &  &  &  &  &  &  &  &  &  &  &  &  &  &  &  &  &  &  &  &  &  \\ \hline
 & 4 & 100 &  & 7.7 & 7.2 & 11.1 &  & 5.9 & 5.9 & 11.9 &  & 8.3 & 7.3 & 7.3 &  & 12.7 & 12.8 & 13.1 &  & 10.0 & 8.5 & 9.1 &  & 13.2 & 12.3 & 11.5 \\
 & 4 & 1,000 &  & 6.2 & 4.8 & 41.5 &  & 5.6 & 5.4 & 31.2 &  & 4.2 & 5.0 & 5.1 &  & 18.4 & 16.7 & 17.8 &  & 5.9 & 5.3 & 5.7 &  & 5.2 & 7.0 & 7.8 \\
 & 4 & 5,000 &  & 5.4 & 6.2 & 96.8 &  & 4.9 & 5.8 & 87.7 &  & 5.2 & 5.6 & 6.0 &  & 9.2 & 8.9 & 7.0 &  & 5.0 & 5.2 & 5.8 &  & 5.0 & 9.2 & 17.8 \\
[2mm]
 & 6 & 100 &  & 8.1 & 7.0 & 10.4 &  & 6.6 & 6.0 & 10.2 &  & 15.9 & 13.9 & 16.0 &  & 22.6 & 21.1 & 22.4 &  & 16.9 & 17.0 & 16.6 &  & 21.1 & 20.0 & 20.4 \\
Size $(\times 100)$ & 6 & 1,000 &  & 4.5 & 5.9 & 37.3 &  & 5.9 & 5.2 & 20.3 &  & 6.4 & 6.0 & 7.6 &  & 6.7 & 7.2 & 5.9 &  & 6.4 & 7.1 & 6.8 &  & 6.3 & 7.8 & 8.4 \\
 & 6 & 5,000 &  & 4.9 & 5.1 & 94.0 &  & 5.3 & 4.6 & 70.3 &  & 5.4 & 5.2 & 6.4 &  & 5.2 & 4.9 & 4.2 &  & 6.5 & 5.1 & 5.1 &  & 5.8 & 10.2 & 16.0 \\
[2mm]
 & 10 & 100 &  & 5.1 & 5.6 & 8.6 &  & 6.0 & 5.1 & 8.2 &  & 34.0 & 32.7 & 35.6 &  & 44.3 & 44.8 & 43.7 &  & 39.6 & 37.3 & 38.0 &  & 43.6 & 43.4 & 47.1 \\
 & 10 & 1,000 &  & 5.0 & 5.2 & 23.4 &  & 5.1 & 5.1 & 12.0 &  & 8.5 & 8.1 & 8.2 &  & 8.8 & 9.3 & 8.8 &  & 8.7 & 8.0 & 7.9 &  & 8.6 & 8.3 & 9.8 \\
 & 10 & 5,000 &  & 5.6 & 5.1 & 77.1 &  & 4.7 & 5.1 & 42.2 &  & 5.2 & 5.8 & 5.0 &  & 5.9 & 5.4 & 6.1 &  & 6.2 & 6.1 & 5.3 &  & 5.3 & 7.8 & 8.8 \\ 
\hline\hline
\end{tabular}}}
\end{center}
{\footnotesize 
Notes: The DGP is given by $y_{it}=\mu _{i}(1-\phi _{i})+\phi
_{i}y_{i,t-1}+h_{it}\varepsilon_{it}$, for $i=1,2,...,n$, and $t=-M_{i}+1, -M_{i}+2,...,T$, 
where $\phi_{i} = 0.5$ for all $i$, featuring Gaussian standardized errors with cross-sectional heteroskedasticity without GARCH effects. 
The initial values are given by $(y_{i,-M_{i}} - \mu_{i}) \sim IIDN(b, \kappa \sigma_{i}^{2})$ 
with $b=1$ and $\kappa=2$, where $M_{i} \in \{ 100, 3, 1\}$ for all $i$. See also the notes to Table \ref{tab:homo_a}. }
\end{sidewaystable}

\begin{sidewaystable}
\caption{FDAC, FDLS, AH, AAH, AB, and BB estimators of $\mu_{\phi} = E(\phi_{i}) = 0.4$ in a heterogeneous panel AR(1) model with uniformly distributed $\vert \phi_{i}\vert < 1 $, Gaussian errors without GARCH effects, and different initializations} 
\label{tab:hetro_u1_a_mi}
\begin{center}
\scalebox{0.6}{
{\large\begin{tabular}{l|rrrrrrrrrrrrrrrrrrrrrrrrrr}
   \hline\hline  
& & & &\multicolumn{3}{c}{FDAC} && \multicolumn{3}{c}{FDLS} &&  \multicolumn{3}{c}{AH} && \multicolumn{3}{c}{AAH} 
&& \multicolumn{3}{c}{AB} && \multicolumn{3}{c}{BB} \\  \cline{5-7} \cline{9-11} \cline{13-15} \cline{17-19} \cline{21-23} \cline{25-27}    & $T$ & $n/M_i$ &  & 100 & 3 & 1 &  & 100 & 3 & 1 &  & 100 & 3 & 1 &  & 100 & 3 & 1 &  & 100 & 3 & 1 &  & 100 & 3 & 1 \\ \hline
 & 4 & 100 &  & 0.005 & 0.015 & 0.072 &  & -0.057 & -0.044 & -0.007 &  & -0.124 & -0.052 & 0.013 &  & -0.007 & 0.014 & 0.024 &  & -0.081 & -0.050 & -0.002 &  & -0.041 & -0.046 & -0.031 \\
 & 4 & 1,000 &  & -0.001 & 0.017 & 0.074 &  & -0.061 & -0.050 & -0.014 &  & -0.176 & -0.183 & -0.212 &  & -0.066 & -0.060 & -0.049 &  & -0.061 & -0.021 & 0.018 &  & -0.046 & -0.053 & -0.036 \\
 & 4 & 5,000 &  & 0.000 & 0.017 & 0.073 &  & -0.062 & -0.051 & -0.014 &  & -0.185 & -0.192 & -0.216 &  & -0.073 & -0.066 & -0.050 &  & -0.055 & -0.018 & 0.023 &  & -0.044 & -0.054 & -0.035 \\
[2mm]
 & 6 & 100 &  & 0.001 & 0.011 & 0.044 &  & -0.060 & -0.050 & -0.029 &  & -0.178 & -0.186 & -0.217 &  & -0.025 & -0.018 & -0.004 &  & -0.086 & -0.059 & -0.024 &  & -0.026 & -0.027 & -0.018 \\
Bias & 6 & 1,000 &  & 0.000 & 0.012 & 0.046 &  & -0.061 & -0.054 & -0.033 &  & -0.143 & -0.152 & -0.177 &  & -0.033 & -0.026 & -0.013 &  & -0.052 & -0.022 & 0.011 &  & -0.022 & -0.023 & -0.011 \\
 & 6 & 5,000 &  & 0.000 & 0.012 & 0.046 &  & -0.062 & -0.054 & -0.032 &  & -0.140 & -0.148 & -0.171 &  & -0.033 & -0.027 & -0.013 &  & -0.046 & -0.018 & 0.015 &  & -0.020 & -0.022 & -0.009 \\
[2mm]
 & 10 & 100 &  & 0.000 & 0.007 & 0.025 &  & -0.061 & -0.055 & -0.044 &  & -0.121 & -0.124 & -0.147 &  & -0.031 & -0.026 & -0.019 &  & -0.060 & -0.041 & -0.022 &  & -0.018 & -0.014 & -0.010 \\
 & 10 & 1,000 &  & 0.000 & 0.007 & 0.025 &  & -0.061 & -0.057 & -0.047 &  & -0.092 & -0.101 & -0.118 &  & -0.022 & -0.019 & -0.012 &  & -0.033 & -0.016 & 0.006 &  & -0.002 & -0.002 & 0.008 \\
 & 10 & 5,000 &  & 0.000 & 0.007 & 0.025 &  & -0.061 & -0.058 & -0.046 &  & -0.090 & -0.099 & -0.115 &  & -0.021 & -0.018 & -0.011 &  & -0.029 & -0.012 & 0.010 &  & 0.000 & 0.001 & 0.010 \\ \hline
 &  &  &  &  &  &  &  &  &  &  &  &  &  &  &  &  &  &  &  &  &  &  &  &  &  &  \\ \hline
 & 4 & 100 &  & 0.182 & 0.175 & 0.191 &  & 0.168 & 0.164 & 0.164 &  & 1.930 & 2.949 & 3.145 &  & 0.261 & 0.274 & 0.261 &  & 0.236 & 0.218 & 0.203 &  & 0.157 & 0.151 & 0.150 \\
 & 4 & 1,000 &  & 0.058 & 0.059 & 0.094 &  & 0.079 & 0.071 & 0.055 &  & 0.223 & 0.228 & 0.258 &  & 0.109 & 0.102 & 0.080 &  & 0.092 & 0.069 & 0.064 &  & 0.066 & 0.071 & 0.059 \\
 & 4 & 5,000 &  & 0.026 & 0.031 & 0.078 &  & 0.065 & 0.056 & 0.027 &  & 0.195 & 0.201 & 0.225 &  & 0.077 & 0.071 & 0.056 &  & 0.063 & 0.035 & 0.035 &  & 0.049 & 0.057 & 0.040 \\
[2mm]
 & 6 & 100 &  & 0.114 & 0.111 & 0.123 &  & 0.127 & 0.123 & 0.120 &  & 0.258 & 0.260 & 0.295 &  & 0.123 & 0.121 & 0.116 &  & 0.163 & 0.145 & 0.131 &  & 0.107 & 0.103 & 0.103 \\
RMSE & 6 & 1,000 &  & 0.036 & 0.038 & 0.059 &  & 0.072 & 0.065 & 0.050 &  & 0.155 & 0.164 & 0.188 &  & 0.046 & 0.041 & 0.033 &  & 0.068 & 0.049 & 0.044 &  & 0.039 & 0.039 & 0.034 \\
 & 6 & 5,000 &  & 0.016 & 0.020 & 0.049 &  & 0.064 & 0.056 & 0.036 &  & 0.142 & 0.150 & 0.173 &  & 0.036 & 0.030 & 0.019 &  & 0.050 & 0.027 & 0.024 &  & 0.024 & 0.026 & 0.017 \\
[2mm]
 & 10 & 100 &  & 0.078 & 0.077 & 0.085 &  & 0.102 & 0.100 & 0.099 &  & 0.159 & 0.161 & 0.184 &  & 0.084 & 0.084 & 0.086 &  & 0.110 & 0.099 & 0.093 &  & 0.080 & 0.079 & 0.080 \\
 & 10 & 1,000 &  & 0.025 & 0.026 & 0.036 &  & 0.067 & 0.063 & 0.054 &  & 0.098 & 0.107 & 0.123 &  & 0.033 & 0.031 & 0.027 &  & 0.044 & 0.033 & 0.030 &  & 0.024 & 0.024 & 0.025 \\
 & 10 & 5,000 &  & 0.011 & 0.013 & 0.028 &  & 0.063 & 0.059 & 0.048 &  & 0.091 & 0.100 & 0.116 &  & 0.024 & 0.021 & 0.016 &  & 0.032 & 0.018 & 0.017 &  & 0.011 & 0.011 & 0.015 \\ \hline
 &  &  &  &  &  &  &  &  &  &  &  &  &  &  &  &  &  &  &  &  &  &  &  &  &  &  \\ \hline
 & 4 & 100 &  & 8.2 & 7.6 & 8.9 &  & 8.2 & 8.8 & 8.1 &  & 15.5 & 15.2 & 16.8 &  & 18.2 & 18.8 & 14.9 &  & 14.9 & 11.5 & 10.2 &  & 16.2 & 14.7 & 13.5 \\
 & 4 & 1,000 &  & 6.7 & 5.3 & 26.5 &  & 23.2 & 16.8 & 6.9 &  & 33.6 & 34.8 & 39.9 &  & 30.8 & 26.0 & 18.6 &  & 19.4 & 7.0 & 7.2 &  & 19.8 & 21.6 & 13.9 \\
 & 4 & 5,000 &  & 5.4 & 10.0 & 82.8 &  & 78.1 & 60.1 & 9.2 &  & 85.8 & 86.8 & 89.2 &  & 81.4 & 74.9 & 54.0 &  & 46.8 & 11.8 & 13.9 &  & 56.4 & 70.9 & 37.8 \\
[2mm]
 & 6 & 100 &  & 7.4 & 6.8 & 9.7 &  & 10.3 & 8.8 & 8.3 &  & 32.8 & 34.1 & 37.4 &  & 18.4 & 17.8 & 15.4 &  & 27.7 & 20.7 & 19.4 &  & 24.4 & 23.4 & 24.8 \\
Size $(\times 100)$ & 6 & 1,000 &  & 4.1 & 6.8 & 26.8 &  & 40.8 & 31.4 & 14.4 &  & 71.5 & 75.7 & 81.2 &  & 18.9 & 13.9 & 6.8 &  & 30.0 & 12.4 & 12.1 &  & 15.5 & 16.1 & 10.6 \\
 & 6 & 5,000 &  & 5.0 & 11.8 & 81.6 &  & 96.8 & 91.5 & 49.2 &  & 100.0 & 100.0 & 100.0 &  & 60.1 & 42.4 & 12.4 &  & 70.6 & 20.2 & 19.0 &  & 31.8 & 36.5 & 11.9 \\
[2mm]
 & 10 & 100 &  & 5.2 & 6.0 & 7.7 &  & 13.1 & 11.9 & 11.7 &  & 57.5 & 58.0 & 64.4 &  & 41.5 & 42.3 & 42.0 &  & 48.1 & 43.7 & 39.8 &  & 48.4 & 49.1 & 51.3 \\
 & 10 & 1,000 &  & 4.8 & 5.9 & 17.6 &  & 63.6 & 58.7 & 39.6 &  & 85.2 & 90.2 & 94.6 &  & 17.7 & 16.0 & 9.8 &  & 32.9 & 17.5 & 14.1 &  & 11.5 & 12.6 & 14.9 \\
 & 10 & 5,000 &  & 5.2 & 8.6 & 59.7 &  & 100.0 & 99.8 & 96.0 &  & 100.0 & 100.0 & 100.0 &  & 44.8 & 35.0 & 15.6 &  & 70.9 & 23.2 & 20.2 &  & 9.6 & 8.3 & 23.6\\
\hline\hline
\end{tabular}}}
\end{center}
{\footnotesize 
Notes: The DGP is given by $y_{it}=\mu _{i}(1-\phi _{i})+\phi
_{i}y_{i,t-1}+h_{it}\varepsilon_{it}$, for $i=1,2,...,n$, and $t=-M_{i}+1, -M_{i}+2,...,T$, 
where $\phi_{i} = \mu_{\phi} + v_{i}$ with $v_{i} \sim IIDU[-a,a]$, $a=0.5$ and $\mu_{\phi} = 0.4$, featuring Gaussian standardized errors with cross-sectional heteroskedasticity without GARCH effects. 
The initial values are given by $(y_{i,-M_{i}} - \mu_{i}) \sim IIDN(b, \kappa \sigma_{i}^{2})$ 
with $b=1$ and $\kappa=2$, where $M_{i} \in \{ 100, 3, 1\}$ for all $i$. See also the notes to Table \ref{tab:hetro_u1_a}. }
\end{sidewaystable}

\begin{sidewaystable}
\caption{FDAC, FDLS, AH, AAH, AB, and BB estimators of $\mu_{\phi} = E(\phi_{i})  = 0.5$ in a heterogeneous panel AR(1) model with uniformly distributed $\phi_{i} \in [-1+\epsilon, 1]$ for some $\epsilon>0$, Gaussian errors without GARCH effects, and different initializations} 
\label{tab:hetro_u2_a_mi}
\begin{center}
\scalebox{0.6}{
{\large\begin{tabular}{l|rrrrrrrrrrrrrrrrrrrrrrrrrr}
   \hline\hline  
& & & &\multicolumn{3}{c}{FDAC} && \multicolumn{3}{c}{FDLS} &&  \multicolumn{3}{c}{AH} && \multicolumn{3}{c}{AAH} 
&& \multicolumn{3}{c}{AB} && \multicolumn{3}{c}{BB} \\  \cline{5-7} \cline{9-11} \cline{13-15} \cline{17-19} \cline{21-23} \cline{25-27}    & $T$ & $n/M_i$ &  & 100 & 3 & 1 &  & 100 & 3 & 1 &  & 100 & 3 & 1 &  & 100 & 3 & 1 &  & 100 & 3 & 1 &  & 100 & 3 & 1 \\ \hline
 & 4 & 100 &  & 0.003 & 0.014 & 0.066 &  & -0.053 & -0.040 & -0.003 &  & -0.070 & -0.926 & -0.064 &  & -0.012 & 0.003 & 0.013 &  & -0.109 & -0.075 & -0.032 &  & -0.034 & -0.043 & -0.034 \\
 & 4 & 1,000 &  & -0.001 & 0.015 & 0.069 &  & -0.057 & -0.045 & -0.009 &  & -0.190 & -0.198 & -0.229 &  & -0.062 & -0.059 & -0.050 &  & -0.072 & -0.033 & -0.005 &  & -0.043 & -0.053 & -0.044 \\
 & 4 & 5,000 &  & 0.000 & 0.016 & 0.068 &  & -0.057 & -0.046 & -0.009 &  & -0.204 & -0.210 & -0.236 &  & -0.075 & -0.070 & -0.059 &  & -0.065 & -0.030 & 0.001 &  & -0.043 & -0.054 & -0.044 \\
[2mm]
 & 6 & 100 &  & 0.001 & 0.010 & 0.040 &  & -0.056 & -0.046 & -0.025 &  & -0.224 & -0.232 & -0.267 &  & -0.021 & -0.017 & -0.007 &  & -0.119 & -0.089 & -0.057 &  & -0.022 & -0.026 & -0.022 \\
Bias & 6 & 1,000 &  & 0.000 & 0.011 & 0.043 &  & -0.057 & -0.050 & -0.028 &  & -0.171 & -0.180 & -0.206 &  & -0.034 & -0.029 & -0.019 &  & -0.074 & -0.039 & -0.010 &  & -0.021 & -0.023 & -0.015 \\
 & 6 & 5,000 &  & 0.000 & 0.011 & 0.043 &  & -0.057 & -0.050 & -0.027 &  & -0.166 & -0.174 & -0.197 &  & -0.034 & -0.029 & -0.019 &  & -0.068 & -0.033 & -0.004 &  & -0.020 & -0.021 & -0.012 \\
[2mm]
 & 10 & 100 &  & 0.000 & 0.006 & 0.022 &  & -0.057 & -0.051 & -0.040 &  & -0.164 & -0.166 & -0.189 &  & -0.024 & -0.018 & -0.012 &  & -0.086 & -0.068 & -0.050 &  & -0.016 & -0.015 & -0.012 \\
 & 10 & 1,000 &  & 0.000 & 0.006 & 0.023 &  & -0.057 & -0.053 & -0.042 &  & -0.123 & -0.132 & -0.148 &  & -0.018 & -0.015 & -0.008 &  & -0.059 & -0.036 & -0.015 &  & -0.003 & -0.001 & 0.008 \\
 & 10 & 5,000 &  & 0.000 & 0.006 & 0.023 &  & -0.057 & -0.053 & -0.041 &  & -0.120 & -0.128 & -0.144 &  & -0.017 & -0.014 & -0.007 &  & -0.058 & -0.032 & -0.010 &  & 0.000 & 0.003 & 0.012 \\ \hline
 &  &  &  &  &  &  &  &  &  &  &  &  &  &  &  &  &  &  &  &  &  &  &  &  &  &  \\ \hline
 & 4 & 100 &  & 0.181 & 0.171 & 0.186 &  & 0.171 & 0.165 & 0.167 &  & 2.255 & 42.237 & 8.912 &  & 0.263 & 0.267 & 0.260 &  & 0.305 & 0.265 & 0.236 &  & 0.166 & 0.159 & 0.159 \\
 & 4 & 1,000 &  & 0.057 & 0.057 & 0.089 &  & 0.076 & 0.069 & 0.055 &  & 0.256 & 0.260 & 0.295 &  & 0.125 & 0.115 & 0.107 &  & 0.113 & 0.084 & 0.070 &  & 0.068 & 0.074 & 0.066 \\
 & 4 & 5,000 &  & 0.025 & 0.030 & 0.072 &  & 0.062 & 0.052 & 0.025 &  & 0.216 & 0.223 & 0.249 &  & 0.080 & 0.076 & 0.064 &  & 0.076 & 0.046 & 0.031 &  & 0.050 & 0.058 & 0.049 \\
[2mm]
 & 6 & 100 &  & 0.111 & 0.108 & 0.119 &  & 0.127 & 0.123 & 0.122 &  & 0.314 & 0.314 & 0.355 &  & 0.133 & 0.128 & 0.124 &  & 0.202 & 0.176 & 0.155 &  & 0.114 & 0.107 & 0.107 \\
RMSE & 6 & 1,000 &  & 0.035 & 0.037 & 0.056 &  & 0.069 & 0.062 & 0.048 &  & 0.184 & 0.194 & 0.219 &  & 0.047 & 0.043 & 0.036 &  & 0.092 & 0.064 & 0.049 &  & 0.040 & 0.040 & 0.035 \\
 & 6 & 5,000 &  & 0.016 & 0.019 & 0.045 &  & 0.060 & 0.052 & 0.032 &  & 0.169 & 0.176 & 0.200 &  & 0.037 & 0.032 & 0.023 &  & 0.073 & 0.040 & 0.021 &  & 0.025 & 0.025 & 0.019 \\
[2mm]
 & 10 & 100 &  & 0.077 & 0.076 & 0.083 &  & 0.101 & 0.099 & 0.098 &  & 0.201 & 0.202 & 0.225 &  & 0.087 & 0.087 & 0.089 &  & 0.136 & 0.121 & 0.111 &  & 0.085 & 0.081 & 0.081 \\
 & 10 & 1,000 &  & 0.025 & 0.025 & 0.034 &  & 0.063 & 0.059 & 0.050 &  & 0.129 & 0.138 & 0.153 &  & 0.031 & 0.029 & 0.026 &  & 0.069 & 0.049 & 0.037 &  & 0.025 & 0.024 & 0.025 \\
 & 10 & 5,000 &  & 0.011 & 0.013 & 0.026 &  & 0.058 & 0.055 & 0.043 &  & 0.121 & 0.129 & 0.145 &  & 0.021 & 0.018 & 0.013 &  & 0.061 & 0.035 & 0.018 &  & 0.012 & 0.012 & 0.016 \\ \hline
 &  &  &  &  &  &  &  &  &  &  &  &  &  &  &  &  &  &  &  &  &  &  &  &  &  &  \\ \hline
 & 4 & 100 &  & 8.3 & 6.9 & 9.0 &  & 8.2 & 8.6 & 8.0 &  & 15.2 & 14.9 & 15.8 &  & 16.1 & 16.0 & 14.1 &  & 16.3 & 11.5 & 10.7 &  & 17.2 & 15.0 & 13.9 \\
 & 4 & 1,000 &  & 6.2 & 6.0 & 24.4 &  & 19.8 & 14.5 & 6.3 &  & 30.0 & 30.9 & 34.2 &  & 29.5 & 25.4 & 21.3 &  & 18.5 & 8.8 & 7.1 &  & 16.4 & 19.4 & 16.1 \\
 & 4 & 5,000 &  & 5.1 & 9.8 & 78.8 &  & 70.1 & 51.0 & 6.7 &  & 78.6 & 79.8 & 81.7 &  & 74.1 & 67.7 & 56.1 &  & 40.6 & 16.9 & 5.6 &  & 45.2 & 61.9 & 47.6 \\
[2mm]
 & 6 & 100 &  & 7.2 & 6.7 & 9.7 &  & 9.6 & 8.2 & 8.2 &  & 35.8 & 36.2 & 40.6 &  & 21.0 & 18.1 & 16.4 &  & 30.8 & 24.5 & 21.4 &  & 27.8 & 24.6 & 25.6 \\
Size $(\times 100)$ & 6 & 1,000 &  & 4.4 & 5.8 & 24.8 &  & 35.0 & 27.2 & 11.7 &  & 73.7 & 76.3 & 80.6 &  & 20.5 & 16.0 & 10.5 &  & 38.0 & 19.4 & 10.9 &  & 15.1 & 15.5 & 11.1 \\
 & 6 & 5,000 &  & 5.1 & 10.2 & 77.5 &  & 93.3 & 84.6 & 37.3 &  & 100.0 & 100.0 & 100.0 &  & 61.2 & 47.3 & 23.3 &  & 84.3 & 39.1 & 8.6 &  & 29.1 & 30.2 & 14.9 \\
[2mm]
 & 10 & 100 &  & 5.2 & 6.3 & 7.8 &  & 12.0 & 11.1 & 11.3 &  & 66.4 & 67.8 & 71.5 &  & 41.0 & 42.4 & 41.8 &  & 54.5 & 49.4 & 46.4 &  & 51.5 & 50.4 & 52.0 \\
 & 10 & 1,000 &  & 4.6 & 5.2 & 16.2 &  & 55.3 & 50.1 & 32.6 &  & 94.6 & 96.2 & 97.4 &  & 13.0 & 11.2 & 7.1 &  & 58.1 & 34.0 & 19.2 &  & 11.9 & 12.0 & 15.7 \\
 & 10 & 5,000 &  & 4.9 & 8.0 & 52.9 &  & 99.8 & 99.6 & 91.1 &  & 100.0 & 100.0 & 100.0 &  & 29.5 & 21.7 & 8.6 &  & 97.3 & 67.1 & 20.2 &  & 10.2 & 9.8 & 26.2 \\
\hline\hline
\end{tabular}}}
\end{center}
{\footnotesize 
Notes: The DGP is given by $y_{it}=\mu _{i}(1-\phi _{i})+\phi
_{i}y_{i,t-1}+h_{it}\varepsilon_{it}$, for $i=1,2,...,n$, and $t=-M_{i}+1, -M_{i}+2,...,T$, 
where $\phi_{i} = \mu_{\phi} + v_{i}$ with $v_{i} \sim IIDU[-a,a]$, $a=0.5$ and $\mu_{\phi} = 0.5$, featuring Gaussian standardized errors with cross-sectional heteroskedasticity without GARCH effects. 
The initial values are given by $(y_{i,-M_{i}} - \mu_{i}) \sim IIDN(b, \kappa \sigma_{i}^{2})$ 
with $b=1$ and $\kappa=2$, where $M_{i} \in \{ 100, 3, 1\}$ for all $i$. See also the notes to Table \ref{tab:hetro_u2_a}. }
\end{sidewaystable}

\begin{table}[h!]
\caption{Bias, RMSE, and size of FDAC and MSW estimators of $\protect\mu_{%
\protect\phi} = E(\protect\phi_{i})$ in heterogeneous and homogeneous panel
AR(1) models with Gaussian errors without GARCH effects and different
initializations}
\label{tab:msw_m1_mi}
\begin{center}
\scalebox{0.75}{
\renewcommand{\arraystretch}{1.1}
\begin{tabular}{rrrrrrrrrrrrrrrrrrrr}
\hline \hline
 &  &  & \multicolumn{5}{c}{Bias} &  & \multicolumn{5}{c}{RMSE} &  & \multicolumn{5}{c}{Size $(\times 100)$} \\ \cline{4-8} \cline{10-14} \cline{16-20}
 &  &  & \multicolumn{2}{c}{FDAC} &  & \multicolumn{2}{c}{MSW} &  & \multicolumn{2}{c}{FDAC} &  & \multicolumn{2}{c}{MSW} &  & \multicolumn{2}{c}{FDAC} &  & \multicolumn{2}{c}{MSW} \\ \cline{4-5} \cline{7-8} \cline{10-11} \cline{13-14} \cline{16-17} \cline{19-20}
$T$ & $n/M_{i}$ &  & \multicolumn{1}{c}{100} & \multicolumn{1}{c}{1} &  & \multicolumn{1}{c}{100} & \multicolumn{1}{c}{1} &  & \multicolumn{1}{c}{100} & \multicolumn{1}{c}{1} &  & \multicolumn{1}{c}{100} & \multicolumn{1}{c}{1} &  & \multicolumn{1}{c}{100} & \multicolumn{1}{c}{1} &  & \multicolumn{1}{c}{100} & \multicolumn{1}{c}{1} \\ \hline
\multicolumn{20}{l}{$\mu_{\phi} = 0.4$ with uniformly distributed $\vert\phi_{i} \vert < 1$ for all $i$} \\ \hline
4 & 100 &  & -0.005 & 0.054 &  & -0.145 & -0.092 &  & 0.177 & 0.190 &  & 0.157 & 0.113 &  & 8.0 & 7.8 &  & 82.0 & 42.4 \\
4 & 1,000 &  & 0.000 & 0.062 &  & -0.128 & -0.082 &  & 0.056 & 0.086 &  & 0.130 & 0.086 &  & 4.7 & 18.1 &  & 100.0 & 98.3 \\
[2mm]
6 & 100 &  & -0.004 & 0.038 &  & -0.144 & -0.073 &  & 0.113 & 0.121 &  & 0.155 & 0.097 &  & 5.7 & 8.7 &  & 79.3 & 27.6 \\
6 & 1,000 &  & -0.001 & 0.038 &  & -0.129 & -0.064 &  & 0.037 & 0.053 &  & 0.130 & 0.068 &  & 6.3 & 16.7 &  & 100.0 & 91.5 \\
[2mm]
10 & 100 &  & -0.001 & 0.021 &  & -0.146 & -0.059 &  & 0.079 & 0.084 &  & 0.158 & 0.090 &  & 6.4 & 7.3 &  & 71.2 & 17.5 \\
10 & 1,000 &  & 0.000 & 0.020 &  & -0.141 & -0.055 &  & 0.026 & 0.033 &  & 0.143 & 0.060 &  & 5.7 & 12.4 &  & 100.0 & 76.1\\
 &  &  &  &  &  &  &  &  &  &  &  &  &  &  &  &  &  &  &  \\
\multicolumn{20}{l}{$\mu_{\phi} = 0.5$ with uniformly distributed $\phi_{i} \in [-1+\epsilon, 1]$ for some $\epsilon>0$ and all $i$} \\ \hline
4 & 100 &  & -0.005 & 0.049 &  & -0.207 & -0.130 &  & 0.175 & 0.187 &  & 0.221 & 0.148 &  & 8.7 & 7.7 &  & 84.3 & 54.4 \\
4 & 1,000 &  & 0.000 & 0.057 &  & -0.194 & -0.119 &  & 0.056 & 0.081 &  & 0.196 & 0.122 &  & 5.0 & 15.3 &  & 100.0 & 99.9 \\
[2mm]
6 & 100 &  & -0.004 & 0.031 &  & -0.202 & -0.102 &  & 0.111 & 0.118 &  & 0.215 & 0.124 &  & 5.5 & 7.4 &  & 81.2 & 36.2 \\
6 & 1,000 &  & -0.001 & 0.034 &  & -0.187 & -0.091 &  & 0.036 & 0.050 &  & 0.189 & 0.094 &  & 5.2 & 16.2 &  & 100.0 & 97.9 \\
[2mm]
10 & 100 &  & -0.001 & 0.015 &  & -0.198 & -0.077 &  & 0.079 & 0.083 &  & 0.213 & 0.107 &  & 6.7 & 6.4 &  & 74.3 & 21.2 \\
10 & 1,000 &  & -0.001 & 0.017 &  & -0.192 & -0.072 &  & 0.025 & 0.031 &  & 0.194 & 0.076 &  & 5.9 & 10.4 &  & 100.0 & 87.1\\
 &  &  &  &  &  &  &  &  &  &  &  &  &  &  &  &  &  &  &  \\
\multicolumn{20}{l}{$\phi_{i} = \mu_{\phi} = 0.5$ for all $i$} \\ \hline
4 & 100 &  & 0.002 & 0.067 &  & -0.201 & -0.155 &  & 0.166 & 0.184 &  & 0.208 & 0.164 &  & 7.4 & 10.4 &  & 98.1 & 86.7 \\
4 & 1,000 &  & -0.001 & 0.080 &  & -0.182 & -0.135 &  & 0.054 & 0.097 &  & 0.183 & 0.137 &  & 5.7 & 29.7 &  & 100.0 & 100.0 \\
[2mm]
6 & 100 &  & 0.002 & 0.038 &  & -0.199 & -0.132 &  & 0.097 & 0.105 &  & 0.205 & 0.143 &  & 7.4 & 9.0 &  & 98.5 & 75.0 \\
6 & 1,000 &  & -0.001 & 0.041 &  & -0.185 & -0.121 &  & 0.031 & 0.052 &  & 0.186 & 0.123 &  & 4.4 & 25.8 &  & 100.0 & 100.0 \\
[2mm]
10 & 100 &  & 0.002 & 0.017 &  & -0.198 & -0.111 &  & 0.065 & 0.067 &  & 0.205 & 0.126 &  & 7.7 & 7.4 &  & 95.6 & 50.5 \\
10 & 1,000 &  & -0.001 & 0.020 &  & -0.194 & -0.106 &  & 0.020 & 0.029 &  & 0.195 & 0.108 &  & 4.8 & 16.4 &  & 100.0 & 99.9\\
\hline\hline
\end{tabular}}
\end{center}
\par
{\footnotesize Notes: The DGP is given by $y_{it}=\mu _{i}(1-\phi _{i})+\phi
_{i}y_{i,t-1}+h_{it}\varepsilon_{it}$, for $i=1,2,...,n$, and $t=-M_{i}+1,
-M_{i}+2,...,T$, featuring Gaussian standardized errors with cross-sectional
heteroskedasticity without GARCH effects. The heterogeneous AR(1)
coefficients are generated by uniform distributions: $\phi_{i} = \mu_{\phi}
+ v_{i}$, with $v_{i} \sim IIDU[-a,a]$, $a=0.5$ and $\mu_{\phi} \in \{0.4,
0.5\}$. In the homogeneous case, $\phi_{i} = \mu_{\phi} = 0.5$ for all $i$.
The initial values are given by $(y_{i,-M_{i}} - \mu_{i}) \sim IIDN(b,
\kappa \sigma_{i}^{2})$ with $b=1$ and $\kappa=2$, where $M_{i} \in
\{100,1\} $ for all $i$. For each experiment, $(\alpha_{i}, \phi_{i},
\sigma_{i})^{\prime}$ are generated differently across replications. The
FDAC estimator is calculated by (\ref{Estm1}) in the main paper, and its
asymptotic variance is estimated by the Delta method. \textquotedblleft MSW"
denotes the estimator proposed by \cite{MavroeidisEtal2015}. The estimation
is based on $\{y_{i1}, y_{i2},...,y_{iT}\}$ for $i=1,2,...,n$. The nominal
size of the tests is set to 5 per cent. Due to the extensive computations
required for the implementation of the MSW estimator, the number of
replications is $1,000$. }
\end{table}

\clearpage

\subsection{Empirical results for other sub-periods of the PSID \label%
{subpsid}}

Table \ref{tab:distyear} shows the distribution of cross-sectional
observation numbers by year based on the sample selection criterion in \cite%
{MeghirPistaferri2004}. For different sub-periods, Tables \ref{tab:PSIDm1t5}
and \ref{tab:PSIDm1t10} report the estimates of mean persistence of log real
earnings in a panel AR(1) model with a common linear trend, and Tables \ref%
{tab:PSIDvar}--\ref{tab:PSIDvart5} report the estimates of $%
\sigma_{\phi}^{2} $ of the heterogeneous persistence parameters, $\phi_{i}$.

\begin{table}[htp]
\caption{Distribution of individual observation numbers by year}
\label{tab:distyear}
\begin{center}
\scalebox{0.9}{
\renewcommand{\arraystretch}{1.05}
\begin{tabular}{cc}
\hline\hline
Year & Number of observations \\ \hline
1976 & 1,600 \\
1977 & 1,663 \\
1978 & 1,706 \\
1979 & 1,773 \\
1980 & 1,800 \\
1981 & 1,868 \\
1982 & 1,884 \\
1983 & 1,933 \\
1984 & 1,972 \\
1985 & 2,012 \\
1986 & 2,053 \\
1987 & 2,083 \\
1988 & 2,091 \\
1989 & 2,008 \\
1990 & 1,907 \\
1991 & 1,831 \\
1992 & 1,711 \\
1993 & 1,576 \\
1994 & 1,471 \\
1995 & 1,384 \\\hline
Total & 36,325$\,\,$ \\
\hline\hline
\end{tabular}
}
\end{center}
\par
{\footnotesize Notes: The sample selection criteria of \cite%
{MeghirPistaferri2004} are summarized as the following. (i) Individuals are
from the \textquotedblleft core" sample, i.e., the 1968 SRC cross-section
sample and the 1968 Census sample. (ii) Individuals are continuously heads
of their families. (iii) Over the respective observed period, the range of
individuals' ages is 25 to 55. (iv) Individuals are males. (v) Individuals
have nine years or more observations of usable (non-zero and not top-coded)
money income of labor $earnings_{it}$. (vi) Individuals have no missing
records of education or race over their sample periods. (vii) Observations
with only self-employed status are dropped. (viii) Observations of outcome
variables $y_{it} = log(earnings_{it}/p_{t})$ with outlying deviations $%
\Delta y_{it} >5$ or $\Delta y_{it} <-1$ are dropped. }
\end{table}

\clearpage\newpage 
\begin{sidewaystable}
\caption{Estimates of mean persistence ($\mu_{\phi} = E(\phi_{i})$) of log real earnings in a panel AR(1) model with a common linear trend using the PSID data over the sub-periods 1976--1980, 1981--1985 and 1986--1990}
\label{tab:PSIDm1t5}
\begin{center}
\scalebox{0.8}{
\renewcommand{\arraystretch}{1.05}
\begin{tabular}{cccccccccccccccccc}
\hline \hline 
 & \multicolumn{5}{c}{1976--1980, $T=5$} &  & \multicolumn{5}{c}{1981--1985, $T=5$} &  & \multicolumn{5}{c}{1986--1990, $T=5$} \\ \cline{2-6} \cline{8-12} \cline{14-18}
 & All &  & \multicolumn{3}{c}{Category by education} &  & All &  & \multicolumn{3}{c}{Category by education} &  & All &  & \multicolumn{3}{c}{Category by education} \\ \cline{4-6} \cline{10-12} \cline{16-18}
 & categories &  & HSD & HSG & CLG &  & categories &  & HSD & HSG & CLG &  & categories &  & HSD & HSG & CLG \\ \hline
Homogeneous slopes &  &  &  &  &  &  &  &  &  &  &  &  &  &  &  &  &  \\
AAH & 0.527 &  & 0.545 & 0.489 & 0.560 &  & 0.481 &  & 0.426 & 0.465 & 0.598 &  & 0.499 &  & 0.725 & 0.426 & 0.491 \\
 & (0.051) &  & (0.079) & (0.084) & (0.070) &  & (0.038) &  & (0.083) & (0.046) & (0.072) &  & (0.035) &  & (0.093) & (0.041) & (0.065) \\
AB & 0.326 &  & 0.346 & 0.076 & 0.623 &  & 0.219 &  & 0.286 & 0.178 & -0.066 &  & 0.281 &  & 0.239 & 0.305 & 0.131 \\
 & (0.109) &  & (0.151) & (0.148) & (0.207) &  & (0.071) &  & (0.092) & (0.092) & (0.214) &  & (0.089) &  & (0.303) & (0.100) & (0.171) \\
BB & 0.905 &  & 0.916 & 0.898 & 0.916 &  & 0.957 &  & 0.939 & 0.962 & 1.041 &  & 0.939 &  & 0.897 & 0.929 & 0.978 \\
 & (0.012) &  & (0.015) & (0.015) & (0.028) &  & (0.005) &  & (0.009) & (0.006) & (0.014) &  & (0.011) &  & (0.027) & (0.012) & (0.014) \\
 &  &  &  &  &  &  &  &  &  &  &  &  &  &  &  &  &  \\
Heterogeneous slopes &  &  &  &  &  &  &  &  &  &  &  &  &  &  &  &  &  \\
FDAC & 0.589 &  & 0.567 & 0.595 & 0.607 &  & 0.602 &  & 0.428 & 0.596 & 0.844 &  & 0.675 &  & 0.760 & 0.604 & 0.805 \\
 & (0.037) &  & (0.062) & (0.056) & (0.079) &  & (0.039) &  & (0.076) & (0.053) & (0.056) &  & (0.032) &  & (0.083) & (0.042) & (0.056) \\
MSW & 0.419 &  & 0.388 & 0.434 & 0.452 &  & 0.420 &  & 0.378 & 0.439 & 0.452 &  & 0.429 &  & 0.427 & 0.427 & 0.450 \\
 & (0.060) &  & (0.058) & (0.045) & (0.030) &  & (0.058) &  & (0.055) & (0.031) & (0.031) &  & (0.056) &  & (0.048) & (0.056) & (0.046) \\
 &  &  &  &  &  &  &  &  &  &  &  &  &  &  &  &  &  \\
Common linear trend & 0.023 &  & 0.029 & 0.021 & 0.021 &  & 0.025 &  & 0.036 & 0.019 & 0.032 &  & 0.018 &  & 0.009 & 0.021 & 0.014 \\ \hline
$n$ & 1,312 &  & 363 & 641 & 308 &  & 1,489 &  & 283 & 855 & 351 &  & 1,654 &  & 201 & 994 & 459 \\
\hline\hline
\end{tabular}}
\end{center}
\par
{\footnotesize 
Notes: The estimates are based on the heterogeneous panel AR(1) model with a common linear trend, $y_{it}=\alpha_{i}+g(1-\phi_{i})t+\phi_{i} y_{i,t-1}+u_{it}$, where $y_{it}=log(earnings_{it}/p_{t})$ using the PSID data over the sub-periods 1976--1980, 1981--1985, and 1986--1990. 
``HSD" refers to high school dropouts with less than 12 years of education, ``HSG" refers to high school graduates with at least 12 but less than 16 years of education, and ``CLG" refers to college graduates with at least 16 years of education. 
The common trend, $g$, is estimated by $\hat{g}_{FD} = n^{-1}(T-1)^{-1} \sum_{i=1}^{n} \sum_{t=2}^{T} \Delta y_{it}$. Then the estimation for $\mu_{\phi}$ is based on $\tilde{y}_{it} = y_{it} - \hat{g}_{FD} t$ for $t=1,2,...,T$. 
\textquotedblleft AAH", \textquotedblleft AB", and \textquotedblleft BB" denote different 2-step GMM estimators proposed by \cite{ChudikPesaran2021}, \cite{ArellanoBond1991}, and \cite{BlundellBond1998}. The FDAC estimator is calculated by (\ref{Estm1}) in the main paper, and its asymptotic variance is estimated by the Delta method. \textquotedblleft MSW" denotes the kernel-weighted estimator in \cite{MavroeidisEtal2015} and is calculated based on a parametric assumption that $(\alpha_{i}, \phi_{i}) | y_{i1}$ follows a multivariate normal distribution $N(\boldsymbol{\mu}, \boldsymbol{V})$ with initial values given by $\boldsymbol{\mu}= (5,0.5)$, $\sigma_{\alpha} =2$, $\sigma_{\phi} = 0.4$, $corr(\alpha_{i},\phi_{i}) = 0.5$ with $\sigma_{u} = 0.5$. }
\end{sidewaystable}

\begin{table}[tbp]
\caption{Estimates of mean persistence ($\protect\mu_{\protect\phi} = E(%
\protect\phi_{i})$) of log real earnings in a panel AR(1) model with a
common linear trend using the PSID data over the sub-periods 1976--1985 and
1981--1990}
\label{tab:PSIDm1t10}
\begin{center}
\scalebox{0.8}{
\renewcommand{\arraystretch}{1.05}
\begin{tabular}{cccccccccccc}
\hline \hline
 & \multicolumn{5}{c}{1976--1985, $T=10$} &  & \multicolumn{5}{c}{1981--1990, $T=10$} \\ \cline{2-6} \cline{8-12}
 & All &  & \multicolumn{3}{l}{Category by education} &  & All &  & \multicolumn{3}{l}{Category by education} \\ \cline{4-6} \cline{10-12}
 & categories &  & HSD & HSG & CLG &  & categories &  & HSD & HSG & CLG \\ \hline
Homogeneous slopes &  &  &  &  &  &  &  &  &  &  &  \\
AAH & 0.615 &  & 0.532 & 0.587 & 0.632 &  & 0.579 &  & 0.545 & 0.529 & 0.654 \\
 & (0.044) &  & (0.040) & (0.045) & (0.027) &  & (0.030) &  & (0.038) & (0.027) & (0.043) \\
AB & 0.471 &  & 0.402 & 0.391 & 0.348 &  & 0.265 &  & 0.261 & 0.273 & 0.388 \\
 & (0.048) &  & (0.054) & (0.061) & (0.051) &  & (0.041) &  & (0.053) & (0.038) & (0.059) \\
BB & 0.960 &  & 0.922 & 0.962 & 1.001 &  & 0.958 &  & 0.956 & 0.961 & 0.978 \\
 & (0.002) &  & (0.004) & (0.002) & (0.002) &  & (0.002) &  & (0.002) & (0.002) & (0.002) \\
 &  &  &  &  &  &  &  &  &  &  &  \\
Heterogeneous slopes &  &  &  &  &  &  &  &  &  &  &  \\
FDAC & 0.643 &  & 0.554 & 0.637 & 0.766 &  & 0.628 &  & 0.614 & 0.600 & 0.734 \\
 & (0.028) &  & (0.052) & (0.041) & (0.054) &  & (0.025) &  & (0.057) & (0.033) & (0.042) \\
MSW & 0.443 &  & 0.397 & 0.443 & 0.474 &  & 0.458 &  & 0.453 & 0.446 & 0.541 \\
 & (0.060) &  & (0.047) & (0.067) & (0.062) &  & (0.030) &  & (0.041) & (0.025) & (0.064) \\
 &  &  &  &  &  &  &  &  &  &  &  \\
Common linear trend & 0.024 &  & 0.026 & 0.021 & 0.029 &  & 0.023 &  & 0.031 & 0.019 & 0.025 \\ \hline
$n$ & 885 &  & 201 & 458 & 226 &  & 1,046 &  & 170 & 620 & 256
\\
\hline\hline
\end{tabular}}
\end{center}
\par
{\footnotesize Notes: The estimates are based on the heterogeneous panel
AR(1) model with a common linear trend, $y_{it}=\alpha_{i}+g(1-\phi_{i})t+%
\phi_{i} y_{i,t-1}+u_{it}$, where $y_{it}=log(earnings_{it}/p_{t})$ using
the PSID data over the sub-periods 1976--1985 and 1981--1990. ``HSD" refers
to high school dropouts with less than 12 years of education, ``HSG" refers
to high school graduates with at least 12 but less than 16 years of
education, and ``CLG" refers to college graduates with at least 16 years of
education. The common trend, $g$, is estimated by $\hat{g}_{FD} =
n^{-1}(T-1)^{-1} \sum_{i=1}^{n} \sum_{t=2}^{T} \Delta y_{it}$. Then the
estimation for $\mu_{\phi}$ is based on $\tilde{y}_{it} = y_{it} - \hat{g}%
_{FD} t$ for $t=1,2,...,T$. \textquotedblleft AAH", \textquotedblleft AB",
and \textquotedblleft BB" denote different 2-step GMM estimators proposed by 
\cite{ChudikPesaran2021}, \cite{ArellanoBond1991}, and \cite%
{BlundellBond1998}. The FDAC estimator is calculated by (\ref{Estm1}) in the main paper, and
its asymptotic variance is estimated by the Delta method. \textquotedblleft
MSW" denotes the kernel-weighted estimator in \cite{MavroeidisEtal2015} and
is calculated based on a parametric assumption that $(\alpha_{i}, \phi_{i})
| y_{i1}$ follows a multivariate normal distribution $N(\boldsymbol{\mu}, 
\boldsymbol{V})$ with initial values given by $\boldsymbol{\mu}= (5,0.5)$, $%
\sigma_{\alpha} =2$, $\sigma_{\phi} = 0.4$, $corr(\alpha_{i},\phi_{i}) = 0.5$
with $\sigma_{u} = 0.5$. }.
\end{table}

\begin{table}[tbp]
\caption{Estimates of variance of heterogeneous persistence ($\protect\sigma%
_{\protect\phi}^{2}$) of log real earnings in a panel AR(1) model with a
common linear trend using the PSID data over the sub-periods 1991--1995 and
1986--1995}
\label{tab:PSIDvar}
\begin{center}
\scalebox{0.8}{
\renewcommand{\arraystretch}{1.05}
\begin{tabular}{cccccccccccc}
\hline \hline
 & \multicolumn{5}{c}{1991--1995, $T=5$} &  & \multicolumn{5}{c}{1986--1995, $T=10$} \\ \cline{2-6} \cline{8-12}
 & All &  & \multicolumn{3}{l}{Category by education} &  & All &  & \multicolumn{3}{l}{Category by education} \\ \cline{4-6} \cline{10-12}
 & categories &  & HSD & HSG & CLG &  & categories &  & HSD & HSG & CLG \\ \hline 
FDAC & 0.100 &  & 0.204 & 0.081 & 0.091 &  & 0.129 &  & 0.122 & 0.120 & 0.141 \\
 & (0.042) &  & (0.100) & (0.054) & (0.090) &  & (0.023) &  & (0.060) & (0.031) & (0.036) \\
MSW & 0.012 &  & 0.011 & 0.011 & 0.010 &  & 0.015 &  & 0.010 & 0.011 & 0.014 \\
 & (0.003) &  & (0.009) & (0.004) & (0.007) &  & (0.005) &  & (0.011) & (0.005) & (0.011)\\ \hline 
$n$ & 1,366 &  & 127 & 832 & 407 &  & 1,139 &  & 109 & 689 & 341
\\
\hline\hline
\end{tabular}}
\end{center}
\par
{\footnotesize Notes: The estimates are based on the heterogeneous panel
AR(1) model with a common linear trend, $y_{it}=\alpha_{i}+g(1-\phi_{i})t+%
\phi_{i} y_{i,t-1}+u_{it}$, where $y_{it}=log(earnings_{it}/p_{t})$ using
the PSID data over the sub-periods 1991--1995 and 1986--1995. ``HSD" refers
to high school dropouts with less than 12 years of education, ``HSG" refers
to high school graduates with at least 12 but less than 16 years of
education, and ``CLG" refers to college graduates with at least 16 years of
education. The common trend, $g$, is estimated by $\hat{g}_{FD} =
n^{-1}(T-1)^{-1} \sum_{i=1}^{n} \sum_{t=2}^{T} \Delta y_{it}$. Then the
estimation for $\sigma_{\phi}^{2}$ is based on $\tilde{y}_{it} = y_{it} - 
\hat{g}_{FD} t$ for $t=1,2,...,T$. The The FDAC estimator of $%
\sigma_{\phi}^{2}$ is calculated by (\ref{Estvar}) in the main paper, and its asymptotic
variance is estimated by the Delta method. \textquotedblleft MSW" denotes
the kernel-weighted maximum likelihood estimator in \cite{MavroeidisEtal2015}%
. }
\end{table}

\begin{table}[tbp]
\caption{Estimates of variance of heterogeneous persistence ($\protect\sigma%
_{\protect\phi}^{2}$) of log real earnings in a panel AR(1) model with a
common linear trend using the PSID data over the sub-periods 1976--1985 and
1981--1990}
\label{tab:PSIDvart10}
\begin{center}
\scalebox{0.8}{
\renewcommand{\arraystretch}{1.05}
\begin{tabular}{cccccccccccc}
\hline \hline
 & \multicolumn{5}{c}{1976--1985, $T=10$} &  & \multicolumn{5}{c}{1981--1990, $T=10$} \\ \cline{2-6} \cline{8-12}
 & All &  & \multicolumn{3}{l}{Category by education} &  & All &  & \multicolumn{3}{l}{Category by education} \\ \cline{4-6} \cline{10-12}
 & categories &  & HSD & HSG & CLG &  & categories &  & HSD & HSG & CLG \\ \hline
FDAC & 0.095 &  & 0.139 & 0.100 & 0.001 &  & 0.150 &  & 0.104 & 0.171 & 0.113 \\
 & (0.028) &  & (0.049) & (0.043) & (0.046) &  & (0.022) &  & (0.058) & (0.026) & (0.046) \\
MSW & 0.016   &  & 0.013   & 0.013   & 0.013   &  & 0.011   &  & 0.011   & 0.010   & 0.014   \\
    & (0.007) &  & (0.010) & (0.010) & (0.013) &  & (0.003) &  & (0.008) & (0.003) & (0.012) \\ \hline
$n$ & 885 &  & 201 & 458 & 226 &  & 1,046 &  & 170 & 620 & 256 
\\
\hline\hline
\end{tabular}}
\end{center}
\par
{\footnotesize Notes: The estimates are based on the heterogeneous panel
AR(1) model with a common linear trend, $y_{it}=\alpha_{i}+g(1-\phi_{i})t+%
\phi_{i} y_{i,t-1}+u_{it}$, where $y_{it}=log(earnings_{it}/p_{t})$ using
the PSID data over the sub-periods 1976--1985 and 1981--1990. ``HSD" refers
to high school dropouts with less than 12 years of education, ``HSG" refers
to high school graduates with at least 12 but less than 16 years of
education, and ``CLG" refers to college graduates with at least 16 years of
education. The common trend, $g$, is estimated by $\hat{g}_{FD} =
n^{-1}(T-1)^{-1} \sum_{i=1}^{n} \sum_{t=2}^{T} \Delta y_{it}$. Then the
estimation for $\sigma_{\phi}^{2}$ is based on $\tilde{y}_{it} = y_{it} - 
\hat{g}_{FD} t$ for $t=1,2,...,T$. The FDAC estimator is calculated by (\ref%
{Estvar}) in the main paper, and its asymptotic variance is estimated by the Delta method.
\textquotedblleft MSW" denotes the estimator proposed by \cite%
{MavroeidisEtal2015}. See also the notes to Table \ref{tab:PSIDm1t10}. }
\end{table}

\begin{sidewaystable}
\caption{Estimates of variance of heterogeneous persistence ($\sigma_{\phi}^{2}$) of log real earnings in a panel AR(1) model with a common linear trend using the PSID data over the sub-periods 1976--1980, 1981--1985, and 1986--1990}
\label{tab:PSIDvart5}
\begin{center}
\scalebox{0.8}{
\renewcommand{\arraystretch}{1.05}
\begin{tabular}{cccccccccccccccccc}
\hline \hline 
 & \multicolumn{5}{c}{1976--1980, $T=5$} &  & \multicolumn{5}{c}{1981--1985, $T=5$} &  & \multicolumn{5}{c}{1986--1990, $T=5$} \\ \cline{2-6} \cline{8-12} \cline{14-18}
 & All &  & \multicolumn{3}{c}{Category by education} &  & All &  & \multicolumn{3}{c}{Category by education} &  & All &  & \multicolumn{3}{c}{Category by education} \\ \cline{4-6} \cline{10-12} \cline{16-18}
 & categories &  & HSD & HSG & CLG &  & categories &  & HSD & HSG & CLG &  & categories &  & HSD & HSG & CLG \\ \hline
FDAC & 0.038 &  & 0.072 & 0.025 & 0.013 &  & 0.089 &  & 0.040 & 0.093 & 0.032 &  & 0.126 &  & 0.095 & 0.111 & 0.151 \\
 & (0.056) &  & (0.078) & (0.099) & (0.098) &  & (0.037) &  & (0.068) & (0.052) & (0.072) &  & (0.040) &  & (0.105) & (0.056) & (0.054) \\
MSW & 0.015 &  & 0.014 & 0.013 & 0.009 &  & 0.015 &  & 0.014 & 0.010 & 0.009 &  & 0.015 &  & 0.011 & 0.013 & 0.011 \\
 & (0.004) &  & (0.008) & (0.005) & (0.006) &  & (0.004) &  & (0.008) & (0.002) & (0.005) &  & (0.004) &  & (0.008) & (0.004) & (0.007) \\ \hline
$n$ & 1,312 &  & 363 & 641 & 308 &  & 1,489 &  & 283 & 855 & 351 &  & 1,654 &  & 201 & 994 & 459
\\
\hline\hline
\end{tabular}}
\end{center}
\par
{\footnotesize 
Notes: The estimates are based on the heterogeneous panel AR(1) model with a common linear trend, $y_{it}=\alpha_{i}+g(1-\phi_{i})t+\phi_{i} y_{i,t-1}+u_{it}$, where $y_{it}=log(earnings_{it}/p_{t})$ using the PSID data over the sub-periods 1976--1980, 1981--1985, and 1986--1990. 
``HSD" refers to high school dropouts with less than 12 years of education, ``HSG" refers to high school graduates with at least 12 but less than 16 years of education, and ``CLG" refers to college graduates with at least 16 years of education. 
The common trend, $g$, is estimated by $\hat{g}_{FD} = n^{-1}(T-1)^{-1} \sum_{i=1}^{n} \sum_{t=2}^{T} \Delta y_{it}$. Then the estimation for $\sigma_{\phi}^{2}$ is based on $\tilde{y}_{it} = y_{it} - \hat{g}_{FD} t$ for $t=1,2,...,T$. 
The FDAC estimator is calculated by (\ref{Estvar}) in the main paper, and its
asymptotic variance is estimated by the Delta method. \textquotedblleft MSW"
denotes the estimator proposed by \cite{MavroeidisEtal2015}. 
See also the notes to Table \ref{tab:PSIDm1t5}. 
}
\end{sidewaystable}

\clearpage\newpage \noindent {\Large \textbf{References} }

{\small 
\noindent Anderson, T. W., and Hsiao, C. (1981). \href{https://doi.org/10.2307/2287517}%
{Estimation of dynamic models with error components}. \textit{Journal of the
American Statistical Association} 76, 598-606. 
}

{\small \noindent Anderson, T. W., and Hsiao, C. (1982). \href{https://doi.org/10.1016/0304-4076(82)90095-1}%
{Formulation and estimation of dynamic models using panel data}. \textit{%
Journal of Econometrics} 18, 47-82. }

{\small \noindent Arellano, M. and S. Bond (1991). \href{https://doi.org/10.2307/2297968}%
{Some tests of specification for panel data: Monte Carlo evidence and an
application to employment equations}. \textit{The Review of Economic Studies}
58, 277--297. }

{\small \noindent Blundell, R. and S. Bond (1998). \href{https://doi.org/10.1016/S0304-4076(98)00009-8}%
{Initial conditions and moment restrictions in dynamic panel data models}. 
\textit{Journal of Econometrics}, 87, 115--143. }

{\small \noindent Chudik, A. and M. H. Pesaran (2021). \href{https://www.tandfonline.com/doi/abs/10.1080/07474938.2021.1971388?casa_token=2BwaEs6bQLoAAAAA:nYnfFDroaEJJN6-jZ4cm664YF0L-vppanHVwgPzw65e0Be9uIdHIywoaM4t36lsLXIDy1paeBHl-}%
{An augmented Anderson-Hsiao estimator for dynamic short-$T$ panels.} 
\textit{Econometric Reviews} 1--32. }

{\small \noindent Han, C. and P. C. Phillips (2010). \href{https://doi.org/10.1017/S026646660909063X}%
{GMM estimation for dynamic panels with fixed effects and strong instruments
at unity}. \textit{Econometric Theory} 26, 119--151. }

{\small \noindent Mavroeidis, S., Y. Sasaki, and I. Welch (2015). \href{https://www.sciencedirect.com/science/article/pii/S0304407615001517?casa_token=HPGr9y-Oe7cAAAAA:u5R7ML9kmRODINHPmblJlnUV3z9-NQa4-xYGU9BBxm0dMTVqiwpp4Vh197-wAa-GWYSkgxvgjQ}%
{Estimation of heterogeneous autoregressive parameters with short panel data.%
} \textit{Journal of Econometrics} 188, 219--235. }

{\small \noindent Meghir, C. and L. Pistaferri (2004). \href{https://onlinelibrary.wiley.com/doi/abs/10.1111/j.1468-0262.2004.00476.x?casa_token=45Vh36Scuo8AAAAA:16mrkums6ZB4tQCcGjD8X1ETJCNKBOn-Fr1om-Kt4oywqd7zbvZshPd8FXk3Q4A0JTWSWoDkiIiLVb8}%
{Income variance dynamics and heterogeneity}. \textit{Econometrica} 72,
1--32. }

\end{document}